%% file: Main.tex
\documentclass[acmsmall]{acmart}

\usepackage{natbib}

\usepackage{amsmath}
\usepackage{algorithmic}
\usepackage{graphicx}
\usepackage{textcomp}
\usepackage{color}
\usepackage{xcolor}
\usepackage{makecell}
\usepackage{enumitem}
\usepackage{wrapfig}

\usepackage{amssymb}

\usepackage{amsfonts}

\usepackage{xspace}
\usepackage{multirow}
\usepackage{framed}
\usepackage{tcolorbox}
\usepackage{diagbox}
\usepackage{pifont}
\usepackage{subcaption}

\usepackage{colortbl}
\usepackage{url} 
\usepackage{listings}

\usepackage{tikz}

\usepackage{float}
\newfloat{codelist}{t}{lop}

\AtBeginDocument{%
  \providecommand\BibTeX{{%
    \normalfont B\kern-0.5em{\scshape i\kern-0.25em b}\kern-0.8em\TeX}}}

\setcopyright{acmlicensed}
\copyrightyear{2025}
\acmYear{2025}
\acmDOI{XXXXXXX.XXXXXXX}

\acmISBN{978-1-4503-XXXX-X/XX/XX}

\acmJournal{TOSEM}
\acmVolume{1}
\acmNumber{1}
\acmArticle{1}
\acmMonth{1}

\input{macro}

\begin{document}

\title{A Comprehensive Study on Large Language Models for Mutation Testing}

\author{Bo Wang}
\orcid{0000-0001-7944-9182}
\affiliation{
  \institution{Beijing Jiaotong University}
  \city{Beijing}
  \country{China}
  %\postcode{100044}
}
\email{wangbo\_cs@bjtu.edu.cn}

\author{Mingda Chen}
\orcid{0009-0002-6283-1291}
\affiliation{
  \institution{Beijing Jiaotong University}
  \city{Beijing}
  \country{China}
  %\postcode{100044}
}
\email{23120337@bjtu.edu.cn}

\author{Ming Deng}
\orcid{0009-0007-7193-0087}
\affiliation{
  \institution{Beijing Jiaotong University}
  \city{Beijing}
  \country{China}
}
\email{24120317@bjtu.edu.cn}

\author{Youfang Lin}
\orcid{0000-0002-5143-3645}
\affiliation{
  \institution{Beijing Jiaotong University}
  \city{Beijing}
  \country{China}
  %\postcode{100044}
}
\email{yflin@bjtu.edu.cn}

\author{Mark Harman}
\orcid{0000-0002-5864-4488}
\affiliation{
  \institution{University College London}
  \city{London}
  \country{United Kingdom}
}
\email{mark.harman@ucl.ac.uk}

\author{Mike Papadakis}
\orcid{0000-0003-1852-2547}
\affiliation{
  \institution{University of Luxembourg}
  \country{Luxembourg}
}
\email{michail.papadakis@uni.lu}

\author{Jie M. Zhang}
\orcid{0000-0003-0481-7264}
\affiliation{
  \institution{King's College London}
  \city{London}
  \country{United Kingdom}
}
\email{jie.zhang@kcl.ac.uk}

\renewcommand{\shortauthors}{Bo Wang, Mingda Chen, Ming Deng, Youfang Lin, Mark Harman, Mike Papadakis, and Jie M. Zhang}

\begin{abstract}

Large Language Models (LLMs) have recently been used to generate mutants in both research work and in industrial practice. 
However, there has been no comprehensive empirical study of their performance for this increasingly important LLM-based Software Engineering application. 
To address this, we conduct a comprehensive empirical study evaluating both a wide range of traditional approaches and LLM-based approaches. Particularly, we evaluate BugFarm, LLMorpheus, and our newly designed prompt for mutant generation. The experiments cover both leading open- and closed-source LLMs, on 851 real bugs from two Java real-world bug benchmarks (i.e., 605 bugs from 12 projects of Defects4J 2.0 and 246 bugs of ConDefects).
%To address this, we conduct a comprehensive empirical study evaluating BugFarm and LLMorpheus (the two state-of-the-art LLM-based approaches), alongside seven LLMs using our newly designed prompt, including both leading open- and closed-source models, on 851 real bugs from two Java real-world bug benchmarks.
%(i.e., 605 bugs from 12 projects of Defects4J 2.0 and 246 bugs of ConDefects).
Our results reveal that, compared to existing traditional approaches, LLMs generate more diverse mutants that are behaviorally closer to real bugs and, most importantly, achieve a 1.75$\times$ improvement in real bug detection. Specifically, LLM-based approaches reach a detection rate of 76.47\%, compared to 44.15\% for rule-based techniques, representing an absolute gain of $32.99$ percentage points.
%with $111.29\%$ higher fault detection.
%That is, 87.98\% (for LLMs) vs. 41.64\% (for rule-based); an increase of $46.34$  percentage points.
Nevertheless, our results also reveal that these impressive results for improved effectiveness come at a cost: the LLM-generated mutants have worse non-compilability, duplication, and equivalent mutant rates by $32.73$, $9.38$, and $3.64$ percentage points, respectively.
% (**) TODO: Jie, Bo: please could you add the percentage change for compatibility duplication and equivalence? Also these need to be reported in the boxed out answers to research quetsions. For e.g., RQ2 box out section reports the percentage noncompilable and duplicates for GPT-4o, but does not report the corresponding numbers for Pit or Major. In any case, for a box out answer to research question, we need results _overall_ LLM based approaches compared to aggregate results over _all_ non-LMM based approaches so that we have an answer to the _overall_ research question in eth box out (rather than what the referee might deem to be a "cherry picked" set of numbers). please note that these need to be waited aggregates, since there may be more mutants for one to or another, so you need to compute the overall sum, and then divide by the totals in each case. PLease can I ask you to review the boxed out sections of research questions to add a specifically these changes in percentage. So for example, if it goes from 2% to 23.6% for noncompilable, then that's a 1080% increase, but we should record the number of percentage points as well, which is 21.6 percentage points. This will keep reporting consistent with the reporting for the improvement in effectiveness 
These findings provide actionable insights for both research and practice.
They allow practitioners to have greater confidence in deploying LLM-based mutation, while researchers now have a baseline for the state-of-the-art, with which they can research techniques to further improve effectiveness and reduce cost.
\end{abstract}

\begin{CCSXML}
<ccs2012>
   <concept>   
<concept_id>10011007.10011074.10011099.10011102.10011103</concept_id>
       <concept_desc>Software and its engineering~Software testing and debugging</concept_desc>
<concept_significance>500</concept_significance>
       </concept>
 </ccs2012>
\end{CCSXML}

\ccsdesc[500]{Software and its engineering~Software testing and debugging}

\keywords{Empirical Study, Mutation Testing, Mutation Analysis, Mutation Generation, Large Language Models, Coupling}


\maketitle

\input{sections/intro}
\input{sections/design}

\input{sections/eval}
\input{sections/discussion}
\input{sections/threats}

\input{sections/related}

\input{sections/conclusion}

\begin{acks}
We are grateful to the anonymous reviewers.
This work was supported by the National Natural Science Foundation of China under Grant No.~62202040.
\end{acks}

\bibliographystyle{ACM-Reference-Format}
\bibliography{ref}

\end{document}

%% file: macro.tex
\newcommand{\code}[1]{\texttt{#1}}

\newcommand{\ourtoolraw}{LLMut}
\newcommand{\ourtool}{\ourtoolraw{}\xspace}

\newcommand{\tabref}[1]{Table~\ref{tab:#1}}
\newcommand{\tablabel}[1]{\label{tab:#1}}
\newcommand{\figref}[1]{Figure~\ref{fig:#1}}
\newcommand{\figlabel}[1]{\label{fig:#1}}

\newcommand{\smalltitlecolon}[1]{{\noindent \bf  {#1}:\ }}

\newcounter{num}

\lstdefinestyle{Java}{ % Define a style for your code snippet, multiple definitions can be made if, for example, you wish to insert multiple code snippets using different programming languages into one document
	language=Java,
	basicstyle=\scriptsize\ttfamily, % The default font size and style of the code
	breakatwhitespace=false, % If true, only allows line breaks at white space
	breaklines=true, % Automatic line breaking (prevents code from protruding outside the box)
	captionpos=b, % Sets the caption position: b for bottom; t for top
	commentstyle=\color[rgb]{0.0, 0.5, 0.69},%\color[rgb]{0,0.6,0}, % Style of comments within the code - dark green courier font
	deletekeywords={}, % If you want to delete any keywords from the current language separate them by commas
	escapeinside={<@}{@>},
	firstnumber=1, % Line numbers begin at line 1
	frame=lines, % Frame around the code box, value can be: none, leftline, topline, bottomline, lines, single, shadowbox
	frameround=tttt, % Rounds the corners of the frame for the top left, top right, bottom left and bottom right positions
	keywordstyle={[1]\color{blue!90!black}},
	keywordstyle={[3]\color{red!80!orange}},
	morekeywords={String,int}, % Add any functions no included by default here separated by commas
	numbers=none, % Location of line numbers, can take the values of: none, left, right
	numbersep=-8pt, % Distance of line numbers from the code box
	numberstyle=\tiny\color[rgb]{0.1,0.1,0.1}, % Style used for line numbers
	rulecolor=\color{black}, % Frame border color
	showstringspaces=false, % Don't put marks in string spaces
	showtabs=false, % Display tabs in the code as lines
	stepnumber=1, % The step distance between line numbers, i.e. how often will lines be numbered
	stringstyle=\color[rgb]{0.58,0,0.82},
	tabsize=2, % Number of spaces per tab in the code
	backgroundcolor=\color{white}
}

\newcommand{\codeIn}[1]{{\ttfamily #1}}

\newcommand{\hlc}[2]{{\setlength\fboxsep{0pt}\hspace{-3pt}\colorbox{#1} 
		{\begin{minipage}{\dimexpr\columnwidth-1\fboxsep+0pt\relax}
				\codeIn{\strut\hspace{3pt}#2}
			\end{minipage}}}}

%% file: sections/intro.tex
\section{Introduction}
For many years, mutation testing has been the topic of scientific research~\cite{hamlet1977testing,demillo1978hints,demillo1979program,jia2010analysis,papadakis2019mutation,potter2025evaluating}. 
Nevertheless, until the last five years, it had made little penetration into the industrial sector.
In the most recent five years, it has become increasingly widely deployed in industry.
For example, large tech sector companies such as Meta~\cite{harman2025mutation,beller2021would} and Google~\cite{petrovic2021practical,petrovic2018state} reported on results from the deployment of mutation testing at their companies.
In particular, results recently reported by Meta~\cite{harman2025mutation} highlighted the pivotal role played by Large Language Models (LLMs) in the successful deployment of mutation testing at scale in industry.

This recent trend towards LLM-based mutant generation traces back to work on earlier learning-based approaches, which leverage deep learning models to generate mutants by learning patterns from historical bug fixes~\cite{beller2021would,tufano2019learning,patra2021semantic,tian2022learning,chen2024deep}. These methods aim to introduce mutants that better resemble real faults, potentially increasing their relevance for testing.

However, despite increasing interest in LLM-based mutation testing in both research and practitioner communities, to date, no large-scale comprehensive empirical study exists.
Without such a study, there is no reliable foundation for confidently deploying the technique at an industrial scale.
%Without such a study, we have no basis on which to deploy an industrial scale with confidence.
Furthermore, the research community currently has no reliable baseline results for the state-of-the-art effectiveness and efficiency of LLM-based mutation testing compared to more traditional rule-based mutant generation techniques~\cite{jia2010analysis,papadakis2019mutation}.

The generation of mutants, using any technique, is challenging because of the following three important issues:
\begin{enumerate}
    \item {\bf Generating effective mutants}: 
The ultimate goal of mutation testing is to lead to strong tests that uncover real faults, making it critical that generated mutants closely couple with real faults \cite{ChekamPBTS20,wen2019exposing,petrovic2021does}. While existing methods have shown promise, there is still significant room for improvement in enhancing their fault-revealing potential~\cite{degiovanni2022mubert}.
\item {\bf Generating valid mutants.}: Mutants must be executable against test cases, which allows tests to assess the program’s behavior in the presence of mutants, to judge whether the mutants are killed. 
Ensuring executable mutants is challenging because mutants must remain syntactically and semantically valid, respect type constraints, preserve control and data flow, avoid breaking external dependencies, prevent unintended side effects, and align with test case assumptions to ensure meaningful evaluation~\cite{jia2010analysis}.
\item {\bf Generating diverse and killable mutants}: Determining whether a mutant is equivalent is undecidable~\cite{budd1982two}, making it challenging to avoid the generation of equivalent mutants, leading to poor scalability and waste of resources~\cite{zhang2013faster,papadakis2015trivial,zhang2016predictive,wang2021faster}. At the same time, mutants need to be diverse and cover the entire spectrum of code and program behaviors. 
\end{enumerate}

This paper examines how well Large Language Models (LLMs) address these challenges compared to more traditional rule-based approaches.
We conducted a comprehensive empirical study that studies BugFarm~\cite{ibrahimzada2023automated} and LLMorpheus~\cite{tip2025llmorpheus} (i.e., the state-of-the-art LLM-based mutant generation approaches).
Moreover, we propose a novel prompt template for mutant generation, which is equipped with few-shot examples (referred to as \textit{\ourtool} for ease of presentation).
All three methods are evaluated on two representative LLMs, GPT-4o and DeepSeek-V3-671b~\cite{liu2024deepseek} (DS-671b), denoted as BugFarm (GPT-4o), BugFarm (DS-671b), LLMorpheus (GPT-4o), LLMorpheus (DS-671b), LLMut (GPT-4o), and LLMut (DS-671b), respectively. Furthermore, LLMut is additionally evaluated on five other widely used models: GPT-3.5-Turbo~\cite{achiam2023gpt} (GPT-3.5), GPT-4o-Mini (GPT-4o-M), DeepSeek-Coder-V2-236b~\cite{zhu2024deepseek} (DS-236b), StarChat-$\beta$-16b~\cite{li2023starcoder} (SC-16b), and CodeLlama-Instruct-13b~\cite{roziere2023code} (CL-13b), to provide a more comprehensive analysis across both open- and closed-source LLMs.
Other variants of LLMut follow the same notation, e.g., LLMut (GPT-3.5) and LLMut (CL-13b).
We compare these LLM-based mutant generation approaches with four widely-studied traditional approaches (i.e., Major~\cite{just2011major}, PIT~\cite{coles2016pit}, LEAM~\cite{tian2022learning}, and $\mu$BERT~\cite{degiovanni2022mubert}),
with 851 real-world bugs from two Java benchmarks (i.e., 605 bugs across 12 projects from Defects4J 2.0~\cite{just2014defects4j} and 246 bugs from ConDefects~\cite{wu2023condefects}).
Overall, our primary experiments required the generation of more than 701,400 mutants.

Our evaluation includes both quantitative and qualitative analysis of the mutants generated by LLMs, comparing them with existing approaches in terms of
behavior similarity with real bugs (i.e., real bug detection rate, average Ochiai coefficient score, the number of high-Ochiai score bugs, coupling rate, and coupling categorization) as well as mutant validity (i.e., compilability rate, duplicate mutant rate, equivalent mutant rate, and efficiency).
Additionally, we study how different prompt engineering strategies affect the effectiveness of the task.
%, \todo{and LLMs (i.e., GPT-3.5-Turbo~\cite{achiam2023gpt}, GPT-4o, GPT-4o-mini, DeepSeek-Coder-V2-236b~\cite{zhu2024deepseek}, DeepSeek-V3~\cite{liu2024deepseek}, StarChat-$\beta$-16b~\cite{li2023starcoder}, and CodeLlama-Instruct-13b~\cite{roziere2023code}) affect the effectiveness of the task. 
We also explore the potential of using LLMs for generating high-order mutants.
Moreover, we analyze the root causes and error types for the non-compilable mutants generated by LLMs.

Our results reveal that, compared to existing approaches, LLMs generate more diverse mutants that are behaviorally
closer to real bugs.
This results in a higher fault detection rate compared to traditional approaches.
For instance, LLM-based approaches achieve  1.75$\times$ improvement in real bug detection rate, with an increase of 32.99 percentage points. Specifically, the mutants generated by LLMut (DS-671b) achieve fault detection rates of 91.1\%, whereas those generated by LEAM, PIT, and Major yield substantially lower rates of 63.9\%, 40.1\%, and 66.9\%, respectively. In terms of coupling rate, LLMorpheus (DS-671b) performs best, with 52\% of its mutants coupled with real bugs.
Nevertheless, we also observe that rule-based approaches, PIT and Major, outperform other approaches in terms of compilability rate, duplicate mutant rate, and equivalent mutant rate.
In particular,  Major excels with a 97.6\% compilability rate, 0\% duplicate mutants, and 2.0\% equivalent mutants,
while GPT-4o exhibits a compilability rate of 76.4\%, a duplicate mutant rate of 7.8\%, and an equivalent mutant rate of 3.1\%, respectively.

Another interesting finding of our analysis regards the diversity of the mutants (measured in terms of newly introduced AST node types). 
In particular, our results show that LLMut (DS-236b) and LLMut (DS-671b) exhibit the greatest diversity, newly introducing 49 different AST node types.
In comparison, rule-based approaches like Major~\cite{just2011major} introduce only 2 new node types, while LEAM~\cite{tian2022learning}, a state-of-the-art small model-based approach, introduces 21.
%\todo{The existing LLM-based approach, BugFarm, introduces 31 node types, higher than the traditional approaches as well.}
The greater diversity, which subsumes all existing methods, suggests that LLMs can significantly diversify the mutants.
Moreover, the distributions of AST edit distance of mutants compared to the original code also show that LLMs generate more balanced and expressive mutants.
For instance, almost 70\% of the mutants of Major have an AST edit distance of one, whereas for the LLM-based approach, BugFarm (DS-671b), only 38\% of its mutants have the same distance, indicating that LLM-based approaches tend to conduct more complex code transformations.

We also categorise the mutants generated by GPT models that fail to compile, the most common issue with learning-based approaches generated mutants, and found that they fall into 9 compilation error types, with the \textit{Usage of Unknown Methods} and \textit{Code Structural Destruction} types being the most prevalent ones.
Moreover, we further analyze the code context of the non-compilable mutants generated by LLM-based approaches, and find that \textit{Method Reference} and \textit{Method Invocation} are the most frequent code features.
This highlights the need for improved strategies to guide model-generated mutants toward more syntactically valid and semantically meaningful variations, thereby enhancing the effectiveness of mutation testing using AI models.

We analyze the surviving mutants of each approach, which are important to strengthen the test case. For one surviving non-equivalent mutant, it can be (1) not covered, (2) covered but no state infection, or (3) state infection without detection.
We analyze the distribution of surviving mutants across different approaches and find that LLMs based on our prompt are more likely to generate uncovered mutants, which consequently remain surviving, providing more valuable guidance for enhancing unit tests.

In summary, our paper makes the following main contributions:
\begin{itemize}
    \item We perform extensive and detailed comparative experiments to evaluate LLMs against existing tools/methods. Our study not only covers other existing LLM-based mutant generation approaches, but also introduces a novel, effective prompt template.
    Our findings indicate that LLMs excel in generating diverse mutants that closely mimic real bugs. 
    % \item We compare different LLMs, and find that currently closed-source LLMs are more effective, and more parameters may not consistently lead to better results.
    \item We analyze the error types of non-compilable mutants and determine that member assessment and method invocation are more likely to lead LLMs to generate non-compilable mutants.
    
    \item We construct a high-quality dataset of Java mutants, comprehensively annotated with detailed quality and behavior metrics, which can also serve as a valuable resource for fine-tuning in future research.
\end{itemize}

The remainder of this paper is organized as follows.
Section~\ref{sec:design} presents the design of our study, including the research questions, studied approaches, datasets, and metrics.
Section~\ref{sec:eval} presents the empirical evaluation of these generation approaches and answers to each research question.
Section~\ref{sec:discussion} discusses the influences of different experimental setups and the implications summarized from the findings.
Section~\ref{sec:threats} presents the threats to the validity of our study.
Section~\ref{sec:related} discusses the related work.
Section~\ref{sec:conclusion} concludes this work.

%% file: sections/design.tex
\section{Study Design} \label{sec:design}
In this section, we introduce the design and conduct of our study.

\subsection{Overview and Research Questions}
This paper aims to evaluate existing mutant generation approaches by answering the following research questions.

\begin{itemize}[leftmargin=*]
    \item \textbf{RQ1 (Effectiveness) How well do mutants generated by LLMs mimic real bugs compared to existing methods?}
    This research question investigates the effectiveness of LLM-generated mutants in representing real bugs. Specifically, it evaluates whether these mutants exhibit similar characteristics to real faults in terms of detection capability, behavioral similarity, and diversity.
    This RQ is divided into the following sub-RQs:
    \begin{itemize}
        \item \emph{RQ1.1: (Detectability) To what extent do bug-triggering tests successfully detect LLM-generated mutants?} 
        \item \emph{RQ1.2 (Coupling) To what extent do LLM-generated mutants couple with real bugs?}
        \item \emph{RQ1.3 (Diversity) How diverse are LLM-generated mutants compared to traditional approaches?} 
    \end{itemize}    
% rq1
    \item \textbf{RQ2 (Validity) How do LLMs compare to existing approaches in terms of validity?}
    This research question explores the validity of LLM-generated mutants compared to traditional approaches. Validity is assessed based on key factors such as compilability, duplication, and equivalence of generated mutants.
    By analyzing these aspects, we aim to determine whether LLMs produce valid and meaningful mutants that contribute to practical mutation testing.
    This RQ is divided into the following sub-RQs:
    \begin{itemize}
        \item \emph{RQ2.1 (Compilability) What percentage of generated mutants are compilable across different approaches?}
        \item \emph{RQ2.2 (Duplication) What percentage of the generated mutants are duplicates of the original code across different approaches?}
        \item \emph{RQ2.3 (Equivalence) What percentage of generated mutants are equivalent across different approaches?}
    \end{itemize}

    \item \textbf{RQ3 (Efficiency) What are the general costs of different mutant generation approaches?}
    This research question investigates the general costs of different mutant generation approaches.
    We consider time and token costs in this RQ, which are associated with the efficiency and practicality of mutant generation approaches.

    %This RQ evaluates the diversity of the transformations applied by each generation approach to the original code.
    %A high level of syntactical diversity ensures that the generated mutants cover different types of potential bugs.
    
    \item \textbf{RQ4 (Compilation Error) What are the types of compilation errors of the non-compilable mutants of different approaches?}
    This RQ aims to categorize and analyze the types of compilation errors that arise from non-compilable mutants generated by different approaches.
    Understanding these errors can guide mutant generation approaches to enhance their practical usability.

    \item \textbf{RQ5 (Surviving Mutants) What are the distributions of surviving mutants across different approaches?}
    This RQ aims to investigate the underlying reasons why mutants remain surviving, by categorizing them into three cases following the existing study~\cite{just2014efficient}: (1) uncovered by tests, (2) covered but without infecting the program states, and (3) injecting the states but remaining undetected by tests.
    By examining how these cases are distributed across different approaches, we can better understand the effectiveness of generated mutants and their potential value for guiding test suite improvement.
\end{itemize}

\subsection{Studied Mutant Generation Approaches}
Our study is conducted in Java because: 1) it is one of the most widely used programming languages, shown as the TIOBE ranking\footnote{\url{https://www.tiobe.com/tiobe-index/}}; 2) the majority of existing mutant generation approaches are for Java.
We categorize these approaches into two groups: \textit{traditional approaches} (i.e., those developed before the emergence of LLMs) and \textit{LLM-based approaches} (i.e., approaches that leverage LLMs and are designed for interaction in human-like conversations), as shown below.
The details of the settings for each generation approach are illustrated in Section~\ref{sec:mut-gen}.
\tabref{all-approaches} summarizes the mutant generation approaches covered by this study.

\begin{table}[tp]
  \centering
    \scriptsize
  \caption{Studied Mutant Generation Approaches}
    \begin{tabular}{|l|l|l|}
    \hline
    \textbf{Type} & \textbf{Name} & \textbf{Description} \\
    \hline
    \multirow{4}{*}{\textbf{Traditional}} & \bf PIT & Widely studied mutation testing framework with 29 mutation operators. \\
       & \bf Major & Widely studied mutation testing framework with 11 mutation operators. \\
       & \bf LEAM & Custom-trained models learning from a large corpus of real bugs and patches. \\
       & \bf $\mu$BERT & Mutant generation with the pre-trained model CodeBERT. \\
    \hline
    \multirow{9}{*}{\textbf{LLM-based}}  & \bf BF (GPT-4o) & The existing LLM-based approach BugFarm with GPT-4o. \\
       & \bf BF (DS-671b) & The existing LLM-based approach BugFarm with DeepSeek-V3-671b. \\
\cline{2-3}       & \bf LLMph (GPT-4o) & The existing LLM-based approach LLMorpheus with GPT-4o. \\
       & \bf LLMph (DS-671b) & The existing LLM-based approach LLMorpheus with DeepSeek-V3-671b. \\
\cline{2-3}       & \bf \ourtool (GPT-3.5) & Our prompt with GPT-3.5-Turbo, trained on data before 2021/09, released in 2021/03. \\
       & \bf  \ourtool (GPT-4o) & Our prompt with GPT-4o, trained on data before 2023/10, released in 2024/05. \\
       & \bf \ourtool (GPT-4o-M) & Our prompt with GPT-4o-Mini, trained on data before 2023/10, released in 2024/07. \\
       & \bf \ourtool (SC-16b) & Our prompt with StarChat-$\beta$-16b, whose base model is StarCoder, released in 2023/06. \\
       & \bf \ourtool (CL-13b) &  Our prompt with CodeLlama-Instruct-13b, whose base model is Llama, released in 2023/08. \\
       & \bf \ourtool (DS-236b)  & Our prompt with  DeepSeek-Coder-V2-236b, whose base model is DeepSeek, released in 2024/07. \\
       & \bf \ourtool (DS-671b) & Our approach with DeepSeek-V3-671b, whose base model is DeepSeek, released in 2024/12. \\
    \hline
    \end{tabular}
  \tablabel{all-approaches}
\end{table}

\subsubsection{Traditional Approaches}
For Java, there exist several mutation testing frameworks, such as JavaLanche~\cite{schuler2009javalanche}, MuJava~\cite{ma2005mujava}, PIT~\cite{coles2016pit}, and Major~\cite{just2011major}.
As some tools are no longer maintained, following existing studies~\cite{kaufman2022prioritizing,ojdanic2023syntactic,kushigian2024equivalent}, we employ PIT~\cite{coles2016pit}, Major~\cite{just2011major}, LEAM~\cite{tian2022learning}, and $\mu$BERT~\cite{degiovanni2022mubert}. 
Major and PIT are widely studied rule-based approaches, while LEAM and $\mu$BERT involve deep learning models.

\noindent\textbf{PIT: } PIT~\cite{coles2016pit} is a widely used mutation testing tool on the Java bytecode level, which supports 29 well-designed mutation operators and four groups of active mutation operator sets.
In our study, we employ PIT 1.14.4 and leverage the \code{ALL} group, which activates all operators.

\noindent\textbf{Major: } Major~\cite{just2011major} is another high-impact mutation testing framework that supports nine operators.
In our study, we use Major 1.3.5 and involve all mutation operators.

\noindent\textbf{LEAM: } LEAM~\cite{tian2022learning,tianleam++} is the state-of-the-art custom-trained-model-based approach, which employs sequence-to-sequence models to generate mutants.
Its model is trained on a large corpus of real-world bugs and their corresponding patches, allowing it to learn patterns for generating more effective mutants.

\noindent\textbf{$\mu$BERT: } $\mu$BERT~\cite{degiovanni2022mubert} is the state-of-the-art mutant generation approaches based on the small-scale pre-trained model, BERT~\cite{devlin2018bert}.
Equipped with the pre-trained model, $\mu$BERT transforms the task of mutant generation to code completion under the surrounding context.

\subsubsection{LLM-based Approaches}

Our study incorporates two existing LLM-based mutant generation approaches (i.e., BugFarm~\cite{ibrahimzada2023automated} and LLMorpheus~\cite{tip2025llmorpheus}), and introduces a new prompt design for mutant generation, which will be presented later.

\noindent\textbf{Existing LLM-based Approaches:}
We adopt two the state-of-the-art LLM-based mutant generation approaches, BugFarm~\cite{ibrahimzada2023automated} and LLMorpheus~\cite{tip2025llmorpheus}. 

BugFarm~\cite{ibrahimzada2023automated} trains a small-scale model for selecting lines for mutating, and then utilizes LLMs to mutate the target line.
Based on the prompt template given in the repository of BugFarm\footnote{\url{https://github.com/Intelligent-CAT-Lab/BugFarm}}, we can find that BugFarm numbers all lines of a Java method and uses a lightweight model to predict the target line numbers to mutate. These line numbers are then provided to the LLM.
BugFarm then instructs the LLM to generate three different mutants for the Java method by changing the specified lines. Finally, BugFarm outputs each mutant as a complete mutated method, with the mutation embedded in the full method.

LLMorpheus~\cite{tip2025llmorpheus} is an LLM-based mutation testing approach for JavaScript. Upon examining its prompt from its repository\footnote{\url{https://github.com/githubnext/llmorpheus}}, we found it to be language-agnostic. Therefore, we reimplement the code-extraction component for Java (such as parsing ASTs and extracting the code context for mutants) while retaining its original prompt.
When porting LLMorpheus to Java, we retain its default configuration.
LLMorpheus masks specific syntactic elements (e.g., the conditional expression of an \code{if} statement) with a \code{PLACEHOLDER} to mark the target mutation location, and instructs the LLM to fill in this placeholder.
For each masked location, LLMorpheus generates three mutants.
By default, LLMorpheus provides up to 200 lines of surrounding context and instructs the LLM to return only the code fragment that should replace the placeholder.

%LLMorpheus performs mutation testing given the whole method as context. Note that LLMorpheus was originally designed for JavaScript. Upon examining its prompt, we found it to be language-agnostic. Therefore, we reimplemented the code-extraction component for Java (such as parsing ASTs and extracting the code context for mutants) while retaining the original prompt.

In our study, both BugFarm and LLMorpheus are equipped with GPT-4o~\cite{achiam2023gpt} and DeepSeek-V3-671B~\cite{liu2024deepseek}, representing the current leading closed- and open-source LLMs, respectively.

%%%%%%%%%%%%%%%%%%%%%%%%%%%%%%%%%%%%%%%%%%%%%%%%%%%
\begin{figure}[t]
    \centering
    \includegraphics[width=\linewidth]{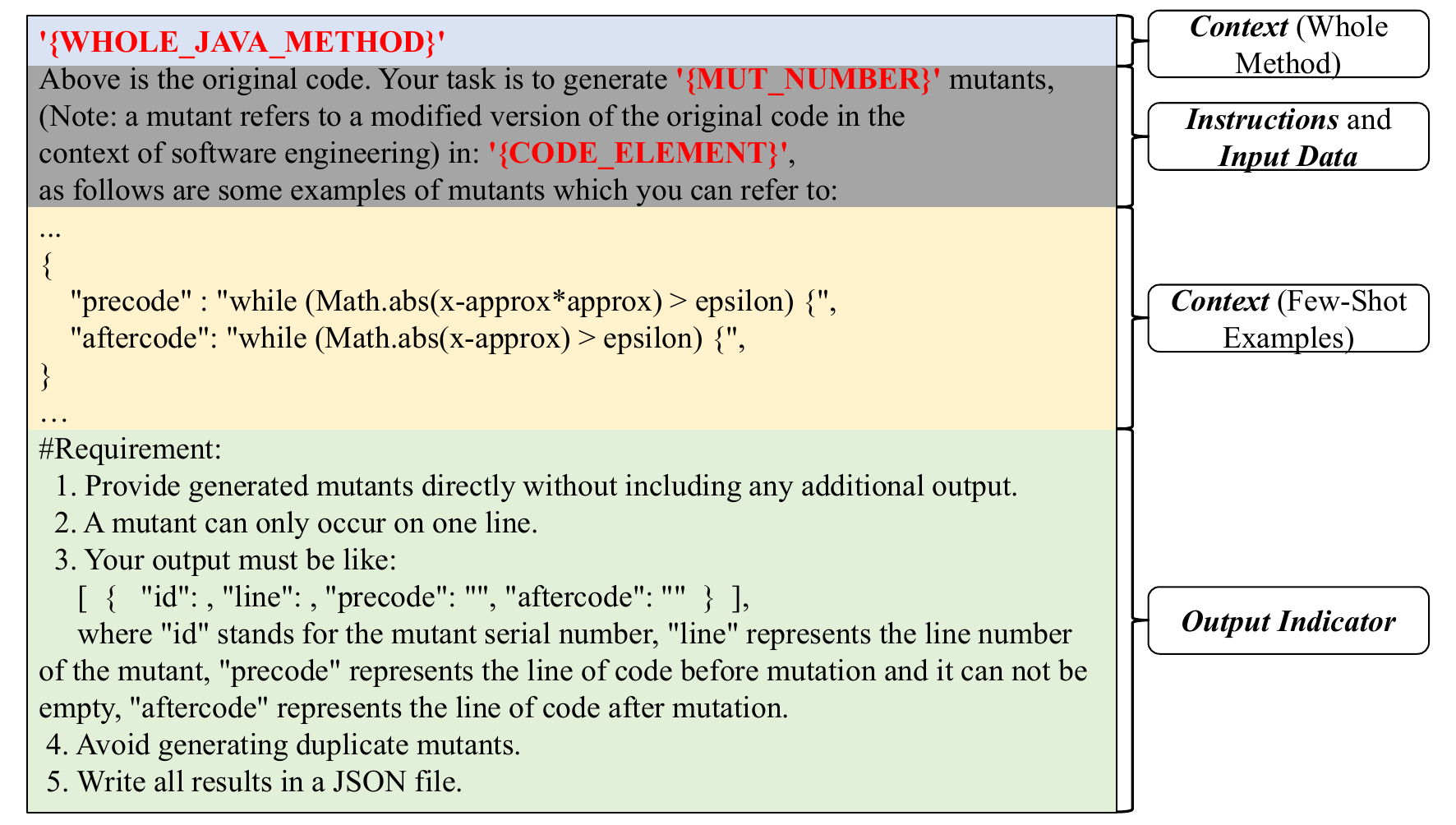}
    \caption{Our New Prompt Template for Mutant Generation}
    \figlabel{prompt}
\end{figure}

\noindent\textbf{LLMs with Our New Prompt:}
Different from the existing LLM-based approaches, we propose a new prompt engineering for mutant generation, shown as \figref{prompt}.
For ease of presentation, in the following part of this paper, we call our prompt \textbf{\ourtool}.

To craft effective prompts, we follow the best practices which suggest that prompts should comprise four aspects: \textit{Instruction}, \textit{Context}, \textit{Input Data}, and \textit{Output Indicator}~\cite{guo2024exploring,ma2024llmparser}.
In the \textit{Instructions}, we direct the LLMs to generate mutants for the target code element. The placeholder \textit{MUT\_NUMBER} specifies the number of mutants required, while the placeholder \textit{CODE\_ELEMENT} specifies the target code snippet to be mutated. In the \textit{Context}, we clearly state that a mutant is the concept of mutation testing, and additionally provide various sources of information to observe the impact on the performance of LLMs, including the whole Java method surrounding the target code element and few-shot examples sampled from real-world bugs from another benchmark.
In the \textit{Input Data}, we specify the code element to be mutated and the number of mutants to generate.
In the \textit{Output Indicator}, we specify the \textit{JSON} file format for outputs, facilitating further experiments.
In addition, our prompts need few-shot examples, which should originate from real bugs, enabling LLMs to learn how to mutate code from real bugs.
We further explore the different prompt settings in Section~\ref{sec:diss-sentivitiy-to-settings}.

To avoid the few-shot examples leaking information, beyond the evaluation datasets (i.e., as shown later, they are Defects4J and ConDefects), we employ another benchmark, QuixBugs~\cite{lin2017quixbugs}, which comprises 40 real-world Java bugs and is commonly used by the testing and debugging community~\cite{zhu2023tare,jiang2023knod,xia2022less,li2023nuances}.
%\todo{Referring to the study by Deng et al.~\cite{deng2024large}, LLMs perform best in generating valid code when using around six few-shot examples.}
As shown later, our experimental results indicate that providing six examples results in the best overall performance, consistent with observations by Deng \emph{et al.}~\cite{deng2024large}.
Therefore, we sample six bugs from QuixBugs for few-shot learning in our prompt.
To guarantee the diversity of the examples, we randomly select one bug from the dataset and assess whether its modification pattern is similar to the examples already collected.
If it is different, we add it to our collection and continue until we have collected all examples, shown as \tabref{few-shots}.
As LLM capabilities continue to advance, the optimal number of few-shot examples may vary. We therefore investigate the impact of different numbers of few-shot examples in Section~\ref{sec:discussion-diff-few-shot-examples}.
Note that adopting few-shot examples may also have drawbacks, which we will discuss later.

\begin{table}[tp]
  \centering
  \scriptsize
  \caption{The Default Few-Shot Examples of \ourtool from QuixBugs}
  \vspace{-2mm}
    \begin{tabular}{l|l}
    \hline
    \bf Correct Version & \bf Buggy Version \\
    \hline
    \code{n = (n \& (n - 1));} & \code{n = (n \^{} (n - 1));} \\
    \hline
    \code{while (!queue.isEmpty())}  & \code{while (true)}  \\
    \hline
    \code{return depth==0;} & \code{return true;} \\
    \hline
    \multicolumn{1}{p{18.61em}|}{\code{ArrayList r = new ArrayList();}\newline{}\code{r.add(first).addll(subset);}\newline{}\code{to\_add(r);}} & \code{to\_add.addAll(subset);} \\
    \hline
    \code{c = bin\_op.apply(b,a);} & \code{c = bin\_op.apply(a,b);} \\
    \hline
    \code{while(Math.abs(x-approx*approx)>epsilon)} & \code{while(Math.abs(x-approx)>epsilon)} \\
    \hline
    \end{tabular}
  \tablabel{few-shots}
\end{table}

To determine the number of mutants generated in a prompt, we adopt the setting used in PIT~\cite{coles2016pit}.
We use PIT-generated mutants with the universal set of operators and find that PIT generates 1.15 mutants per line of code.
Therefore, based on the given code context, we generate \textit{one mutant per line}.
We explore the influence of context length on mutant generation in Section~\ref{sec:discussion}.
Additionally, for the temperature, we use the default value for each LLM.

Each LLM-based approach in our study adopts a different prompt-engineering strategy. For example, LLMut incorporates few-shot examples and uses a structured JSON output format, whereas LLMorpheus masks high-value code fragments and instructs the LLM to fill in the masked content to ensure the quality of the output mutants. Note that the different prompt-engineering strategies are complementary, with different approaches emphasizing different aspects of the mutants.

%Compared with the prompts used in BugFarm\footnote{\url{https://github.com/Intelligent-CAT-Lab/BugFarm}} and LLMorpheus\footnote{\url{https://github.com/githubnext/llmorpheus}}, our prompt design offers several advantages:
%\begin{itemize}
%    \item it incorporates real-world bugs as few-shot examples, which have been shown to significantly enhance LLM performance in many code-related tasks;
%    \item it determines the number of generated mutants based on the size of the target code, rather than on a fixed number;
%    \item it specifies the mutant location using the target statement itself rather than line numbers, enabling more precise localization~\cite{fan2023automated}; and
%    \item it employs JSON output formatting, which is widely used in LLM fine-tuning and downstream tasks~\cite{wang2026assessing}.
%\end{itemize}

Based on \ourtool, our study comprehensively covers the most widely studied LLMs.
Specifically, we select a representative set of LLMs with code capabilities, both from commercial closed-source and open-source models.
Following recent studies on evaluating the code capabilities of LLMs~\cite{wang2024software,guo2024exploring,yuan2024evaluating}, we adopt the closed-source LLMs from the GPT family, including GPT-3.5-Turbo (GPT-3.5)~\cite{achiam2023gpt}, GPT-4o~\cite{achiam2023gpt}, and GPT-4o-Mini (GPT-4o-M)~\cite{achiam2023gpt}), and open-source LLMs, including StarChat-$\beta$-16b (SC-16b), CodeLlama-Instruct-13b (CL-13b), DeepSeek-Coder-V2-236b (DS-236b), and DeepSeek-V3-671b (DS-671b).
The details of these models are shown as~\tabref{all-approaches}.
Note that OpenAI has not officially released the parameter size of their closed-source models.
The parameter size of the open-sourced models we studied ranges from 13 billion to 671 billion.
Moreover, the selected LLMs represent a balance between general-purpose models (e.g., GPT-4o and DeepSeek-V3-671b), which are trained for a wide range of natural language and reasoning tasks, and code-specialized models (e.g., CodeLlama-Instruct-13b and DeepSeek-V2-Coder-236b), which are further fine-tuned on large-scale code corpora. This selection allows us to evaluate mutant generation across both broadly capable models and those explicitly optimized for programming tasks, ensuring the conclusions are representative of the current landscape of LLMs.

\subsection{Datasets} \label{sec:datasets}

\begin{table}[tp]
  \centering
  \scriptsize
  \caption{Real-world Bugs Used in Our Study}
    \begin{tabular}{|l|l|r|r|}
    \hline
    \textbf{Dataset} & \textbf{Project} & \multicolumn{1}{l|}{\textbf{\# of Bugs}} & \multicolumn{1}{l|}{\textbf{Time Span}} \\
    \hline
    \multirow{12}{*}{\bf Defects4J (D4J)} & Math & 106 & 2006/06/05 - 2013/08/31 \\
\cline{2-4}       & Lang & 65 & 2006/07/16 - 2013/07/07 \\
\cline{2-4}       & Chart & 26 & 2007/07/06 - 2010/02/09 \\
\cline{2-4}       & Time & 27 & 2010/10/27 - 2013/12/02 \\
\cline{2-4}       & Closure & 133 & 2009/11/12 - 2013/10/23 \\
\cline{2-4}       & Mockito & 38 & 2009/06/20 - 2015/05/20 \\
    %\hline
    %\multirow{6}{*}{\bf Defects4J 2.0} & Cli & 39 & 2007/05/15 - 2018/02/26 \\
\cline{2-4}       & Cli & 39 & 2007/05/15 - 2018/02/26 \\
\cline{2-4}       & Codec & 18 & 2008/04/27 - 2017/03/26 \\
\cline{2-4}       & Csv & 16 & 2012/03/27 - 2018/05/18 \\
\cline{2-4}       & Gson & 18 & 2010/11/02 - 2017/09/21 \\
\cline{2-4}       & JacksonCore & 26 & 2013/08/28 - 2019/04/05 \\
\cline{2-4}       & Jsoup & 93 & 2011/07/02 - 2019/07/04 \\
    \hline
    \bf ConDefects (CD) & —  & 246 & 2024/03/01 - 2024/06/30 \\
    \hline
    \textbf{Total} & —  & 851 & 2006/07/16 - 2024/06/30 \\
    \hline
    \end{tabular}
  \tablabel{bugs}
\end{table}

As one of the major tasks of our study is to evaluate the behavioral similarity between mutants and real-world bugs, we selected datasets that include real-world bugs along with their corresponding fault-revealing test cases.
Specifically, we employ the datasets with the following properties:
\begin{itemize}%[leftmargin=*]
    \item The datasets should comprise Java programs, as existing methods are primarily based on Java, and we need to compare with them.
    \item The bugs of the datasets should be real-world bugs so that we can compare the difference between mutants and real bugs.
    \item Every bug in datasets has the correctly fixed version provided by developers so that we can mutate the fixed version and compare them with the corresponding real bugs.    
    \item Every bug is accompanied by at least one bug-triggering test because we need to measure whether the mutants affect the execution of bug-triggering tests.
\end{itemize}
To this end, we employ the Defects4J 2.0~\cite{just2014defects4j} and ConDefects~\cite{wu2023condefects} to evaluate the mutant generation approaches, shown as~\tabref{bugs}.
In total, we conducted experiments on 851 bugs.

Defects4J 2.0 is a widely used benchmark in the field of mutation testing~\cite{ojdanic2023syntactic,just2014mutants,kaufman2022prioritizing,degiovanni2022mubert,tian2022learning,kim2022predictive}, which contains historical bugs from open-source projects of diverse domains,  ensuring a broad representation of real-world bugs.
%Apache Commons-Math (Math) is a scientific computing library.
%Apache Commons-Lang (Lang) is a supplementary library for enhancing the original JDK.
%JFreeChart (Chart) is a toolset for drawing charts.
%Joda-Time (Time) is a library designed to enhance and replace the time and date handling capabilities of JDK.
%Closure compiler (Closure) is the Google implementation for the language.
%Mockito is a mocking framework for unit tests written in Java.
We select the most widely used 12 projects from Defects4J 2.0, which contain 605 bugs in total.
However, from~\tabref{all-approaches} and~\tabref{bugs}, we observe that the periods of the Defects4J bugs are earlier than the LLMs' training time, which may introduce data leakage.
Therefore, we supplement another dataset, ConDefects~\cite{wu2023condefects}, designed to address the data leakage concerns.
ConDefects consists of tasks from AtCoder\footnote{\url{https://atcoder.jp}} programming contest.
To prevent data leakage, we use data from the four most recent months following the release of the LLMs (i.e., from March 1, 2024, to June 30, 2024), during which we collected 246 Java bugs.
Among them, only the most recent model, DeepSeek-V3-731b, cannot guarantee data leakage-free, since it was released after the ConDefects dataset.

\subsection{Evaluation Metrics}

In this subsection, we present the evaluation metrics used in our study to comprehensively assess the quality of the generated mutants.
We evaluate the mutants from multiple perspectives, as mentioned before.
Note that only relying on the mutation score is insufficient for a meaningful assessment. 
A higher mutation score does not necessarily indicate better quality mutants, as it only reflects the proportion of killed mutants rather than their effectiveness.
However, we still report the mutation score of each mutant generation approach later in Section~\ref{sec:eval}.

To better understand the metrics, we first illustrate the categories of mutants.
Let all generated mutants be the set $A$, where LLMs may generate syntactically incorrect mutants that can not pass compilation.
Thus, we denote the compilable mutants as the set $C$.
Note that the set $C$ contains two types of mutants that should be removed, the duplicate mutants (denoted as $D$) and equivalent mutants (denoted as $E$).
Therefore, we have $(C \supseteq D) \wedge (C \supseteq E) \wedge (D \cap E = \emptyset)$.
The duplicate mutants refer to those identical to the original code or other mutants.
Ideally, we should perform mutation testing against the mutant set of $(C-D-E)$, excluding all duplicated mutants ($D$) and equivalent mutants ($E$), which have no contribution to discovering the weakness of tests. Based on our definition, the duplicated mutants are syntactically identical mutants that are trivially decidable. However, determining the equivalent mutants set $E$ is undecidable, and $E$ is negligible compared to the universal mutant set (i.e., $|E| \ll |C-D|$). Therefore, we use the set of mutants $(C-D)$ as an approximation to compute the mutation score (i.e., $|C - D|  \approx |C - D - E|$), which is the common practice of existing studies~\cite{diniz2021dissecting,tian2022learning,zhang2022mutant,wang2025systematic}.
%In mutation testing, the ideal set of mutants to employ is from the set $(C-D-E)$, excluding equivalent mutants that have no contribution to discovering the weakness of tests.
%However, determining the equivalent mutants set $E$ is undecidable, so we follow the common practice and use the set $(C-D)$ as a practical approximation to perform mutation testing~\cite{diniz2021dissecting,tian2022learning,zhang2022mutant,wang2025systematic}.
Therefore, let $K$ denote the set of killed mutants among the compilable and non-duplicate ones (i.e., $|C - D|$). The mutation score $MS$ is then computed as follows:
\begin{equation}
    MS = \frac{| K | }{ | C - D | } \approx \frac{| K | }{ | C - D - E |} (|E| \ll |C - D|)
\end{equation}

Note that traditional mutant generation approaches, such as Major and PIT, produce very few duplicate mutants (i.e., $|D|$ is 0). Hence, early studies~\cite{kintis2017detecting} defined equivalent mutants as those semantically equivalent to the original code, which may implicitly include $D$, which is an empty set. However, LLM-based approaches tend to generate a large number of non-mutated duplicates. Moreover, while duplicate mutants can be easily filtered, identifying equivalent mutants remains challenging~\cite{tian2024large}. Therefore, in this paper, we refine the definition of equivalent mutants to be those that are syntactically different but semantically equivalent to the original code.

\subsubsection{RQ1: Effectiveness} \label{subsec:coulping-metrics}
%Mutation testing relies on the \textit{fault coupling effect}, which suggests that although mutations are simple syntactic changes, they can effectively mimic more complex real bugs~\cite{offutt1992investigations,jia2010analysis}.
%To evaluate this relationship, researchers have proposed the concept of \textit{coupling}, which quantifies the behavioral similarity between generated mutations and real bugs.
%Assessing the degree of coupling is crucial for determining the effectiveness of different mutation generation approaches.
Evaluating the effectiveness of mutant generation approaches requires assessing how well the generated mutants mimic real bugs.
A key aspect of this evaluation is understanding whether LLM-generated mutants exhibit fault characteristics that align with real-world defects.

To quantify this effectiveness, we employ multiple metrics that capture different aspects of behavioral similarity between mutants and real bugs.
These metrics assess whether bug-triggering tests can detect LLM-generated mutants (i.e., \textit{RQ1.1}), how closely mutants resemble real faults based on test case failures (i.e., \textit{RQ1.2}), and the diversity of syntactic changes introduced by different mutant generation techniques (i.e., \textit{RQ1.3}).
The metrics of these sub-RQs are listed below.

%To comprehensively evaluate the degree of coupling between generated mutations and real bugs across different approaches, we adopt well-established metrics from existing studies~\cite{just2014mutants,papadakis2018mutation,khanfir2023ibir,ojdanic2023syntactic,gay2023closely}. Specifically, we use \textit{Real Bug Detectability}~\cite{just2014mutants,khanfir2023ibir}, \textit{Average Ochiai Coefficient}~\cite{ojdanic2023syntactic}, \textit{High Ochiai Bug Count}~\cite{ojdanic2023syntactic}, and \textit{Coupling Rate}~\cite{just2014mutants} to provide a thorough assessment of how well the generated mutations reflect real-world bugs.
%Furthermore, to gain deeper insights, we categorize each generated mutation into distinct coupling scales based on their similarity to real bugs, following the classification proposed by~\cite{gay2023closely}, and analyze their distribution across different approaches.

%Behavior similarity refers to the degree of similarity in functionality and execution between a mutation and its corresponding real bug.
%Because capturing all program behaviors is undecidable, following existing studies, we thus employ the following basic metrics as a proxy to measure behavior similarity~\cite{papadakis2018mutation,khanfir2023ibir,ojdanic2023syntactic}.

\smalltitlecolon{RQ1.1: Real Bug Detection Rate}
Following the approach of Just et al.~\cite{just2014mutants}, we evaluate the effectiveness of mutant generation by measuring its ability to detect real bugs using developer-written tests.
For a given real bug (which may span multiple lines), we use a mutant-generation approach to create a set of mutants on the fixed version of the program. If the test cases that kill a mutant also fail on the real bug, we consider that the mutant generation approach has successfully ``detected'' this real bug.
To quantify this effectiveness, we compute the \textit{Real Bug Detection Rate}, which represents the proportion of real bugs for which at least one generated mutant is detected by the same test cases that reveal the real bug. 
In other words, it measures whether the generated mutants can reproduce the failing behaviors of a set of real bugs under tests.
This metric provides insight into how well LLM-generated mutants align with actual software defects in a testing context.

\smalltitlecolon{RQ1.2: Average Ochiai Coefficient}
This metric estimates the semantic similarity between a mutant and its corresponding real bug based on their passing and failing test cases~\cite{offutt2001mutation}.
\textit{Ochiai coefficient} measures how a mutation resembles its corresponding real bug.
The \textit{Ochiai coefficient} was originally introduced as a general similarity measure but later became widely adopted in spectrum-based fault localization (SBFL)~\cite{zou2019empirical}.
In this study, we use the Ochiai coefficient in its original form to assess how closely a mutant resembles its corresponding real bug, which is consistent with the findings of Ojdanic et al.~\cite{ojdanic2023syntactic}.

More concretely, let $m$ and $b$ be a mutant and its corresponding real bug, and let $fT_m$ and $fT_b$ be the set of their failing test cases, respectively. 
The coefficient of $m$ is computed by the following formula:
\begin{equation}
    Ochiai(m,b) = \frac{| fT_m \cap fT_b | }{ \sqrt{| fT_m | \times |fT_b|} }
\end{equation}
Therefore, if the Ochiai coefficient of $m$ is $1$, it is semantically equal to its corresponding bug $b$.
Given a bug $b$, the Ochiai coefficient for $b$ is calculated as the mean Ochiai coefficient across all its corresponding de-duplicated mutants (i.e., $(C_b - D_b)$), shown as the following formula:
\begin{equation}
    Ochiai_b = \frac{ \sum_{m \in (C_b-D_b)} Ochiai(m,b) }{ | C_b - D_b | }
\end{equation}
Given a mutant generation approach and a set of bugs $B$, the Average Ochiai Coefficient of the approach is calculated as the mean Ochiai coefficient across all bugs, shown as the following formula:
\begin{equation}
    AOC = \frac{\sum_{b \in B} {Ochiai_b} }{|B|}
\end{equation}

\smalltitlecolon{RQ1.2: High Ochiai Bug Count}
Following the study~\cite{ojdanic2023syntactic}, a bug $b$ and a set of mutants are considered to have high Ochiai if they satisfy $(Ochiai_b \geq 0.8)$.
This threshold indicates a strong semantic similarity between the generated mutants and the real bug.
For each mutant generation approach, we compute the number of bugs that meet this criterion.

%\subsubsection{RQ4: Degree of Coupling to Real Bugs} \label{subsec:coulping-metrics}
%Mutation testing relies on the \textit{fault coupling effect}, which suggests that although mutations are simple syntactic changes, they can effectively represent more complex real bugs~\cite{offutt1992investigations,jia2010analysis}. To quantify this relationship, researchers have introduced the concept of \textit{coupling}, which measures the semantic similarity between generated mutations and actual faults.

%To comprehensively assess the degree of coupling for mutations generated by different approaches, we adopt established metrics from existing studies~\cite{just2014mutants,ojdanic2023syntactic,gay2023closely}. 
%Specifically, we employ the \textit{coupling rate}~\cite{just2014mutants} to provide an overall evaluation of coupling effectiveness. 
%Additionally, to gain deeper insights, we categorize each generated mutation into different scales of the existing study~\cite{gay2023closely} based on their degree of similarity to real bugs and analyze their distribution.

\smalltitlecolon{RQ1.2: Coupling Rate}
Given a real bug, one mutant is considered \textit{coupling} if the set of test cases that kill the mutant overlaps with the bug-triggering tests.
Note that each bug we analyzed has at least one failing test to reveal the bug, called \textit{bug-triggering test}.
The \textit{Coupling Rate} refers to the proportion of coupled mutants within the set of de-duplicated mutants (i.e., $(C - D)$).

Note that \textit{Real Bug Detection Rate} and \textit{Coupling Rate} assess effectiveness from different granularities. \textit{Real Bug Detection Rate} evaluates effectiveness at the bug level—\textbf{the proportion of bugs} for which at least one generated mutant reproduces the failing behavior. \textit{Coupling Rate} evaluates effectiveness at the mutant level—\textbf{the proportion of mutants} whose failing tests overlap with the bug-triggering tests.
These two metrics capture different phenomena.
For example, an approach may generate many high-quality mutants for only a few bugs while producing very few mutants for the remaining bugs.
In this case, the \textit{Coupling Rate} can be high (because many mutants align well with the few covered bugs), but the \textit{Real Bug Detection Rate} remains low (because many bugs never receive any useful mutants).
Conversely, an approach might generate mutants that touch a wide range of bugs, achieving a high \textit{Real Bug Detection Rate}.
However, if per-bug mutants only weakly reflect the bug-triggering behavior, the overall \textit{Coupling Rate} may remain low.
Therefore, \textbf{the two metrics are complementary}: \textit{Real Bug Detection Rate} measures coverage across bugs, while \textit{Coupling Rate} measures behavioral precision within mutants.
Neither metric alone can fully represent the effectiveness of mutation generation.

\smalltitlecolon{RQ1.2: Coupling Categorization}
We categorize each generated mutant into distinct coupling scales based on their similarity to real bugs, following the classification proposed by~\cite{gay2023closely}, and analyze their distribution across different approaches.
Specifically, given a compilable mutant (i.e., an element from the set $C$) on the fixed version of a real bug (which can be uncovered by at least one triggering test), we can classify it into the following six scales based on its behavior:
\begin{itemize}
    \item \textbf{Strong Substitution:} All triggering tests fail and no additional tests fail, which represents the mutant is a semantic replacement of the real bug under the given test suite.
    \item \textbf{Strong Substitution + Additional Tests Fail:} All triggering tests fail, but additional non-triggering tests fail.
    \item \textbf{Partial Substitution:} Some triggering tests fail and no additional tests fail.
    \item \textbf{Partial Substitution + Additional Tests Fail:} Some triggering tests fail, and some additional non-triggering tests fail.
    \item \textbf{Not Substitution:} Only non-triggering tests fail.
    \item \textbf{Not Detected:} No tests fail (i.e., the mutant is surviving under the given test suite).
\end{itemize}
Later, we will classify the mutants of each approach and analyze their distributions.

\smalltitlecolon{RQ1.3: AST Node Diversity}
We prefer the set of mutants with diverse syntactic forms, i.e., method invocations, field accesses, etc.
Therefore, we parse the code before and after mutation, and check the diversity of newly introduced AST nodes.
In other words, we identify the unique AST node types introduced by each approach’s generated mutants, relative to the set of node types present in the original program.
The approach with a greater variety of AST node types is considered superior, as it indicates a broader exploration of the syntactic space and enhances the likelihood of detecting weaknesses in the test suite.

\smalltitlecolon{RQ1.3: AST Edit Distance}
The AST edit distance measures the structural difference between the original and mutated code by quantifying the number of tree transformation operations (i.e., insertions, deletions, and modifications) needed to convert one AST into another.
This metric evaluates the extent of syntactic changes introduced by mutations.
A higher edit distance suggests more significant syntactic changes.
The AST Edit Distance metric is well-suited for structured code, as it captures syntactic differences at the tree level rather than merely comparing token sequences, as in \textit{BLEU} and similar metrics.
However, it still has inherent limitations, such as its inability to fully reflect the impact of mutants on program behavior.
We further discuss these limitations later in Section~\ref{sec:threats}.

%\subsubsection{RQ2: Cost, Compilability Rate, Useless Mutation rate, and Equivalent Mutation Rate}
\subsubsection{RQ2: Validity}
We consider three key metrics to evaluate the validity of different mutant generation approaches: generation efficiency, compilability, and the proportion of duplicate and equivalent mutants, providing a basic understanding of the validity of each approach.

\smalltitlecolon{RQ2.1: Compilability Rate}
the proportion of mutants that successfully compile, which is computed as:
\begin{equation}
    CR = \frac{\enspace | C | \enspace }{ | A | } 
\end{equation}
Because LLMs cannot guarantee the generation of entirely correct code and the compilation process can be time-consuming, we prefer the approach with a higher compilability rate.

\smalltitlecolon{RQ2.2: Duplicate Mutant Rate}
the proportion of duplicate mutants, computed as:
\begin{equation}
    DMR = \frac{\enspace | D | \enspace}{ | A | } 
\end{equation}
LLMs sometimes generate duplicate mutants, i.e., mutants that are either identical to the original code or identical to each other.
While traditional approaches rarely generate duplicate mutants, 
LLMs can produce them due to their probabilistic text-generation mechanisms and learned patterns from large-scale training data.
After eliminating duplicate mutants, all remaining mutants are guaranteed to be syntactically distinct both from the original code and from each other.

%\subsubsection{RQ3: Equivalent mutations}
%Equivalent mutation ~\cite{kushigian2024equivalent,tian2024large} are notoriously harmful in mutation testing as they lead to deflated mutation scores.
%We evaluate the approaches by the proportion of equivalent mutations they generated.

\smalltitlecolon{RQ2.3: Equivalent Mutant Rate}
the proportion of equivalent mutants, computed as:
\begin{equation}
    EMR = \frac{\enspace | E | \enspace }{ | C - D | } 
\end{equation}
Equivalent mutants are semantically equivalent to the original program, which impacts the accuracy of the mutation score.
We prefer the approach with a lower \textit{EMR}.
Note that while equivalent mutants are also considered semantically duplicate, they are not easily identifiable. Therefore, we distinguish them from other duplicate mutants, which can be filtered syntactically.

\subsubsection{RQ3: Efficiency} 
Efficiency describes how fast a mutation approach produces mutants.
Mutation testing requires numerous mutants to evaluate the robustness of a test suite effectively. 
Therefore, the efficiency of mutant generation directly impacts the feasibility and scalability of mutation testing.
To assess this, we compare these approaches based on time costs.

\smalltitlecolon{RQ3.1: Average Generation Time}
measures the average time required to generate each mutant, providing a quantitative evaluation of the efficiency of each approach.

\smalltitlecolon{RQ3.2: Average Token}
measures the average number of consumed tokens per mutant, reflecting both the financial cost (due to token-based pricing) and the computational time cost (since longer token sequences typically require more processing time). Since input and output token costs differ across models, we calculate them separately.

%n

\subsection{Mutant Generation Setups} \label{sec:mut-gen}
In this subsection, we discuss the setups for mutant generation approaches, especially the setups for LLM-based generation.

%\subsubsection{Mutation Generation via Traditional Approaches}
For the well-established mutant generation approaches (i.e., PIT, Major, LEAM, and $\mu$BERT), we activate all their mutation operators.
Since we try to compare the similarity between the generated mutations and real bugs, we filter the mutants within the context of real-world bugs.
Note that we supply the same code snippets as context across all mutant generation approaches to ensure a fair comparison.

For all LLM-based approaches, we use the default settings of each model (e.g., token limits and temperature). For models accessible via Web API (e.g., GPT series and DeepSeek models), we rely on their official APIs. For the remaining open-source LLMs, we perform inference on four rented cloud servers, each equipped with dual NVIDIA GeForce RTX 3090 Ti GPUs, 168 GB of memory, and a 64-core Intel(R) Xeon(R) Platinum 8350C processor.

%% file: sections/eval.tex
\section{Evaluation Results} \label{sec:eval}

In this section, we aim to comprehensively evaluate these approaches via various aspects.
\tabref{phase1} presents the overall performance of each generation approach.
In terms of the number of mutants generated (Mutant Count), PIT produces the largest number of mutants at 340,728, significantly more than other approaches, while the remaining approaches range from about twenty thousand to thirty thousand.
In terms of mutation score, the traditional rule-based approaches (PIT and Major) are over 0.5, LEAM and $\mu$BERT are over 0.6, while most LLM-based approaches are over 0.7. Particularly, BugFarm (DS-671b) achieves the highest mutation score, which is nearly 0.79.

Note that the number of generated mutants and mutation scores are not quality metrics for evaluating these approaches.
A large number of mutants may include redundant or equivalent mutants that do not contribute to mutation scores, while a high mutation score may be influenced by the ease with which mutants are detected rather than their resemblance to real faults.

\input{sections/results/TAll-new}

%%%%%%%%%%%%%%%%%%%%%%%%%%%%%%%%%%%%%%%%%%%%%%%%%%%%%%%%%%%%%%%%%%%%%%%%%%%%%%%%%%%
\subsection{RQ1: Performance on Effectiveness} \label{subsec:rq3}
To measure the effectiveness of mutant generation approaches, we evaluate them with three sub-RQs, which estimate their effectiveness via real bug detectability, coupling, and diversity, respectively.

\subsubsection{RQ1.1 Detectability}
As illustrated in Section~\ref{subsec:coulping-metrics}, we estimate real bug detectability via the following metric.

\noindent\textbf{Real Bug Detection Rate:}
The column \textit{Real Bug Detec.} of \tabref{phase1} presents the proportion of real bugs that can be detected across different mutant generation approaches.

Among the traditional approaches, PIT achieves a detection rate of 40.1\%, while Major, LEAM, and $\mu$BERT perform significantly better with a detection rate of 66.9\%, 63.9\%, and 60.5\%, respectively.

For LLM-based approaches, the results are significantly more effective than those of traditional ones.
Across all models, LLMorpheus and LLMut constantly outperform all traditional approaches. For example, LLMut (DS-671b) achieves the highest detection rate at 91.1\%, closely followed by LLMut (GPT-4o-M) and LLMut (GPT-4o), with detection rates of 90.8\% and 89.9\%, respectively. Moreover, we can find that all DS-671b outperforms GPT-4o in terms of bug detection rate, across all three LLM-based approaches.

\input{sections/results/TRealBugDet-new}

To better understand the detection of the generation approaches, we analyze their project-wise detectability, shown as \tabref{real-bug-detect}.
For individual projects, LLM-based approaches consistently achieve top performance across multiple projects.
Notably, LLMut (DS-671b) detects the largest number of bugs in most Defects4J projects and achieves the highest overall bug detection count. Similarly, LLMut (DS-236b), LLMut (GPT-4o), and LLMut (GPT-4o-M) demonstrate strong performance, achieving the best results in projects such as Mockio, Csv, and ConDefects.

\subsubsection{RQ1.2 Coupling}
To assess the degree of coupling, we perform comprehensive experiments that include both an overall metric evaluation and a project-wise analysis.  

\noindent\textbf{Average Ochiai Coefficient:}
The column \textit{Avg. Ochiai} of \tabref{phase1} presents the Ochiai Coefficient values across the approaches, which measure the semantic distance between mutants and real bugs.
Among the traditional approaches, Major achieves an Ochiai coefficient of 35.5\%, higher than the others.
LEAM and $\mu$BERT achieve coefficients of 32.4\% and 29.3\%, respectively, indicating that the best-performing rule-based approach, Major, outperforms both of them.

The average Ochiai coefficients of LLM-based approaches vary significantly, ranging from 17.7\% to 59.8\%, indicating diverse semantic distances of their mutants. Among them, LLMut (DS-671b) achieves the highest value at 59.8\%, followed by LLMut (GPT-4o) at 56.2\% and LLMut (GPT-4o-M) at 54.1\%, indicating that these mutants capture fault-triggering behaviors more closely aligned with real bugs compared to traditional approaches.

\noindent\textbf{High Ochiai Bug Count:}
Following the study of Ojdanic et al.~\cite{ojdanic2023syntactic}, the mutant set with an Ochiai coefficient over 0.8 is considered to be of high semantic similarity with its corresponding real bug.
For each project, we count the number of bugs that achieved a high Ochiai coefficient using each approach, as shown in \tabref{ochiai}.
The overall trend is similar to \textit{Real Bug Detectability}, which indicates that the LLMut (DS-671b) achieves the highest number of bugs across most projects. 
Moreover, LLMut (GPT-4o) and LLMut (GPT-4o-M) also exhibit strong performance, closely following LLMut (DS-671b) in many cases.
Compared with our approach, all traditional approaches exhibit significantly fewer high Ochiai bugs. Interestingly, among traditional approaches, the rule-based approach Major outperforms the other learning-based ones, LEAM and $\mu$BERT, across most projects.

\input{sections/results/TOchiai-new}

\noindent\textbf{Coupling Rate:}
The overall coupling rate for each mutant generation approach is shown in the column \textit{Coup. Rate} of \tabref{phase1}.
The traditional approaches, PIT, Major, LEAM, and $\mu$BERT, exhibit lower coupling rates than LLM-based approaches, with coupling rates of 23.5\%, 38.9\%, 38.1\%, and 47.3\%, respectively.
Among the LLM-based approaches, LLMorpheus (DS-671b) and LLMorpheus (GPT-4o) achieve the highest coupling rates, reaching 52.0\% and 51.7\%, respectively, suggesting that LLMorpheus produces mutants whose failing behaviors align most closely with those of real bugs. LLMut (DS-671b) and LLMut (GPT-4o-M) closely follow LLMorpheus, with coupling rates of 50.8\% and 50.6\%, respectively.
Overall, the trends are similar to the behavior similarity discussed in Section~\ref{subsec:rq3}, indicating the superiority of the LLM-based approaches, particularly LLMorpheus (DS-671b), in generating overall highly coupled mutants that better mimic real bugs.

\input{sections/results/TCoupling-new}

\noindent\textbf{Coupling Categorization:}
As mentioned in Section~\ref{subsec:coulping-metrics}, we classify each mutant of each generation approach into different scales, i.e., \textit{Strong Substitution}, \textit{Strong + Additional Tests Fail}, \textit{Partial Substitution}, \textit{Partial + Additional Tests Fail}, \textit{Not Substitution}, and \textit{Not Detected}.
Note that the first four categories are used to calculate the \textit{Coupling Rate}, indicating these mutants semantically overlap with their corresponding real bugs.

\tabref{coupling} presents the distribution of each generation approach.
Among the traditional approaches, PIT and Major exhibit a significantly higher proportion of mutants in the \textit{Not Detected} category, with rates of 45.9\% and 47.9\%, respectively, indicating that a substantial number of their generated mutants fail to couple with real bugs.
However, they achieve relatively higher proportions in the \textit{Strong} category, with PIT at 10.1\% and Major at 9.6\%, suggesting that while their coupling ability is limited, they can still generate a small number of strongly coupled mutants.
In contrast, LLM-based approaches overall demonstrate stronger coupling capabilities across multiple categories.
For instance, LLMorpheus (DS-671b) achieves the highest proportion in the \textit{Strong} category at 11.8\%, closely followed by LLMut (DS-671b) with the value of 11.2\%. BugFarm (DS-671b) also achieves a notable 10.0\% of strongly coupled mutants. In terms of the \textit{Strong + Additional Tests Fail} category, LLMut (GPT-4o-M) leads with a proportion of 12.1\%.
Additionally, BugFarm and LLMorpheus exhibit the lowest proportion of \textit{No Substitution} and \textit{Not Detected}, indicating their mutation strategies target to.
Overall, LLM-based approaches generate a larger proportion of both strongly coupled and partially coupled mutants, demonstrating a stronger ability to generate diverse and highly coupled mutants.
%while the traditional rule-based approaches generate a larger proportion of strong substitution mutants,

\subsubsection{RQ1.3 Diversity}
We explore the syntactic diversity of mutants generated by each approach.
Note that PIT works at the Java bytecode level, we thus skip it in this RQ.

\begin{figure}[t]
    \centering
    \includegraphics[width=0.70\linewidth]{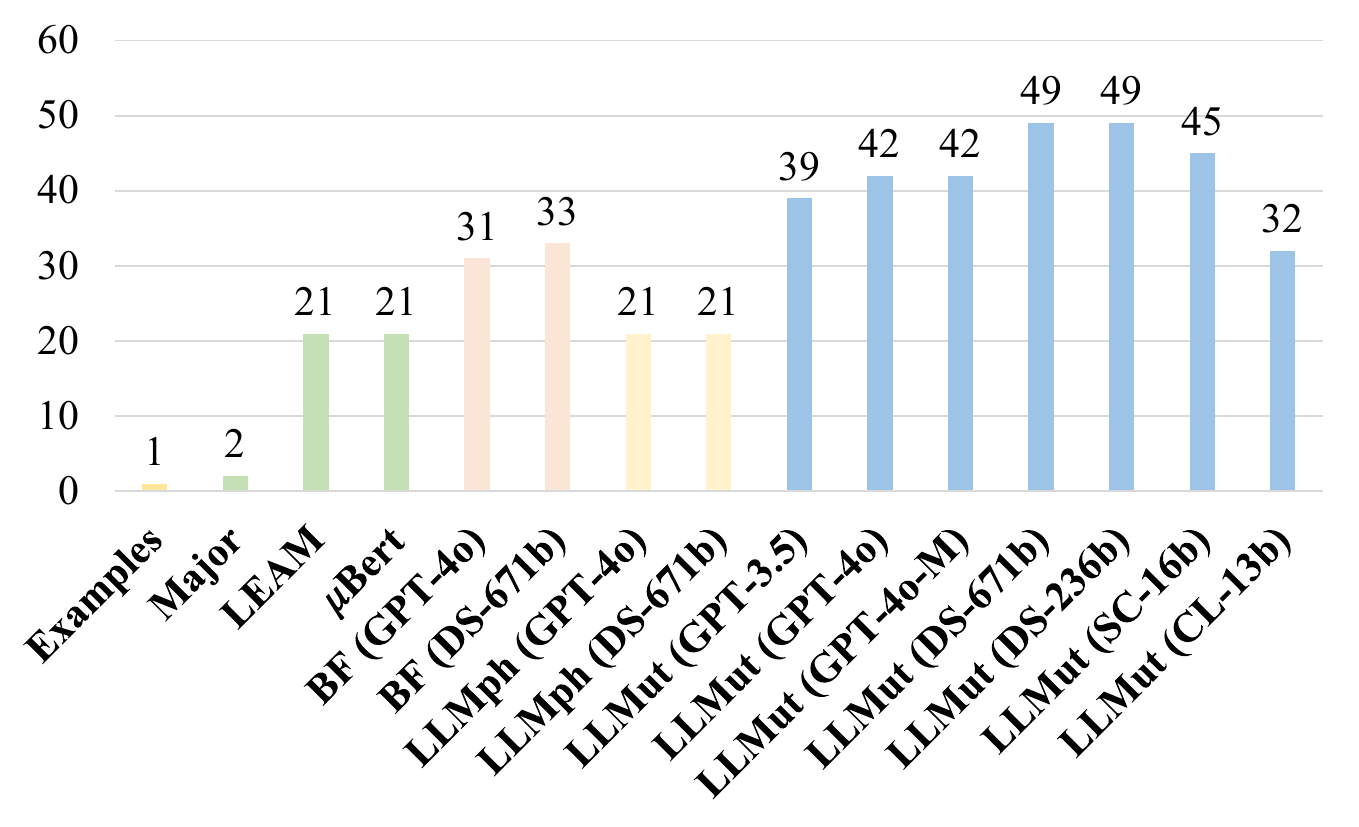}
    \caption{Number of Newly Introduced AST Nodes of the Studied Approaches and the Few-Shot Examples}
    \figlabel{diversity}
\end{figure}

%\begin{wrapfigure}[11]{r}{0.5\linewidth}
%    \centering
%    \includegraphics[width=0.95\linewidth]{figs/diversity-bar1.pdf}
%    \caption{Number of newly introduced AST nodes of the studied approaches and the few-shot examples.}
%    \figlabel{diversity}
%\end{wrapfigure}

\noindent\textbf{AST Node Diversity:}
To assess the diversity of AST node types, we first determine if a mutant involves \textit{deletion} (i.e., removing a code element).
For the \textit{non-deletion mutants}, we parse the code before and after mutation by the parser Javalang\footnote{\url{https://pypi.org/project/javalang}}, to observe which new AST node types are introduced.
A larger number of newly introduced AST node types indicates richer diversity.

\figref{diversity} illustrates the number of introduced AST node types, with traditional approaches highlighted in green and LLM-based approaches in blue.
The figure reveals that LLM-based approaches present more types of new AST nodes than traditional approaches.
The best LLM-based approaches, LLMut (DS-671b) and LLMut(DS-236b), exhibit the greatest diversity, introducing 49 new AST node types, followed by GPT-4o-Mini, which introduces 45 new types.
Additionally, all models with LLMut introduce significantly more AST node types than traditional approaches. Meanwhile, BugFarm (DS-671b) and BugFarm (GPT-4o) present 33 and 31 new AST node types, respectively, which also outperform other traditional approaches.
%Additionally, \todo{all six LLMs based on our prompt are significantly higher than all other approaches in AST node types.
%The existing LLM-based approaches, BugFarm and LLMorpheus, introduce 31 and 21 new AST node types, respectively. While BugFarm outperforms traditional approaches, it still lags behind ours.}
We also check the AST node diversity of the examples we provided in the LLM prompts (see~\tabref{few-shots}), as shown by the ``Examples'' in \figref{diversity}.
Interestingly, we find that these examples introduce only one new type of AST node and are subsumed by all other approaches. 
This indicates that LLMs are creative in generating mutants, producing significantly more diverse mutants than the examples provided to them.

\input{sections/results/TDiversity-new}

We further analyzed the distribution of deletions and AST node types across different approaches, as shown in~\tabref{diversity}, which highlights the proportion of the mutants that delete the target code element and the top three most frequent AST node types.
For the proportion of deletions, the traditional approach, LEAM, shows the highest deletion mutant ratio, with 48.9\%, followed by Major at 15.3\%.
In contrast, all LLM-based approaches generate deletion mutants at a rate below 6\%, with three models producing less than 1.0\%.
This indicates that, compared to existing approaches, LLMs are less inclined to generate deletion mutants.
As shown later, we find that deletions are one of the key factors contributing to the lower compilation success rate.

Regarding AST node types, the most frequent new AST node type of all the LLM-based approaches is \code{Literal} (LT) (e.g., $a>b \mapsto \text{true}$).
However, for the traditional approach Major, the most frequent node type is Statement (e.g., replacing a statement with an empty statement).
Note that although these mutants are not with the deletion operator, a significant portion of the production of empty statements exists, leading to a large proportion.
For the second most frequent new node types, \code{BinaryOperation} (BO) appears most often among LLM-based approaches, indicating that LLMs frequently modify or insert binary expressions such as arithmetic or comparison operators.
For the third most frequent category, LLM-based approaches introduce a variety of AST nodes, including \code{MethodInvocation} (MI), \code{MemberReference} (MR), and \code{ClassCreator} (CC), reflecting their broader syntactic diversity in mutant generation.

\noindent\textbf{AST Edit Distance:}
To quantify the extent of syntactic modifications introduced by each mutant generation approach, we compute the AST edit distance for each mutant using Gumtree~\cite{falleri2014fine}.
The distribution of edit distances for each approach is presented in \tabref{ast-distance}, providing insights into the complexity and diversity of the generated mutants.

\input{sections/results/TASTDistance-new}

From the table, we can find that all traditional approaches predominantly generate small-scale changes.
Specifically, the proportion of mutants with an edit distance of 1 for Major, LEAM, and $\mu$BERT is 69.2\%, 58.5\%, and 73.8\%, respectively, indicating they involve minimal syntactic changes to the original code in the majority of cases.
In contrast, LLM-based approaches exhibit a broader range of edit distances, reflecting their capability to introduce more complex transformations.
For example, only 38.3\% of the mutants of BugFarm (DS-671b) have an edit distance of 1, and LLMut (DS-671b) generates 45.1\% of mutants with an edit distance of 1 and 44.2\% with a distance of 2, with an additional 6.9\% corresponding to distances of 3. This pattern suggests that LLM-based methods, unlike traditional approaches, are capable of performing multi-node or compositional edits, leading to a richer diversity of syntactic structures.

\begin{tcolorbox}
\underline{\textbf{Answer to RQ1:}} 
In most cases, LLM-based approaches outperform traditional methods in effectiveness. For example, the weighted averages of the rule-based approach in terms of real bug detection rate and coupling rate are 41.64\% and 24.37\%, respectively. In contrast, LLM-based approaches reach 76.47\% and 51.54\%, representing a significant improvement in both metrics. Specifically, LLMut (DS-671b) achieves the highest real bug detection rate (91.1\%) compared to PIT (40.1\%) and Major (66.9\%). LLMorpheus (DS-671b) achieves the highest coupling rate (51.7\%, versus 17.9\% for PIT and 42.7\% for Major). Moreover, LLM-generated mutants exhibit markedly greater syntactic diversity, introducing a wider range of new AST node types (e.g., 49 for LLMut (DS-671b) and LLMut (DS-236b)) and a lower proportion of deletion mutations (e.g., 0.3\% for LLMorpheus (GPT-4o) versus 15.3\% for Major).
\end{tcolorbox}

\subsection{RQ2: Performance on Validity} \label{sec:rq1}
To evaluate the validity of mutant generation approaches, we consider the aspects that may impact their validity, i.e., compilability rate, duplicate mutant rate, and equivalent mutant rate.
This RQ is divided into the following sub-RQs.

\subsubsection{RQ2.1: Compilability}
The column \textit{Comp. Rate} of \tabref{phase1} presents the overall Compilability Rate ($CR$) of each approach.
Among the traditional approaches, the rule-based approaches, PIT and Major, demonstrate superior compilability, with PIT achieving a perfect 100\% $CR$, and Major following closely at 97.60\%.
Note that although Major operates under simple syntax rules, it can still generate non-compilable mutants.
For example, the Java compiler's analyzer rejects constructs like \code{while(true)}.
PIT uses the mutation operators more restrictively and previously filters out non-compilable mutants.
In contrast, LEAM and $\mu$BERT, have rates at 35.0\% and 22.5\%, respectively.

For LLM-based approaches, there is a considerable variation in compilability. Among all LLM-based approaches, LLMut (DS-671b) and BugFarm (DS-671b) attain the highest $CR$ of 77.6\% and 77.5\%, respectively. LLMut (GPT-4o), LLMut (DS-236b), and LLMut (GPT-4o-M) closely follow by achieving 76.4\%, 75.8\%, 75.3\%, respectively. Most LLM-based approaches show significantly higher compilation success rates compared to the traditional learning-based approaches, such as LEAM and $\mu$BERT. For example, the $CR$s of BugFarm, LLMorpheus, and LLMut with GPT and DeepSeek models are over 30\% higher than those of LEAM and $\mu$BERT.

%\begin{tcolorbox}
%\underline{\textbf{Answer to RQ2.2:}} 
%Both rule-based approaches, PIT and Major, outperform LLMs in compilability rate, producing minimal non-compilable or useless mutations.
%In contrast, GPT-4o, the best-performing LLM, generates 23.6\% non-compilable mutations.
%\end{tcolorbox}

\subsubsection{RQ2.2: Duplicate Mutant Rate}
As aforementioned, duplicate mutants refer to those that are redundant or identical to the original code, providing no meaningful contribution to the mutant testing process.
Therefore, a lower Duplicate Mutant Rate ($DMR$) is desirable, as it indicates a more efficient mutant generation process with minimal redundancy.
The column \textit{Dup. Mut. Rate} of \tabref{phase1} presents the results.

Rule-based approaches, PIT and Major, achieve a perfect 0\% $DMR$.
This is expected, as these approaches apply well-defined mutation rules that ensure each generated mutant introduces a meaningful change to the program.
Other traditional approaches, i.e., LEAM and $\mu$BERT, show relatively low $DMR$s of 2.3\% and 1.7\%, respectively.
In contrast, LLMs generate a significant portion of duplicate mutants, exhibiting a higher $DMR$.
For the LLM-based approaches evaluated with the best-performing model, DS-671b, the duplicate mutation rates ($DMR$s) are as follows: BugFarm (DS-671b): 7.5\%, LLMorpheus (DS-671b): 6.9\%, and LLMut (DS-671b): 7.5\%. The smaller models on LLMut exhibit much worse performance, such as LLMut (SC-16b) and LLMut (CL-13b). These results indicate that even equipped with the state-of-the-art LLM, these approaches still generate a non-negligible proportion of duplicate mutants.
%For the LLMs from OpenAI, GPT-3.5, GPT-4o, and GPT-4o-Mini exhibit moderate $DMR$s of 10.3\%, 7.8\%, and 7.0\%, respectively.
%For the existing LLM-based approaches, BugFarm and LLMorpheus, exhibit similar $DMR$s of 7.5\% and 7.1\%, respectively.
%The open-source LLMs such as StarCoder and DeepSeek-Coder-V2-236b demonstrate higher $DMR$s of 12.8\% and 7.7\%, respectively, while CodeLlama shows the highest $DMR$ at 37.1\%.

\subsubsection{RQ2.3: Equivalent Mutants}
Given the undecidable nature of identifying equivalent mutants, we manually evaluate each mutant's equivalence, following the methodology of the existing study~\cite{kushigian2024equivalent}.
For each mutant generation approach, we randomly sample a subset of compilable mutants, ensuring a 95\% confidence level with a 10\% margin of error.

Two authors independently evaluate the equivalence of the selected mutants to minimize subjectivity.
The procedure for checking a mutant involves first transplanting it into the original project and executing all tests. If it passes all tests, the author then manually verifies whether the mutant is semantically equivalent.
Note that PIT does not generate source-level mutants, so its mutants must be manually extracted from Java bytecode and transformed back into source-level representations, which is significantly more time-consuming.

In total, 1437 mutants were labeled, 860 of which were filtered by tests (i.e., killed), and the remaining 577 were manually checked.
The entire process took approximately thirteen hours per person.
The manual verification requires checking the code context, which may involve multiple classes and methods.
For example, to check a mutant that replaces an integer with a method invocation, where the method is invoked by the \textit{this} object but defined in the parent class, we should review the parent class file and verify whether its return value is consistently equal to the integer.
We measure the consistency between the two authors by Cohen's kappa coefficient ($\kappa$)~\cite{viera2005understanding}, achieving an overall $\kappa$ value of 0.8, which indicates a high level of agreement.
In cases of disagreement, both authors conducted thorough discussions to resolve contradictions and finally reach consistent labels.
The labeling results are available at our repo.

\input{sections/results/TSample-new}

%\noindent\textbf{Equivalent Mutation Rate:}
\tabref{sample-eq} presents the estimated Equivalent Mutant Rate ($EMR$), reporting the number of sampled mutants (\#Sampled), the number of equivalent mutants we identified (\#Eq. Mut), and the $EMR$ for each approach.
Among the traditional approaches, PIT achieves the lowest $EMR$ of 1.0\%, followed by Major with an $EMR$ of 2.1\%, indicating that rule-based techniques produce fewer equivalent mutants.
In contrast, the LLM-based approaches exhibit higher $EMR$s, with variations across different models.
LLMut (SC-16) performs the worst by attaining the highest $EMR$s of 10.6\%, while LLMut (GPT-4o), LLMut (GPT-4o-M), and  LLMut (DS-236b) show relatively lower $EMR$s, ranging from 3.1\% to 5.2\%. BugFarm and LLMorpheus exhibit $EMR$s ranging from 3.1\% to 5.2\%. Overall, the LLM-based approaches with advanced models (e.g., GPT-4o or DS-671b) tend to have slightly higher $EMR$s than PIT and Major, the increase remains modest and within an acceptable range.
%StarChat has the highest $EMR$s of 10.6\%, while GPT-4o, GPT-4o-Mini, and DeepSeek-Coder-V2-236b show relatively lower $EMR$s, ranging from 3.1\% to 5.2\%.
%Overall, although the OpenAI models such as GPT-4o and GPT-4o-Mini tend to have slightly higher $EMR$s than PIT and Major, the increase remains modest and within an acceptable range. 

Note that the estimated $EMR$ of Major given by the recent study of Kushigian et al.~\cite{kushigian2024equivalent} is 2.97\%, where the authors sampled a subset of mutants for manual judgment.
We adopt the same experimental process, and the result of Major is 2.0\%, which is consistent with theirs, considering the randomness introduced by sampling.
Moreover, the definition of equivalent mutants in this paper is the set of syntactically different but semantically equivalent mutants, which may differ from existing studies~\cite{tip2025llmorpheus}.

%\begin{tcolorbox}
%\underline{\textbf{Answer to RQ2.4:}} Rule-based approaches, PIT and Major, achieve the lowest equivalent mutant ratios at 1.0\% and 2.0\%, respectively. In contrast, LLM-based approaches generally produce higher equivalent mutant rates, with StarChat generating the highest at 10.6\% and GPT-4o the lowest at 3.1\%.
%\end{tcolorbox}

\begin{tcolorbox}
\underline{\textbf{Answer to RQ2:}} 
Rule-based approaches (i.e., PIT and Major) outperform LLMs in compilability rate, duplicate mutant rate, and equivalent mutant rate, with a weighted average of 99.90\%, 0.00\%, and 1.10\%, respectively.
While the weighted average values of LLM-based approaches are 67.17\%, 9.38\%, and 4.74\%. On average, the rule-based approaches produce about 1.5$\times$ more compilable mutants and 4-9$\times$ fewer redundant or equivalent mutants than LLM-based approaches.
Although rule-based approaches are slightly better in terms of equivalent mutant rates, the equivalent mutant rates across approaches remain comparable.
\end{tcolorbox}
%On average, PIT and Major generate a single mutant in just 0.02s and 0.08s, respectively, whereas the best-performing LLM, GPT-4o, requires 1.31s. Additionally, 

\subsection{RQ3: Performance on Efficiency}
To evaluate the performance on efficiency, we record the average generation time for each approach.

\noindent\textbf{Average Generation Time:}
The column \textit{Time} of \tabref{phase1} presents the results of \textit{Average Generation Time} in seconds.
Among the traditional approaches, PIT demonstrates the fastest generation time at 0.02 seconds per mutant, followed by Major, which takes 0.08 seconds, making them the most efficient approaches.
In contrast, LEAM and $\mu$BERT take 3.06 seconds and 2.34 seconds, respectively.
The LLM-based approaches present much slower generation speed, ranging from 1.31 to 9.06 seconds. In particular, all LLM-based approaches require more time when adopting open-source models. For example, LLMut (SC-16b), LLMut (CL-13b), and LLMut (DS-236b) require 7.53, 9.06, and 4.25 seconds, respectively.
The results suggest that the rule-based approaches are the most time-efficient for generating mutants, which are tens to hundreds of times faster than other approaches.
Note that the time costs of self-deployed open-source models typically depend on the types of GPUs used for execution.
Additionally, the response time of models accessed via network APIs may vary across service providers and can be further affected by network fluctuations.
We further discuss these potential impacts in Section~\ref{sec:threats}.

\input{sections/results/TToken}

\noindent\textbf{Average Token:}
\tabref{tokens} presents the average number of input and output tokens used by each LLM-based approach to generate a single mutant.
All the token usage is computed as the official library of OpenAI,  Tiktoken\footnote{\url{https://github.com/openai/tiktoken}}, which is the standard practice for estimating model usage.

From the table, we observe substantial differences across prompting strategies. 
BugFarm uses relatively short prompts (200 input tokens on average), but its output token usage is high (over 360 tokens) because it returns the \emph{code of the entire mutated method}.
Consequently, its overall token consumption needs over 2.7M input tokens and about 5M output tokens.
In contrast, LLMorpheus requires the largest number of input tokens (370 on average, 7.9M in total), as its prompt includes up to 200 lines of surrounding context, yet it produces moderate output (around 74 tokens on average, 1.5M in total).
Moreover, LLMut requires a moderate amount of input and output tokens, which cost 126 input tokens and 58-63 output tokens to generate one mutant.
In total, LLMut needs 4.3M input tokens and 1.6M-2.1M output tokens, which are also moderate.
Since LLMut asks the models to generate more mutants within a single conversation and wraps the output mutants directly with JSON tags, the average cost is significantly lower than BugFarm and LLMorpheus.
These results demonstrate that various prompt engineering strategies significantly impact token costs.

%For example, BugFarm (GPT-4o) requires 200 input tokens and over 360 output tokens on average, while LLMorpheus (GPT-4o) requires 370 input tokens and about 74 output tokens. LLMut is more cost-efficient, requiring 126 input tokens and 58-63 output tokens. These results demonstrate that various prompt engineering strategies significantly impact token costs.}
%The existing LLM-based approaches, BugFarm and LLMorpheus, consume 200 and 370 input tokens, and 368 and 74 output tokens per mutant, respectively. LLMs using our prompt consume substantially fewer tokens, averaging 126 input tokens and 59–63 output tokens per mutant. As all LLMs share identical inputs, their input token counts remain the same.

\begin{tcolorbox}
\underline{\textbf{Answer to RQ3:}} 
Rule-based traditional approaches, such as PIT and Major, outperform LLMs in time cost. On average, PIT and Major generate a single mutant in just 0.02s and 0.08s, respectively, whereas the best-performing LLM, GPT-4o, requires 1.31s.
In terms of token efficiency, prompt engineering plays an important role.
%The LLMs based on our prompt only require 126 input tokens and 58-63 output tokens per mutant.
\end{tcolorbox}

\subsection{RQ4: Categorizations of Non-Compilable Mutants} \label{sec:rq6}
To answer this RQ, we first identify the error types of non-compilable mutants and then analyze the surrounding code contexts that are error-prone in generating them.

\subsubsection{Categorizations of compilation error types}
Non-compilable mutants require a compilation step to filter out, which results in wasted computational resources.
As mentioned in Section~\ref{sec:rq1}, LLMs generate many non-compilable mutants.
This RQ analyzes the types of errors and potential root causes of these non-compilable mutants.
Following the setting of the previous steps, we first sample 384 non-compilation mutants from the outputs of GPT models, ensuring the confidence level is 95\% and the margin of error is 5\%~\cite{linares2017enabling,tufano2019learning,macklon2023taxonomy,de2015investigating}.
From the manual analysis of these non-compilable mutants, we identified nine distinct error types, shown as follows.
\begin{enumerate}
    \item \textit{Usage of Unknown Methods}: This type of error is due to invoking unknown methods under the code context.
    \item \textit{Code Structural Destruction}: This type of error is due to illegal code structures, such as unmatched parentheses or braces.
    \item \textit{Incorrect Method Parameters}: This type of error is due to invoking a method with illegal parameters, including invoking with the wrong number of parameters or mismatched types.
    \item \textit{Usage of Unknown Variables}: This type of error is due to using unknown variables in the code context.
    \item \textit{Usage of Unknown Types}: This type of error is due to using unknown type names in the code context.
    \item \textit{Type Mismatch}: This type of error is due to using illegal types when applying a certain operation, such as using objects to perform arithmetic operations.
    \item \textit{Incorrect Initialization}: This type of error is due to using illegal forms of constructors or deleting necessary initialization statements.
    \item \textit{Incorrect Location}: This type of error is due to a mutant being applied to an incorrect code location. 
    \item \textit{Incorrect Exceptions}: This type of error is due to throwing incorrect exceptions.
\end{enumerate}

\input{sections/results/TCompilationErrorTypes}

\tabref{error-type} presents the results.
From the table, we observe that the most frequent error type, \textit{Usage of Unknown Methods}, accounts for 27.3\% of total errors, highlighting the challenge of hallucination in LLM-generated code~\cite{huang2023survey}.
The second most common error, \textit{Code Structural Destruction}, makes up 22.9\%, indicating that maintaining syntactic correctness remains a persistent difficulty for LLMs.
These findings underscore the need for further improvements in LLMs to enhance their reliability and accuracy in code generation.

\begin{figure}[t]
  \centering
  
  \begin{subfigure}[t]{0.32\linewidth}
    \centering
    \includegraphics[width=\linewidth]{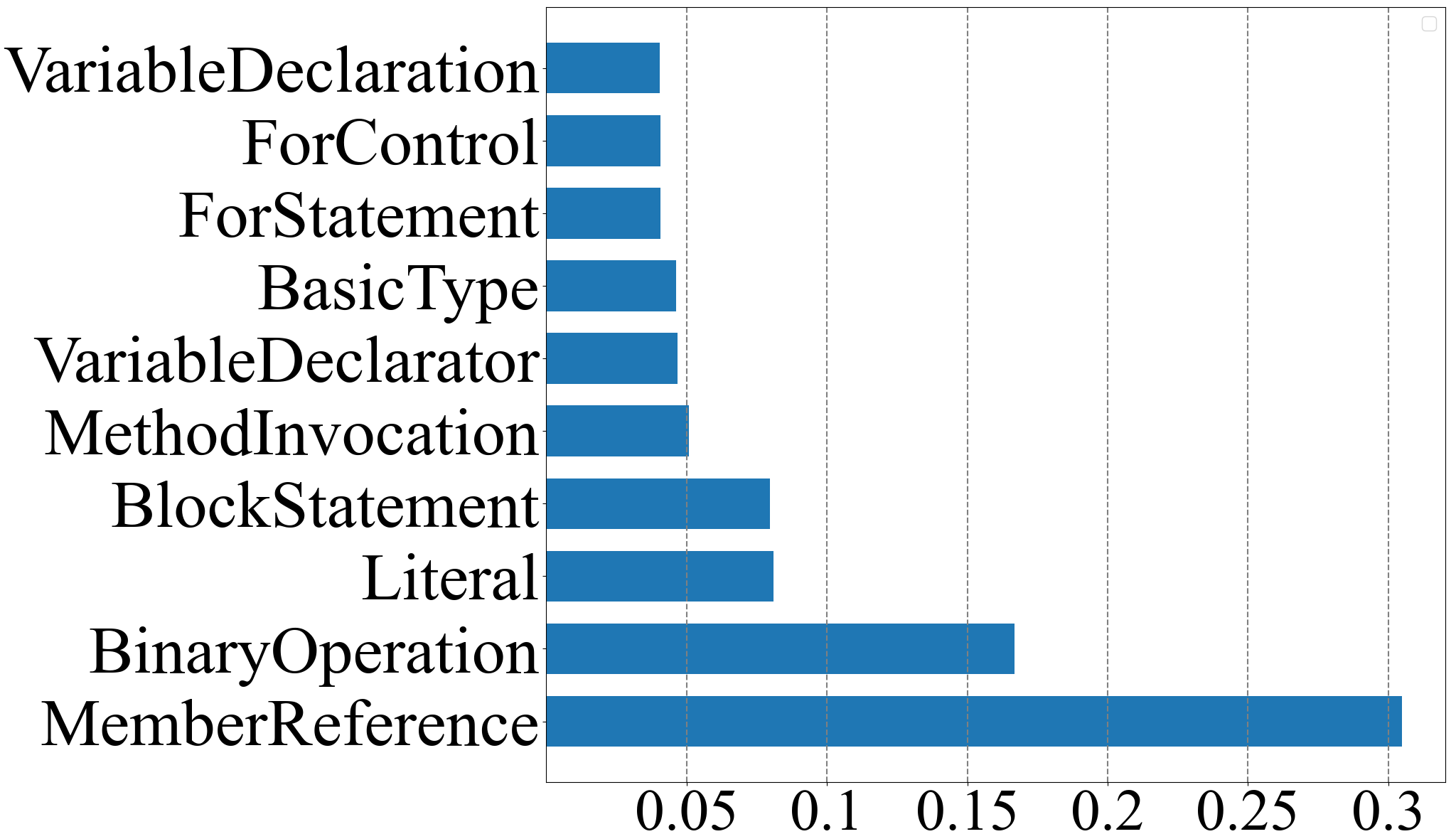}
    \caption{Major}
    \figlabel{non-compilable-major}
  \end{subfigure}
  \begin{subfigure}[t]{0.32\linewidth}
    \centering
    \includegraphics[width=\linewidth]{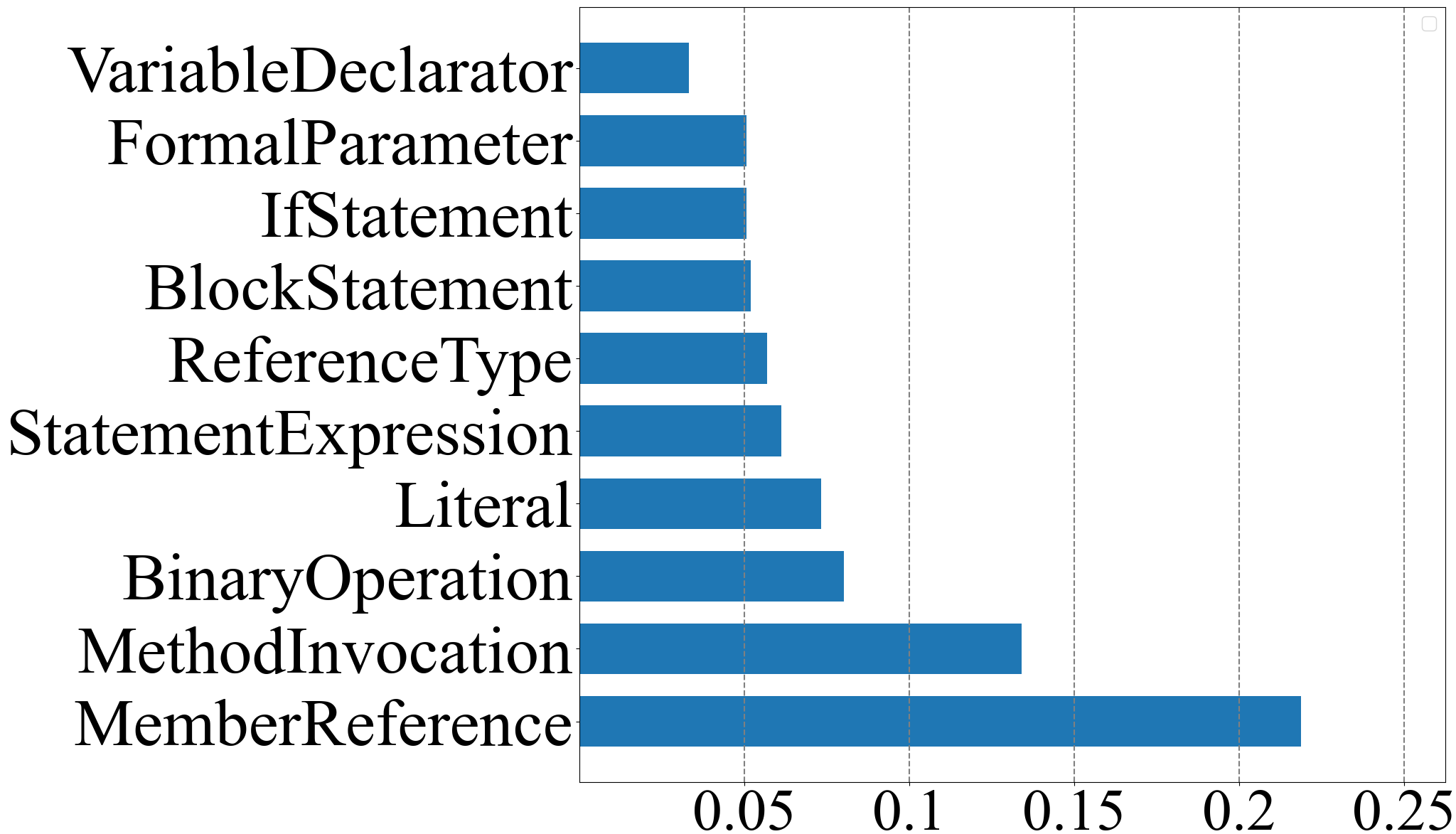}
    \caption{LEAM}
    \figlabel{non-compilable-leam}
  \end{subfigure}
  \begin{subfigure}[t]{0.32\linewidth}
    \centering
    \includegraphics[width=\linewidth]{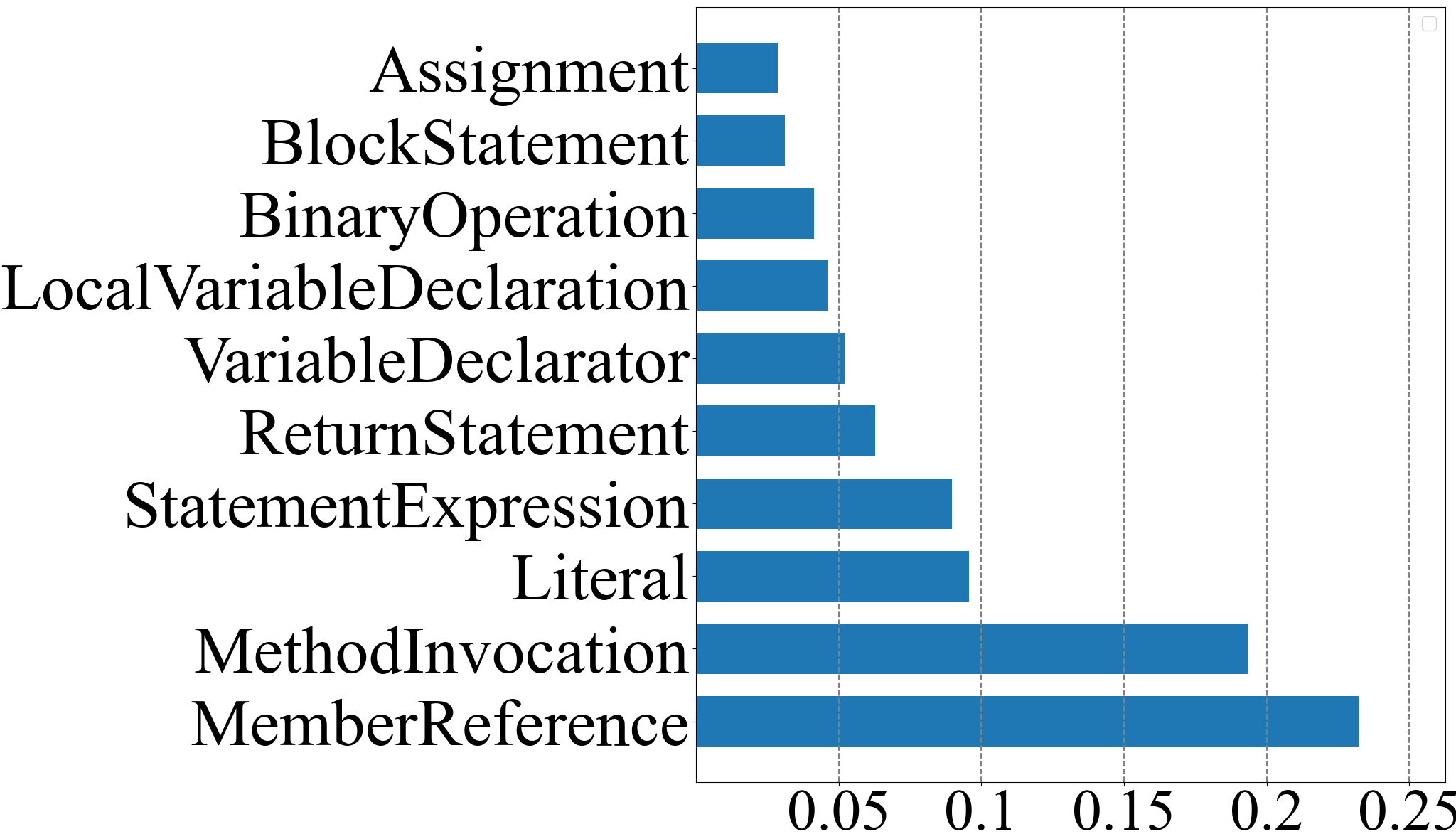}
    \caption{$\mu$BERT}
    \figlabel{non-compilable-mubert}
  \end{subfigure}

  \begin{subfigure}[t]{0.32\linewidth}
    \centering
    \includegraphics[width=\linewidth]{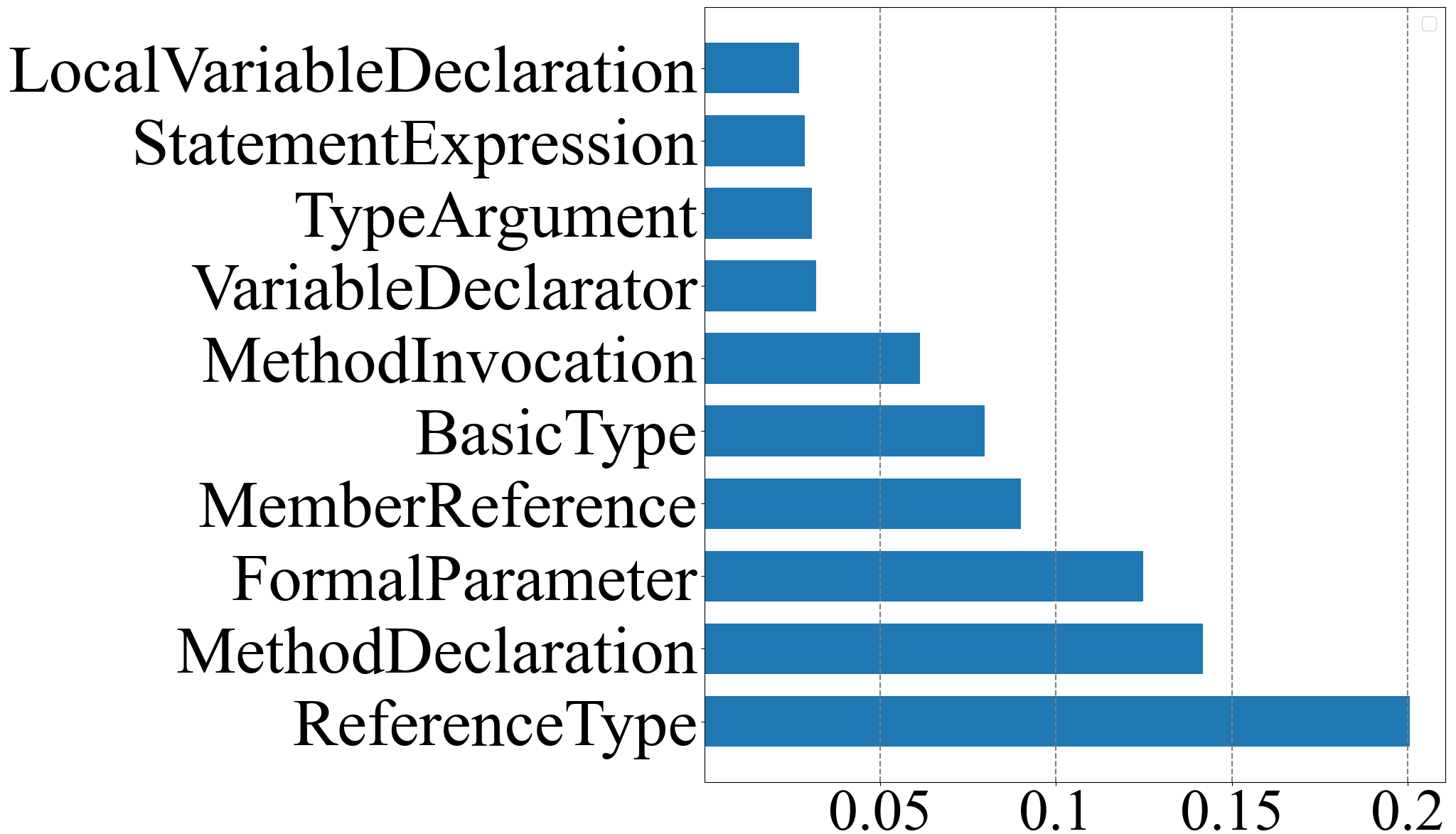}
    \caption{BugFarm (GPT-4o)}
    \figlabel{non-compilable-bugfarm-gpt4o}
  \end{subfigure}
  \begin{subfigure}[t]{0.32\linewidth}
    \centering
    \includegraphics[width=\linewidth]{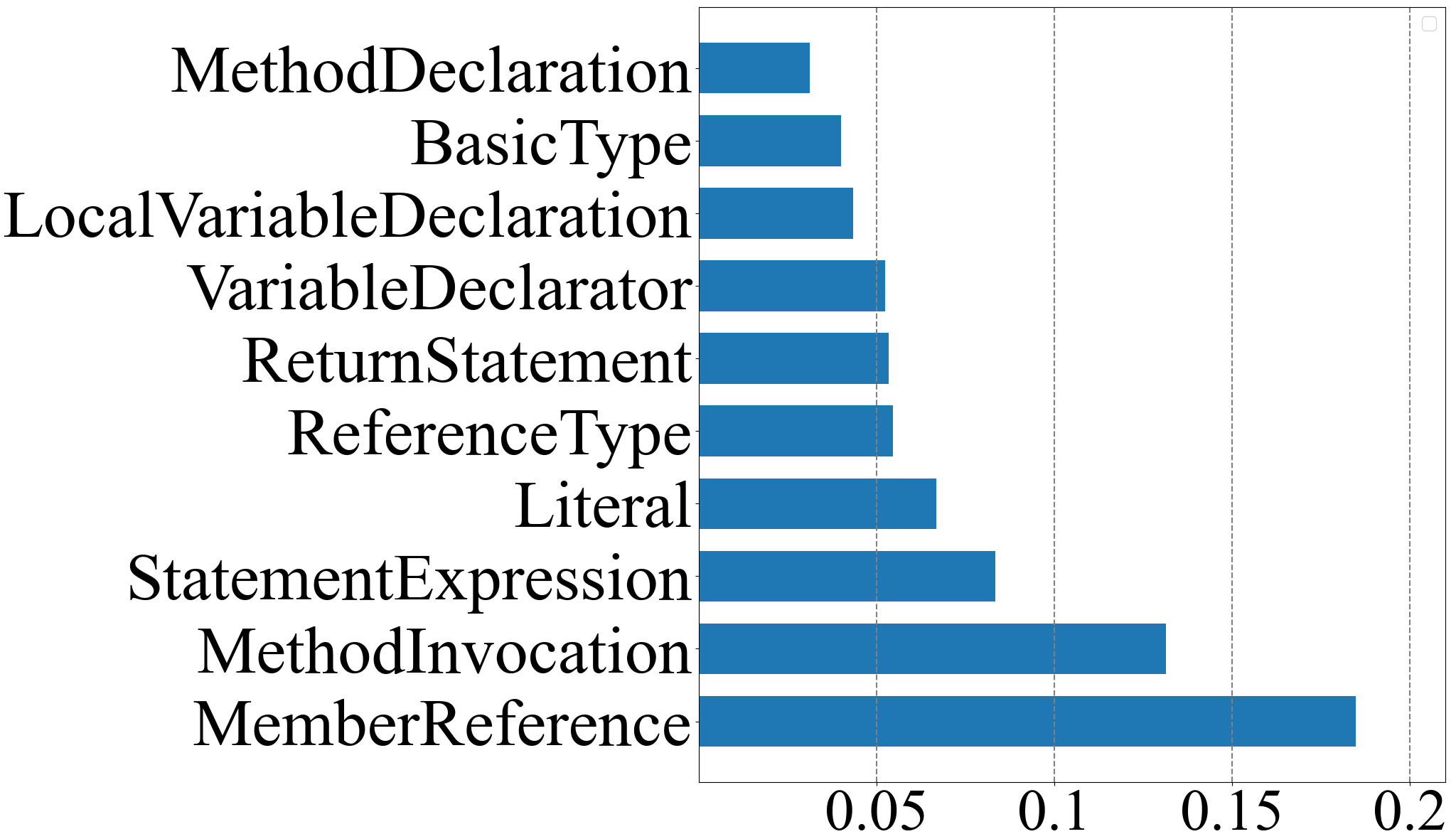}
    \caption{BugFarm (DS-671b)}
    \figlabel{non-compilable-bugfarm-ds671b}
  \end{subfigure}
  \begin{subfigure}[t]{0.32\linewidth}
    \centering
    \includegraphics[width=\linewidth]{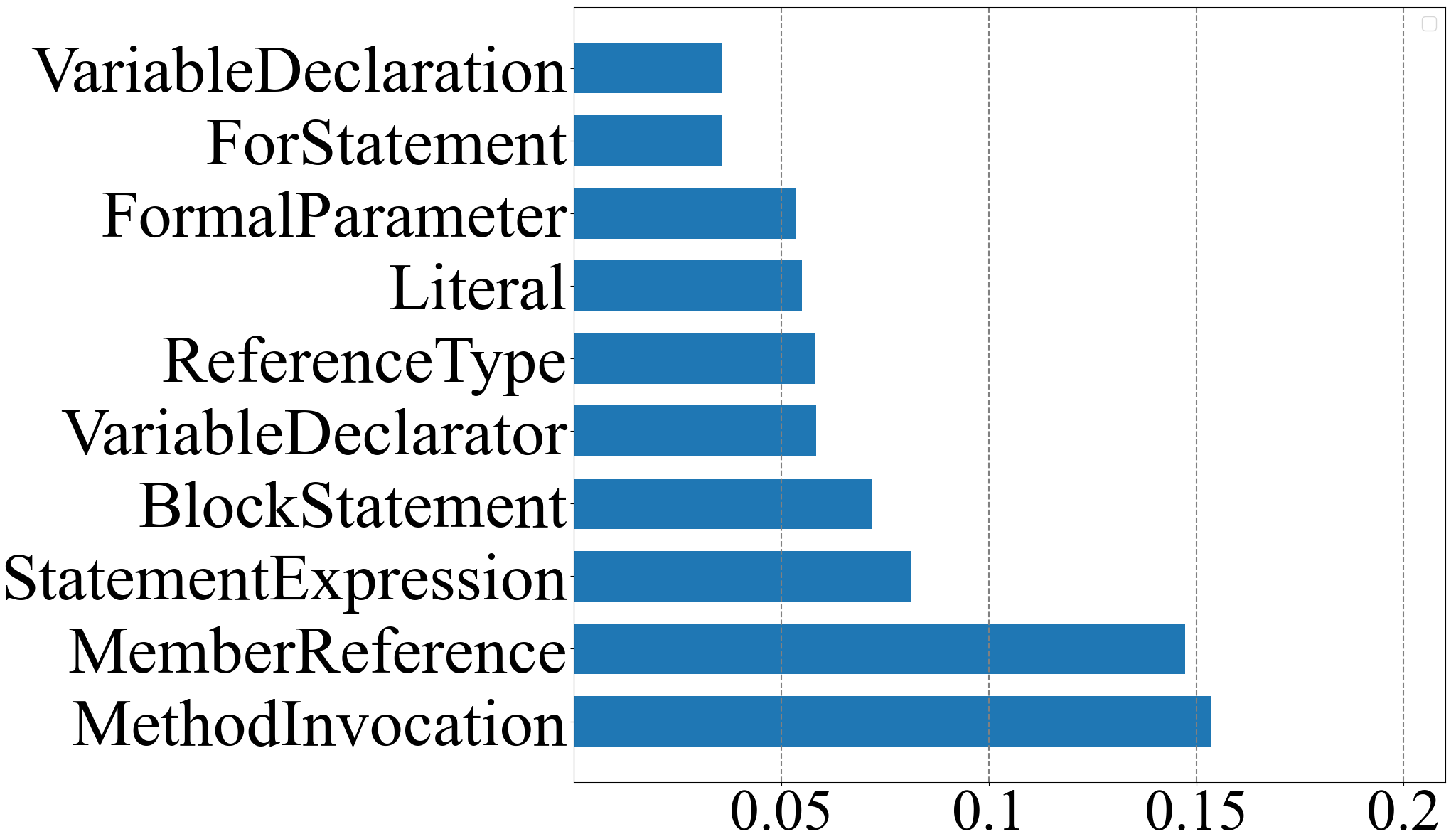}
    \caption{LLMorpheus (GPT-4o)}
    \figlabel{non-compilable-llmorpheus-gpt4o}
  \end{subfigure}
  
  \begin{subfigure}[t]{0.32\linewidth}
    \centering
    \includegraphics[width=\linewidth]{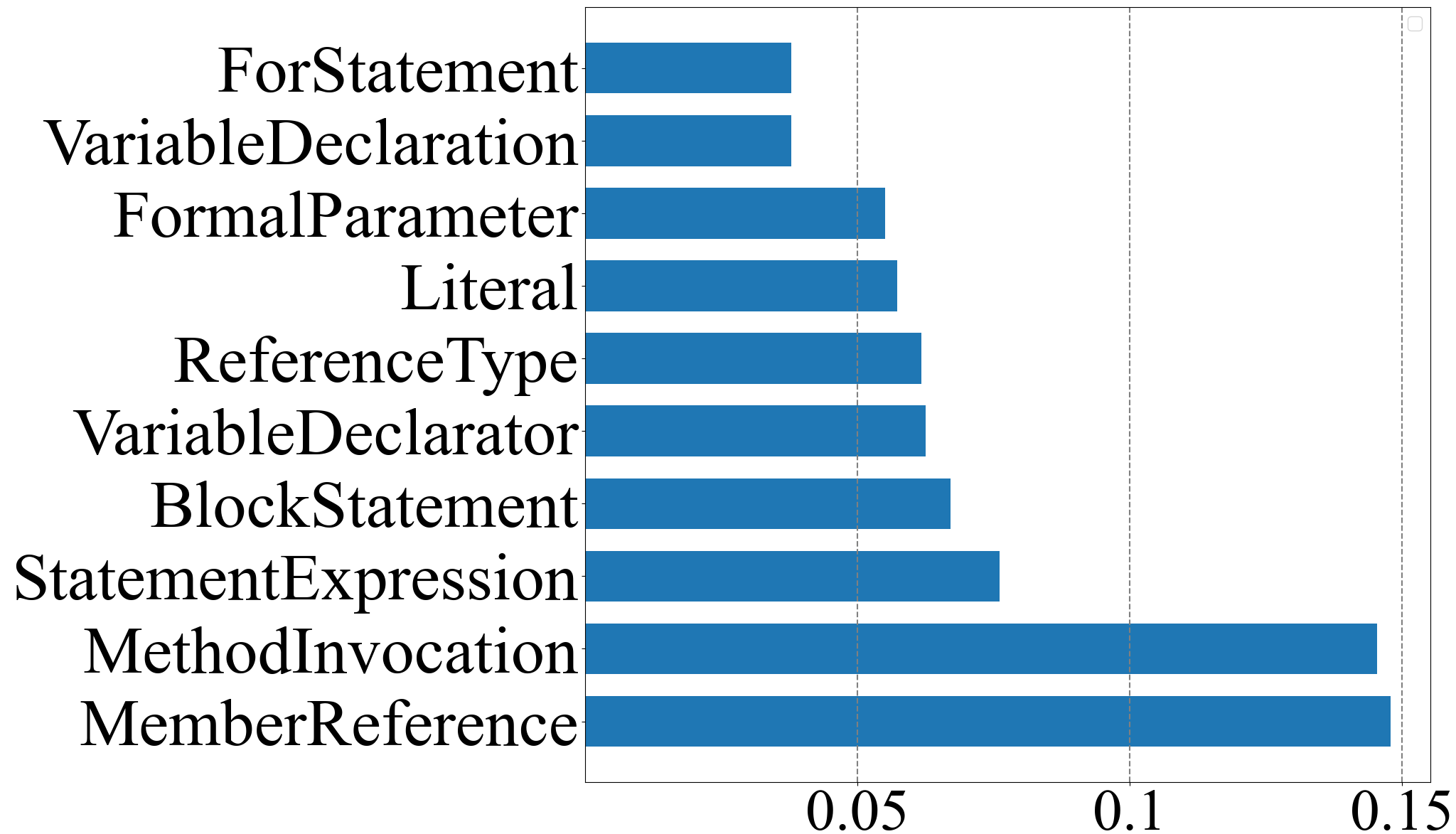}
    \caption{LLMorpheus (DS-671b)}
    \figlabel{non-compilable-llmorpheus-ds671b}
  \end{subfigure}
  \begin{subfigure}[t]{0.32\linewidth}
    \centering
    \includegraphics[width=\linewidth]{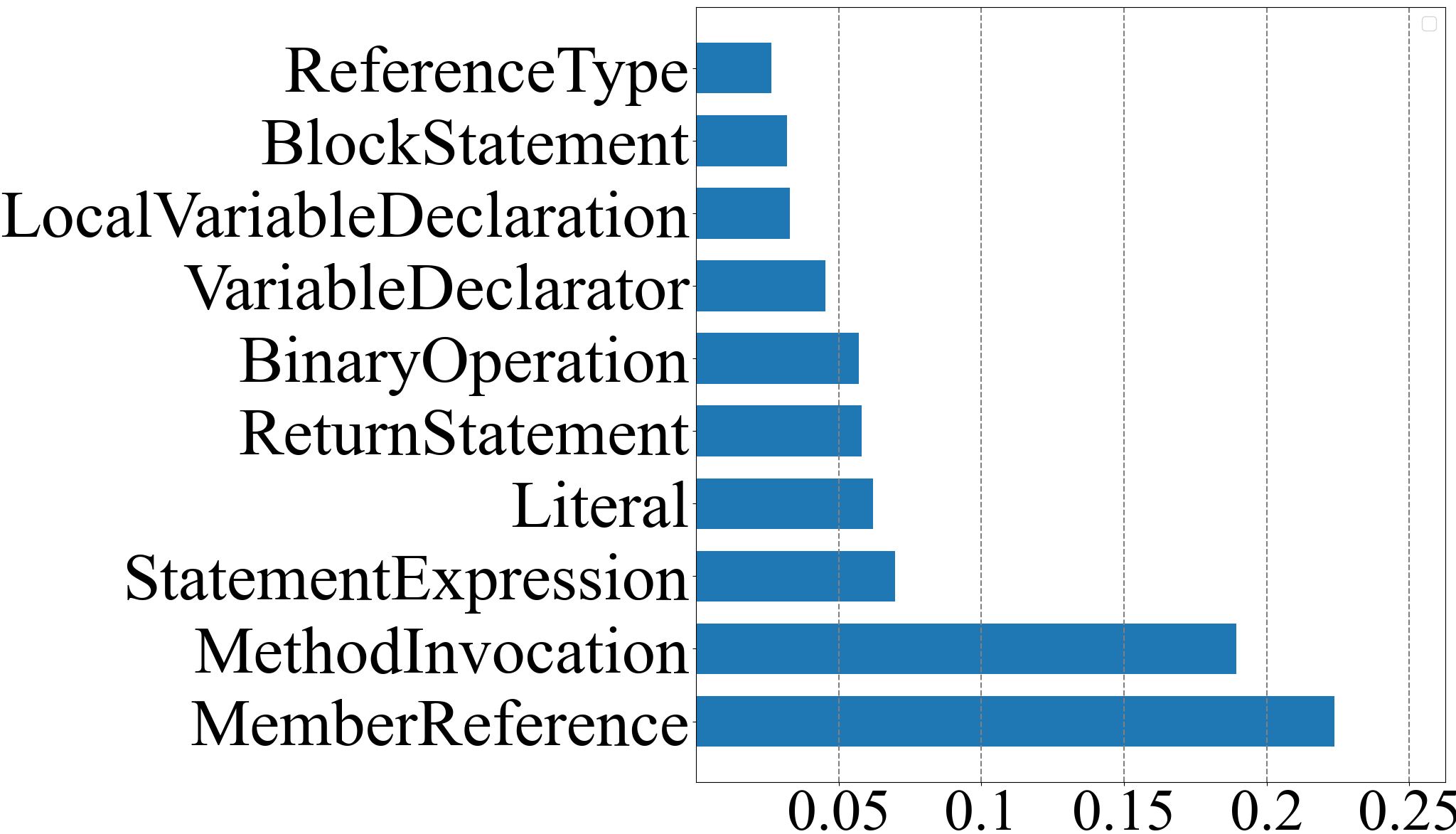}
    \caption{\ourtool (GPT-3.5)}
    \figlabel{non-compilable-gpt35}
  \end{subfigure}
  \begin{subfigure}[t]{0.32\linewidth}
    \centering
    \includegraphics[width=\linewidth]{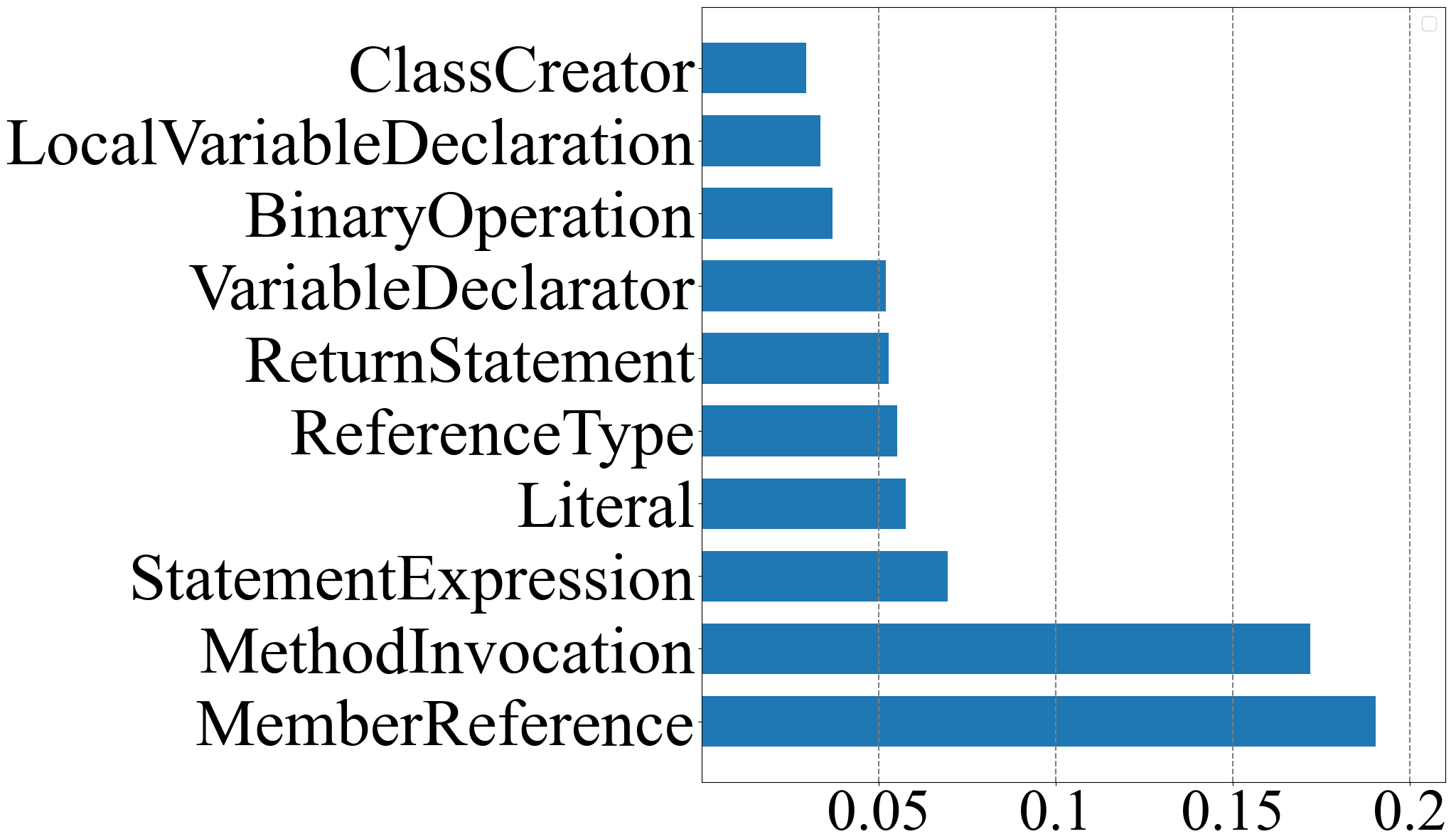}
    \caption{\ourtool (GPT-4o)}
    \figlabel{non-compilable-gpt4o}
  \end{subfigure}
  
  \begin{subfigure}[t]{0.32\linewidth}
    \centering
    \includegraphics[width=\linewidth]{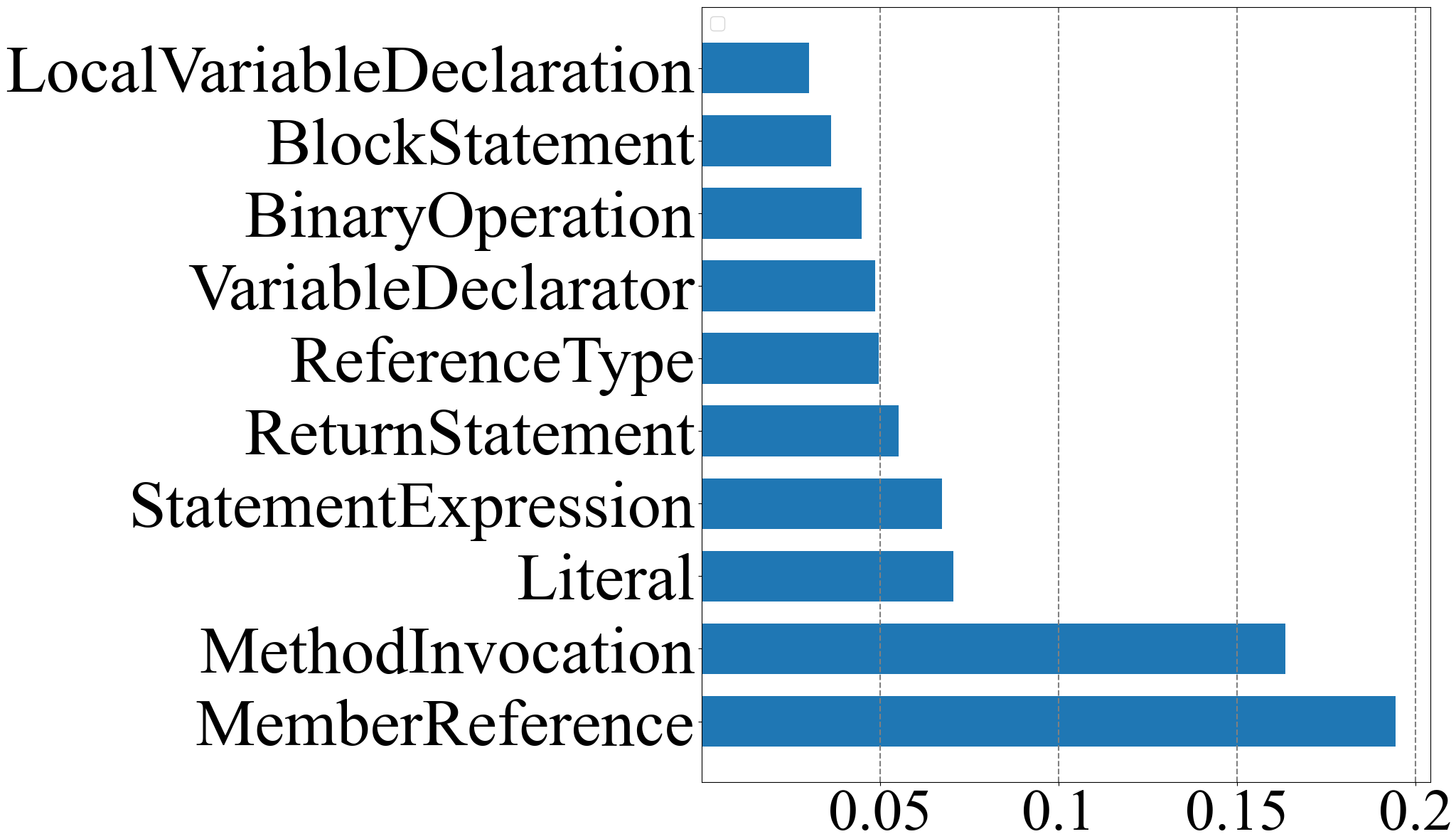}
    \caption{\ourtool (GPT-4o-M)}
    \figlabel{non-compilable-gpt4om}
  \end{subfigure}
  \begin{subfigure}[t]{0.32\linewidth}
    \centering
    \includegraphics[width=\linewidth]{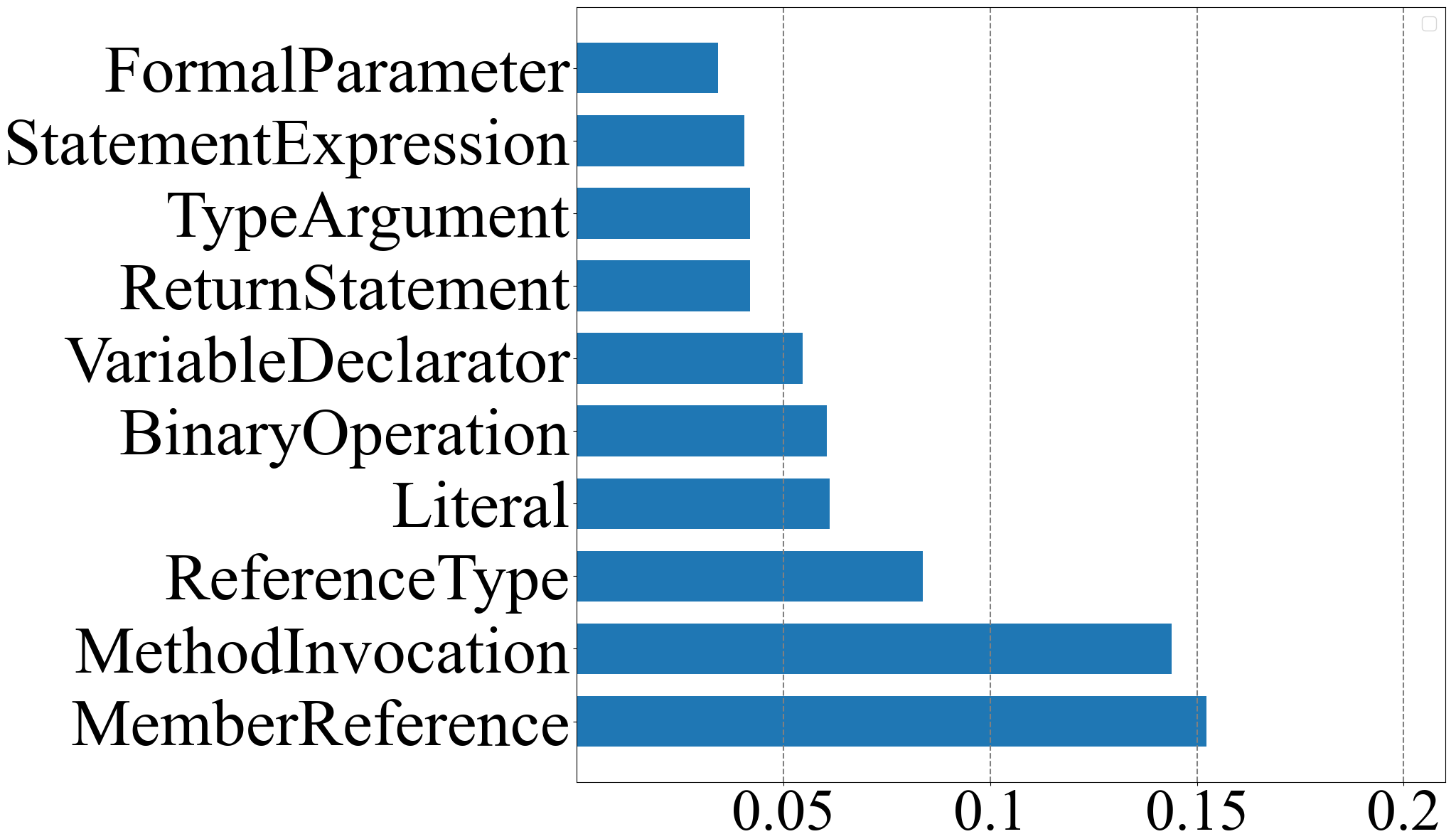}
    \caption{\ourtool (DS-671b)}
    \figlabel{non-compilable-deepseek-v3}
  \end{subfigure}
  \begin{subfigure}[t]{0.32\linewidth}
    \centering
    \includegraphics[width=\linewidth]{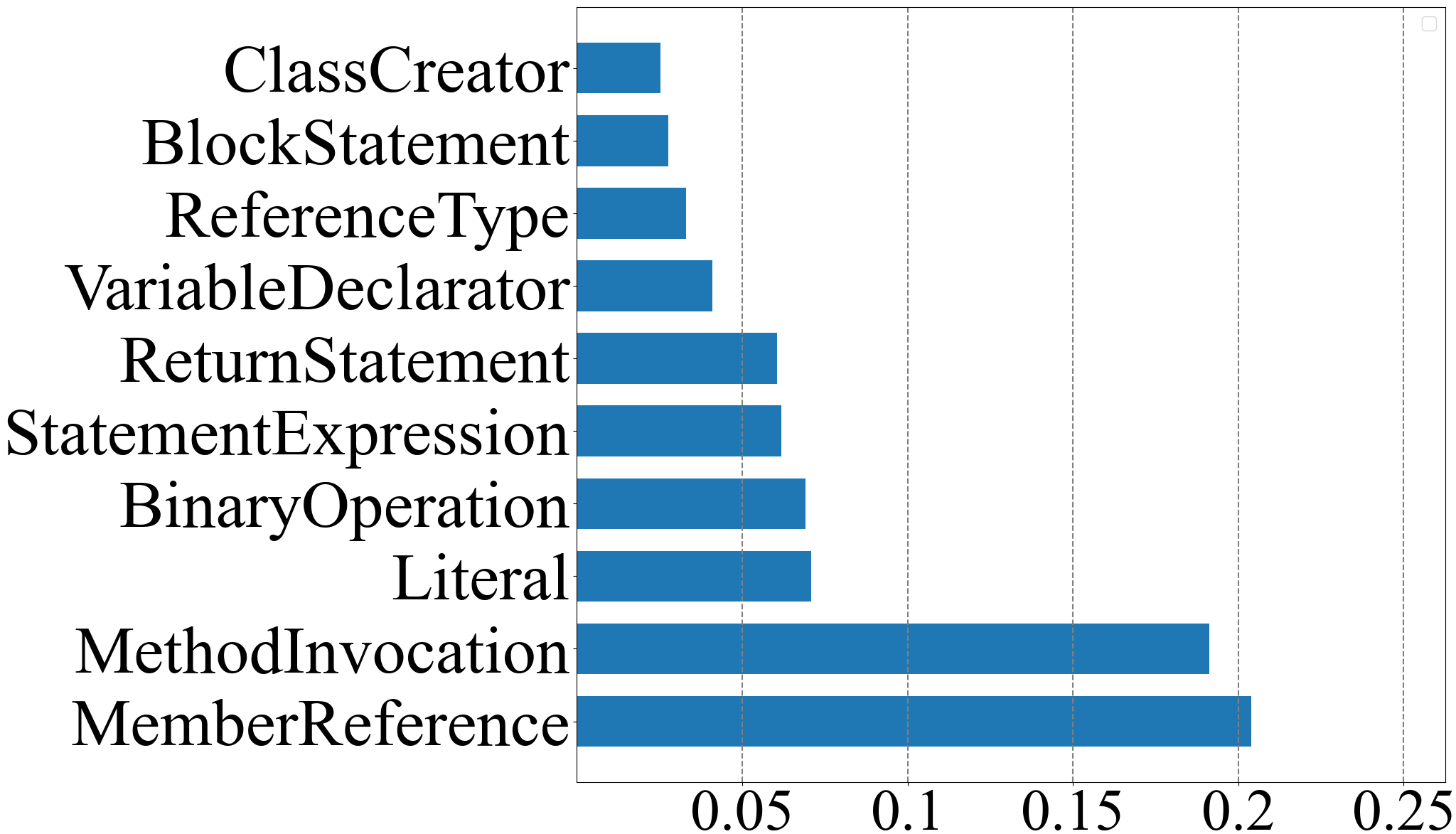}
    \caption{\ourtool (DS-236b)}
    \figlabel{non-compilable-deepseek}
  \end{subfigure}
  
  \begin{subfigure}[t]{0.32\linewidth}
    \centering
    \includegraphics[width=\linewidth]{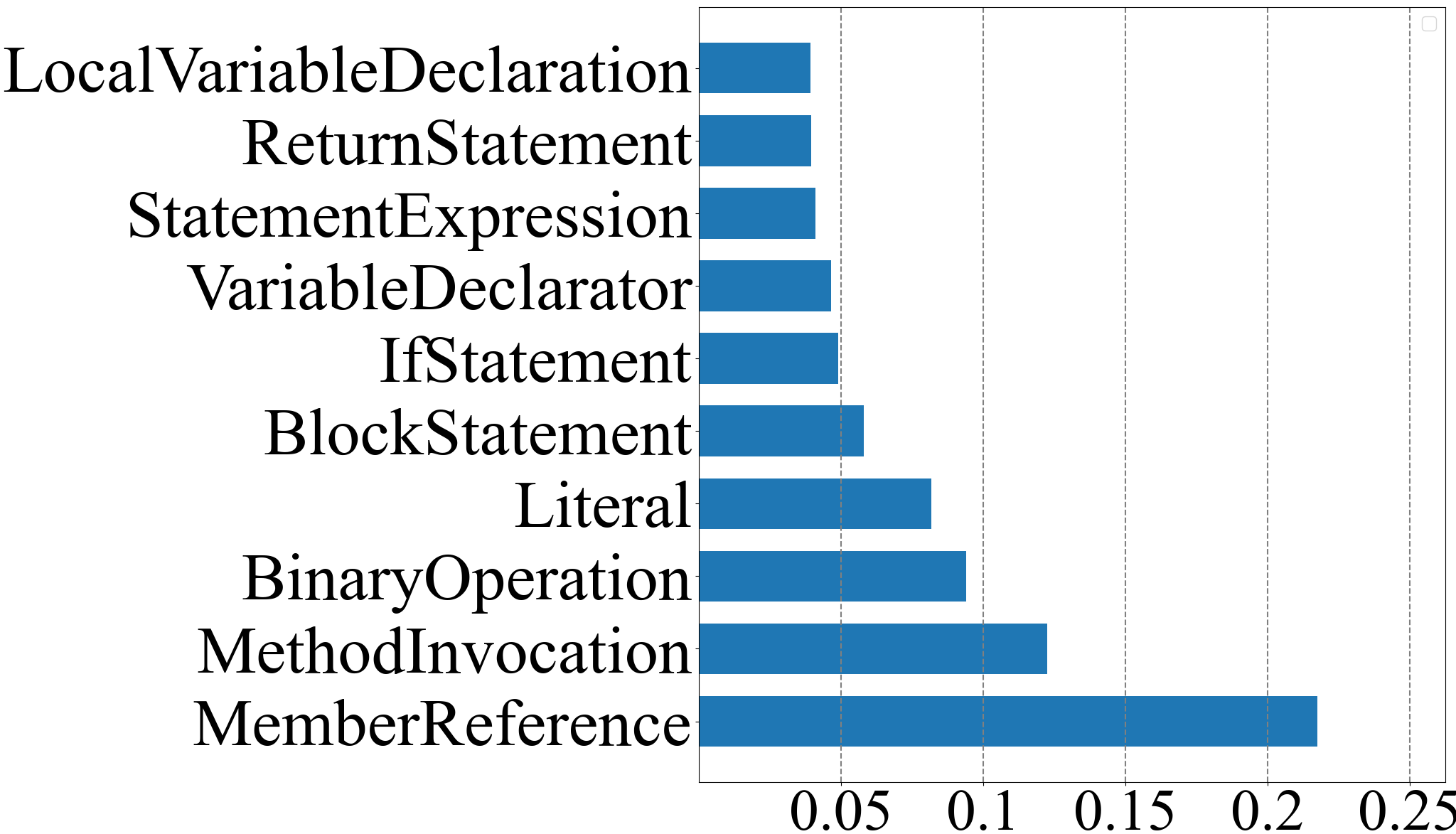}
    \caption{\ourtool (SC-16b)}
    \figlabel{non-compilable-starchat}
  \end{subfigure}
  \begin{subfigure}[t]{0.32\linewidth}
    \centering
    \includegraphics[width=\linewidth]{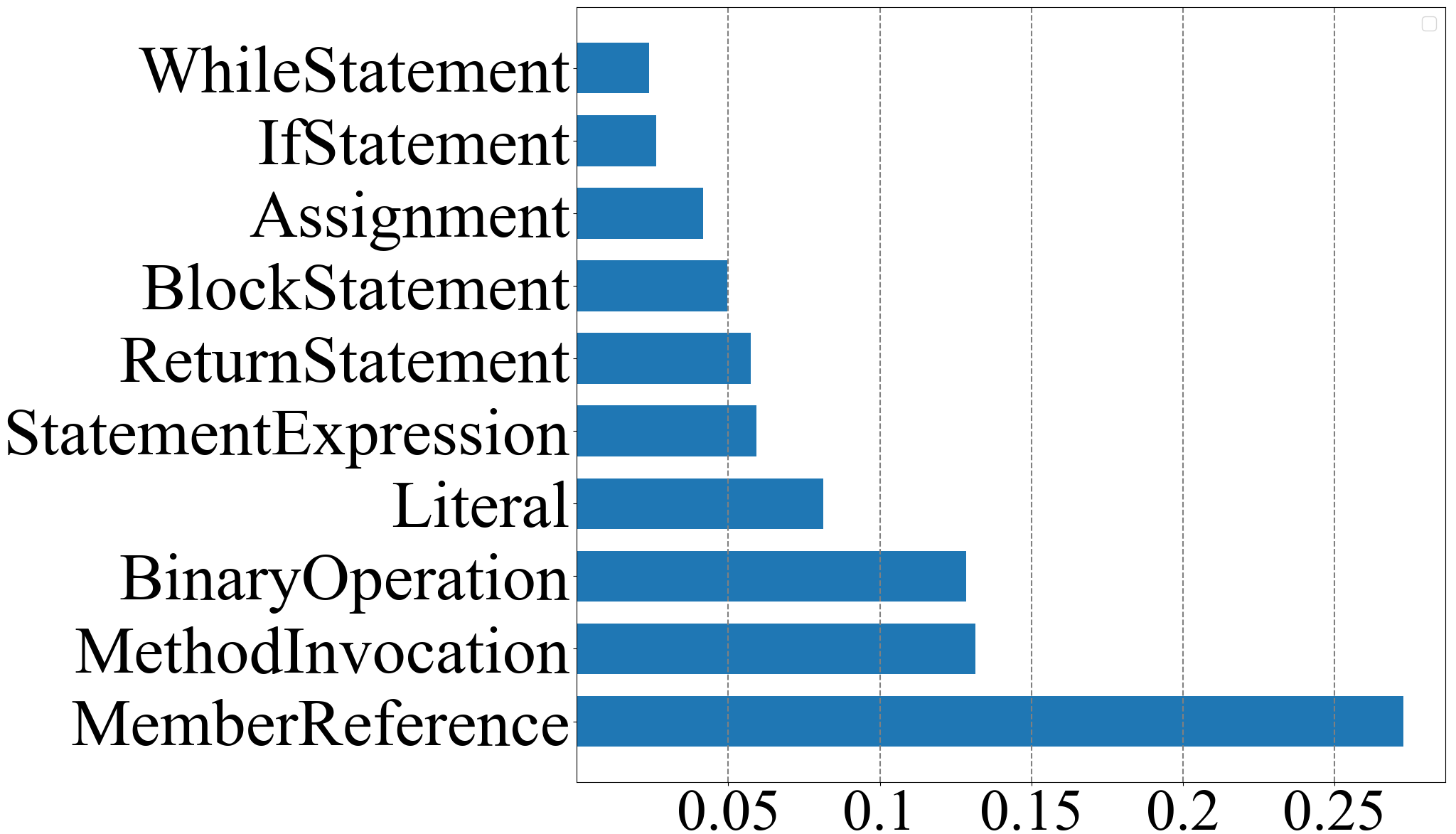}
    \caption{\ourtool (CL-13b)}
    \figlabel{non-compilable-codellama}
  \end{subfigure}
  \caption{Distribution of the AST Node Types of Origin Code for Non-Compilable Mutants}
  \figlabel{non-compilable}
  %\vspace{-10pt}
\end{figure}

\subsubsection{Categorizations of code context of non-compilable mutants}
To analyze which types of code are prone to causing LLMs to generate non-compilable mutants, we examined the code locations of all non-compilable mutants generated by all nine mutant generation approaches in \tabref{phase1}, as shown in \figref{non-compilable}.
For most approaches (except Major and BugFarm), the code locations with \textit{MemberReference} and \textit{MethodInvocation} are the top-2 most-prevent AST node types.
For example, for GPT and DeepSeek models, about 40\% of non-compilable mutants occur at the location with \textit{MemberReference} or \textit{MethodInvocation}.
This is potentially caused by the inherent complexity of these operations, which often involve multiple dependencies and references.
If any required method or member is missing or incorrectly specified, it can easily lead to non-compileable mutants.
The errors highlight a need for better context-aware mutant generation, ensuring that method calls and member references align with the intended program structure.
Additionally, we inspect the deletion mutants rejected by the compiler and find that they account for 7.48\% (GPT-3.5), 1.69\% (GPT-4o), 0.41\% (GPT-4o-Mini), 3.4\% (DeepSeek-V3-671b), 1.11\% (DeepSeek-Coder-V2-236b),  3.4\% (BugFarm), 0.8\% (LLMorpheus), 0.02\% (StarChat), 0.14\% (CodeLlama), 39.99\% (LEAM), 0.11\% ($\mu$BERT), and 16.61\% (Major) of all their non-compilable mutants, respectively. Thus, for LLMs, deletion is not the major reason for non-compilation.

\begin{tcolorbox}
\underline{\textbf{Answer to RQ4:}}
The mutants generated by GPTs have nine main types of compilation errors, with Usage of Unknown Methods and Code Structural Destruction being the most prevalent.
Deletion mutants are not the primary cause of non-compilable code.
An analysis of the surrounding code context where GPTs produce non-compilable mutants reveals that member references and method invocations are the most frequently occurring code features associated with these errors.
\end{tcolorbox}

\subsection{RQ5: Distributions of Surviving Mutants}

As shown in \tabref{phase1}, the mutation scores of rule-based approaches (i.e., PIT and Major) are lower than those of LLMs, indicating their mutants are harder to detect. Therefore, we intend to analyze the reasons for surviving mutants.
To address this RQ, we adopt a setting similar to the analysis of equivalent mutants.
Specifically, for each approach, we randomly sample its surviving mutants and manually classify them. We employ the categories of surviving mutant causes defined in the study by Just \emph{et al.}~\cite{just2014efficient} as follows.
\begin{itemize}
    \item (1) \textit{Not covered}, i.e., the mutant is located in code never executed by the test suite. We refer to the mutant set of this category as $NC$.
    \item (2) \textit{Covered but no state infection}, i.e., the mutant is executed but does not propagate any state change. We refer to the mutant set of this category as $NI$.
    \item (3) \textit{State infection without detection}, i.e., the mutant alters the program state but any test oracle does not observe the deviation. We refer to the mutant set of this category as $ND$.
\end{itemize}

These categories are mutually exclusive, with the following relationships: $NC~\cap~NI = \varnothing$, $NC~\cap~ND = \varnothing$, and $NI~\cap~ND = \varnothing$. This classification enables us to analyze the underlying reasons for surviving mutants and further evaluate the extent to which different approaches contribute to strengthening the test suite.
Given a surviving mutant, we first determine whether it is equivalent; if so, it is discarded. For non-equivalent mutants, we then check whether they are uncovered. If a mutant is covered, we further examine whether it alters the program state, i.e., whether any memory values are modified. Finally, the remaining surviving mutants that are covered but do not lead to detectable differences are classified as not detected. This stepwise process ensures that every surviving mutant is uniquely assigned to one category (equivalent, uncovered, not injecting state, or not detected).

\input{sections/results/TSurviving}

As shown in \tabref{surviving}, the distributions of surviving mutants differ significantly across approaches, where the columns \textbf{NC\%}, \textbf{NI\%}, and \textbf{ND\%} represent the proportions of $NC$, $NI$, and $ND$ within all non-equivalent mutants, respectively.

From the table, we can make the following observations.
First, BugFarm and LLMut exhibit a substantially higher proportion of $NC$ (i.e., \textit{not covered}) mutants, while LLMopheus shows a higher proportion of $NI$ (i.e., \textit{not injected}) mutants. 
For instance, among the surviving mutants generated by BugFarm (GPT-4o) and BugFarm (DS-671b), 79.52\% and 72.94\% fall into the $NC$ category, respectively. In contrast, the traditional approaches PIT, Major, and $\mu$Bert exhibit much lower proportions of 46.24\%, 45.65\%, and 46.25\%, respectively.
This suggests that LLM-based approaches either tend to generate mutants located in previously untested code regions or, in the case of LLMorpheus, often produce mutations that do not alter the program state.
This characteristic highlights the potential of LLMs in guiding test suite augmentation by directing attention to untested program areas.
Notably, a recent study guides LLMs to generate unit test cases to kill surviving mutants~\cite{harman2025mutation,wang2025mutation}, indicating the usefulness of these mutants for other applications.
Second, traditional approaches, especially rule-based methods such as PIT and Major, produce a larger fraction of $NI$ (i.e., \textit{not injected}) and $ND$ (i.e., \textit{not detected}) mutants compared to LLM-based methods.
For instance, PIT and Major exhibit $NI$ proportions of 29.03\% and 32.61\%, respectively, whereas both LLMut and BugFarm across different models remain below 20\%.
These types of surviving mutants require not only generating inputs that execute the mutated code but also propagating and detecting subtle semantic differences, which is inherently more challenging than increasing coverage.
Therefore, the overall implication is that LLM-generated surviving mutants are more directly beneficial for strengthening unit tests, as they emphasize extending coverage, while rule-based approaches tend to stress the harder task of state propagation and detection.

\begin{tcolorbox}
\underline{\textbf{Answer to RQ5:}}
Compared with rule-based approaches, LLM-based approaches tend to generate mutants in code lines that are not covered or not injected by existing test suites, providing valuable guidance for enhancing unit tests.
\end{tcolorbox}

%% file: sections/results/TAll-new.tex
\begin{table}[tp]
  \centering
  \scriptsize
  \caption{Overall Performance of Each Mutation Generation Technique}
    \begin{tabular}{l|l|l|l|l|l|l|l|l}
    \hline
    \textbf{Name} & \textbf{\#Mut. } & \textbf{Mut. S.} & \textbf{R. B. Detec.} & \textbf{Avg. Ochiai} & \textbf{Coup. Rate} & \textbf{Comp. Rate} & \textbf{Dup. Rate} & \textbf{Time}  \\
    \hline
    \textbf{PIT} & \cellcolor{gray!40}340728 & 0.541 & 40.1\% & 23.0\% & 23.5\% & \cellcolor{gray!40}100\% & \cellcolor{gray!40}0.0\% & \cellcolor{gray!40}0.02  \\
    \textbf{Major} & 20831 & 0.521 & 66.9\% & 35.5\% & 38.9\% & 97\% & \cellcolor{gray!40}0.0\% & 0.08  \\
    \textbf{LEAM} & 36505 & 0.648 & 63.9\% & 32.4\% & 38.1\% & 44.5\% & 2.3\% & 3.06  \\
    \textbf{$\mu$BERT} & 12946 & 0.677 & 60.5\% & 29.3\% & 47.3\% & 26.5\% & 1.7\% & 2.34 \\
    \hline
    \textbf{BF (GPT-4o)} & 13068 & 0.762 & 58.6\% & 17.7\% & 47.9\% & 74.1\% & 7.5\% & 5.35 \\
    \textbf{BF (DS-671b)} & 13896 & \cellcolor{gray!40}0.787 & 58.9\% & 17.9\% & 48.0\% & 77.5\% & 4.5\% & 7.29 \\
    \hline
    \textbf{LLMph (GPT-4o)} & 20043 & 0.754 & 68.5\% & 24.1\% & 51.7\% & 67.8\% & 7.1\% & 3.41 \\
    \textbf{LLMph (DS-671b)} & 21356 & 0.759 & 69.3\% & 26.0\% & \cellcolor{gray!40}52.0\% & 68.1\% & 6.9\% & 5.44 \\
    \hline
    \textbf{\ourtool (GPT-3.5)} & 31656 & 0.702 & 87.0\% & 47.7\% & 46.9\% & 62.5\% & 10.3\% & 1.79 \\
    \textbf{\ourtool (GPT-4o)} & 31945 & 0.737 & 89.9\% & 56.2\% & 50.6\% & 76.4\% & 7.8\% & 1.31 \\
    \textbf{\ourtool (GPT-4o-M)} & 34417 & 0.708 & 90.8\% & 54.1\% & 49.5\% & 75.3\% & 7.0\% & 1.65 \\
    \textbf{\ourtool (DS-671b)} & 33596 & 0.749 & \cellcolor{gray!40}91.1\% & \cellcolor{gray!40}59.8\% & 50.8\% & 77.6\% & 7.5\% & 2.57 \\
    \textbf{\ourtool (DS-236b)} & 34741 & 0.714 & 89.4\% & 50.9\% & 49.5\% & 75.8\% & 7.7\% & 4.25 \\
    \textbf{\ourtool (SC-16b)} & 28058 & 0.507 & 38.5\% & 21.7\% & 34.0\% & 16.4\% & 12.8\% & 7.53 \\
    \textbf{\ourtool (CL-13b)} & 27614 & 0.669 & 71.2\% & 33.3\% & 45.6\% & 69.3\% & 37.1\% & 9.06 \\
    \hline
    \end{tabular}
    \\ {\scriptsize Note: The mutation generation approach achieving the best performance in terms of a metric is highlighted in a grey background color. \\
    \textbf{BF}: BugFarm. \textbf{LLMph}: LLMorpheus. \textbf{\ourtoolraw}: Our new prompt.
     \textbf{GPT-4o-M}: GPT-4o-Mini. \textbf{DS-671b}: DeepSeek-V3-671b. \textbf{DS-236b}: DeepSeek-Coder-V2-236b. \textbf{SC-16b}: StarChat-$\beta$-16b. \textbf{CL-13b}: CodeLlama-Instruct-13b.}
  \tablabel{phase1}
\end{table}%

%% file: sections/results/TRealBugDet-new.tex
\begin{table}[tp]
  \centering
  \scriptsize
  \caption{Number of Real Bugs Detected by Each Mutation Generation Approach}
    \begin{tabular}{l|llllllllllll|l|l}
    \hline
    \textbf{Proj.} & \textbf{Chart} & \textbf{Lang} & \textbf{Time} & \textbf{Math} & \textbf{Moc.} & \textbf{Clo.} & \textbf{Cli} & \textbf{Co.} & \textbf{Csv} & \textbf{Gs.} & \textbf{Ja.} & \textbf{Js.} & \textbf{ConD.} & \textbf{All} \\
    \hline
    \textbf{\#Bug} & 26  & 65  & 27  & 106  & 38  & 133  & 39  & 18  & 16  & 18  & 26  & 93  & 246  & 851  \\
    \hline
    \textbf{PIT} & 19 & 23 & 12 & 40 & 14 & 38 & 6  & 6  & 7  & 6  & 1  & 48 & 121 & 341 \\
    \textbf{Major} & 17 & 51 & 23 & 82 & 28 & 92 & 11 & 11 & 13 & 9  & 17 & 77 & 138 & 569 \\
    \textbf{LEAM} & 16 & 27 & 18 & 44 & 27 & 122 & 27 & 12 & 13 & 11 & 17 & 90 & 120 & 544 \\
    \textbf{$\mu$BERT} & 10 & 38 & 15 & 65 & 5  & 103 & 29 & 11 & 7  & 12 & 18 & 86 & 116 & 515 \\
    \hline
    \textbf{BF (GPT-4o)} & 14  & 37  & 15  & 57  & 28  & 93  & 23  & 6  & 8  & 11  & 7  & 83  & 117  & 499  \\
    \textbf{BF (DS-671b)} & 16 & 37  &  15 &   49& 25  &  100 &  22 & 7  & 8  & 9  & 7  & 88  & 118  & 501  \\
    \hline
    \textbf{LLMph (GPT-4o)} & 17  & 51  & 20  & 61  & 31  & 121  & 30  & 12  & 8  & 12  & 9  & 87  & 124  & 583  \\
    \textbf{LLMph (DS-671b)} & 17 & 50  & 20  &  63 &  33 &  123 &  29 & 9  &  10 & 13  & 9  &  86 &  128 &  590 \\
    \hline
    \textbf{\ourtool (GPT-3.5)} & 22 & 52 & 25 & 92 & 30 & 130 & 37 & 14 & 15 & 13 & 23 & \cellcolor{gray!40}93 & 194 & 740 \\
    
    \textbf{\ourtool (GPT-4o)} & 22 & 55 & 24 & 94 & 29 & 131 & \cellcolor{gray!40}39 & 16 & \cellcolor{gray!40}16 & \cellcolor{gray!40}16 & 23 & 92 & \cellcolor{gray!40}208 & 765 \\
    
    \textbf{\ourtool (GPT-4o-M)} & 23 & 56 & 26 & 96 & \cellcolor{gray!40}36 & 131 & \cellcolor{gray!40}39 & \cellcolor{gray!40}17 & \cellcolor{gray!40}16 & \cellcolor{gray!40}16 & 21 & \cellcolor{gray!40}93 & 203 & 773 \\
    
    \textbf{\ourtool (DS-671b)} & \cellcolor{gray!40}24  & \cellcolor{gray!40}61  & \cellcolor{gray!40}27  & \cellcolor{gray!40}101  & 35  & 131  & \cellcolor{gray!40}39  & 13  & \cellcolor{gray!40}16  & \cellcolor{gray!40}16  & \cellcolor{gray!40}25  & 92  & 195  & \cellcolor{gray!40}775  \\
    
    \textbf{\ourtool (DS-236b)} & 23 & 57 & 22 & 95 & 34 & \cellcolor{gray!40}132 & \cellcolor{gray!40}39 & \cellcolor{gray!40}17 & \cellcolor{gray!40}16 & 15 & 21 & 90 & 200 & 761 \\
    
    \textbf{\ourtool (SC-16b)} & 9  & 11 & 10 & 58 & 16 & 79 & 2  & 1  & 2  & 1  & 1  & 57 & 81 & 328 \\
    \textbf{\ourtool (CL-13b)} & 8  & 41 & 17 & 77 & 21 & 114 & 34 & 11 & 13 & 10 & 17 & 87 & 156 & 606 \\
    \hline
    \end{tabular}
  \tablabel{real-bug-detect}
  \\{\scriptsize Note: The mutation generation approaches that achieve the best performance in one project are highlighted in grey background color. \\
  \textbf{BF}: BugFarm. \textbf{LLMph}: LLMorpheus. \textbf{\ourtool}: our prompt. \textbf{Moc.}: Mockito. \textbf{Clo.}: Closure. \textbf{Co.}: Codec. \textbf{Gs.}: Gson. \textbf{Ja.}: JacksonCore. \textbf{Js.}: Jsoup. \textbf{ConD}: ConDefects.}
\end{table}

%% file: sections/results/TOchiai-new.tex
\begin{table}[tp]
  \centering
  \scriptsize
  \caption{Number of Bugs with High Ochiai Scores (Over 0.8) Achieved by Each Mutation Generation Approach}
    \begin{tabular}{l|llllllllllll|l|l}
    \hline
    \textbf{Proj.} & \textbf{Chart} & \textbf{Lang} & \textbf{Time} & \textbf{Math} & \textbf{Moc.} & \textbf{Clo.} & \textbf{Cli} & \textbf{Co.} & \textbf{Csv} & \textbf{Gs.} & \textbf{Ja.} & \textbf{Js.} & \textbf{ConD.} & \textbf{All} \\
    \hline
    \textbf{\#Bug} & 26  & 65  & 27  & 106  & 38  & 133  & 39  & 18  & 16  & 18  & 26  & 93  & 246  & 851  \\
    \hline
    \textbf{PIT} & 9  & 15 & 8  & 20 & 1  & 12 & 3  & 3  & 4  & 2  & 2  & 16 & 101 & 196 \\
    \textbf{Major} & 9  & 34 & 14 & 42 & 12 & 25 & 3  & 1  & 2  & 1  & 3  & 6  & 150 & 302 \\
    \textbf{LEAM} & 3  & 13 & 8  & 13 & 5  & 40 & 13 & 3  & 1  & 4  & 9  & 32 & 132 & 276 \\
    \textbf{$\mu$BERT} & 5  & 18 & 4  & 27 & 3  & 14 & 8  & 3  & 2  & 3  & 8  & 20 & 134 & 249 \\
    \hline
    \textbf{BF (GPT-4o)} & 9  & 15  & 3  & 13  & 1  & 13  & 3  & 2  & 1  & 4  & 2  & 14  & 71  & 151  \\
    \textbf{BF (DS-671b)} & 6 & 10  & 3  & 11  & 1  & 13  & 2  &  2 & 0  & 3  & 2  & 15  & 83  &  152 \\
    \hline
    \textbf{LLMph (GPT-4o)} & 6  & 25  & 7  & 24  & 6  & 22  & 9  & 2  & 3  & 3  & 5  & 12  & 81  & 205  \\
    \textbf{LLMph (DS-671b)} & 6 &  21 & 8  &  22 & 5  & 26  & 11  & 2  & 3  & 5  &  6 &  14 & 92  &  221 \\
    \hline
    \textbf{\ourtool (GPT-3.5)} & 10 & 32 & 8  & 46 & 14 & 38 & 18 & 2  & 4  & 2  & 11 & 46 & 175 & 406 \\
    
    \textbf{\ourtool (GPT-4o)} & \cellcolor{gray!40}13 & 35 & 12 & 56 & 13 & 69 & 22 & \cellcolor{gray!40}8  & \cellcolor{gray!40}10 & 8  & 14 & 39 & 179 & 478 \\
    
    \textbf{\ourtool (GPT-4o-M)} & 12 & 37 & 12 & 48 & 17 & 56 & 23 & 6  & \cellcolor{gray!40}10 & 7  & 12 & 39 & \cellcolor{gray!40}181 & 460 \\
    
    \textbf{\ourtool (DS-671b)} & \cellcolor{gray!40}13  & \cellcolor{gray!40}40  & \cellcolor{gray!40}15  & \cellcolor{gray!40}62  & \cellcolor{gray!40}21  & \cellcolor{gray!40}79  & \cellcolor{gray!40}27  & \cellcolor{gray!40}8  & \cellcolor{gray!40}10  & \cellcolor{gray!40}12  & \cellcolor{gray!40}17  & \cellcolor{gray!40}59  & 146  & \cellcolor{gray!40}509  \\
    
    \textbf{\ourtool (DS-236b)} & 12 & 34 & 11 & 47 & 14 & 45 & 20 & 7  & 8  & 6  & 11 & 41 & 177 & 433 \\
    \textbf{\ourtool (SC-16b)} & 7  & 12 & 4  & 19 & 3  & 21 & 4  & 3  & 3  & 3  & 5  & 25 & 112 & 221 \\
    \textbf{\ourtool (CL-13b)} & 3  & 23 & 7  & 33 & 3  & 26 & 10 & 2  & 1  & 2  & 6  & 19 & 148 & 283 \\
    \hline
    \end{tabular}
  \\{\scriptsize Note: The mutation generation approaches that achieve the best performance in one project are highlighted in grey background color. \\
  \textbf{BF}: BugFarm. \textbf{LLMph}: LLMorpheus. \textbf{LLMut}: our prompt. \textbf{Moc.}: Mockito. \textbf{Clo.}: Closure. \textbf{Co.}: Codec. \textbf{Gs.}: Gson. \textbf{Ja.}: JacksonCore. \textbf{Js.}: Jsoup. \textbf{ConD}: ConDefects.}
  \tablabel{ochiai}
\end{table}

%% file: sections/results/TCoupling-new.tex
\begin{table}[tp]
  \centering
  \scriptsize
  \caption{Distribution of Coupling Categories Across Mutation Generation Approaches}
      \begin{tabular}{l|l|l|l|l||l|l}
    \hline
    \textbf{Category} & \textbf{Strong} & \textbf{Strong Addi} & \textbf{Partial} & \textbf{Partial Addi} & \textbf{Not Subs.} & \textbf{Not Detected} \\
    \hline
    \textbf{PIT} & 10.1\% & 4.7\% & 1.8\% & 6.9\% & 30.6\% & 45.9\% \\
    \textbf{Major} & 9.6\% & 5.8\% & 1.1\% & 22.4\% & 13.2\% & 47.9\% \\
    \textbf{LEAM} & 4.9\% & 4.3\% & 0.5\% & 28.4\% & 26.7\% & 35.2\% \\
    \textbf{$\mu$BERT} & 7.8\% & 5.2\% & 0.8\% & 33.5\% & 20.4\% & 32.3\% \\
    \hline
    \textbf{BugFarm (GPT-4o)} & 8.4\% & 5.4\% & 0.6\% & 33.6\% & 28.3\% & 23.8\% \\
    \textbf{BugFarm (DS-671b)} & 10.0\% & 4.8\% & 0.7\% & 32.4\% & 30.7\% & 21.3\% \\
    \hline
    \textbf{LLMorpheus (GPT-4o)} & 9.3\% & 4.9\% & 2.1\% & \cellcolor{gray!40}35.4\% & 23.6\% & 24.6\% \\
    \textbf{LLMorpheus (DS-671b)} & \cellcolor{gray!40}11.8\% & 4.6\% &  1.8\% & 33.8\%  & 23.9\%  & 24.1\%  \\
    \hline
    \textbf{\ourtool (GPT-3.5)} & 7.0\% & 6.7\% & 1.2\% & 32.0\% & 23.3\% & 29.8\% \\
    \textbf{\ourtool (GPT-4o)} & 6.8\% & 10.2\% & 1.2\% & 32.5\% & 23.1\% & 26.3\% \\
    \textbf{\ourtool (GPT-4o-M)} & 6.2\% & \cellcolor{gray!40}12.1\% & 0.8\% & 30.4\% & 21.3\% & 29.2\% \\
    \textbf{\ourtool (DS-671b)} & 11.2\% & 4.6\% & 1.9\% & 33.1\% & 24.1\% & 25.1\% \\
    \textbf{\ourtool (DS-236b)} & 6.4\% & 10.3\% & 1.0\% & 31.8\% & 21.9\% & 28.6\% \\
    \textbf{\ourtool (SC-16b)} & 8.0\% & 5.7\% & \cellcolor{gray!40}2.7\% & 17.6\% & 16.7\% & 49.3\% \\
    \textbf{\ourtool (CL-13b)} & 6.8\% & 6.0\% & 0.9\% & 31.9\% & 21.3\% & 33.1\% \\
    \hline
    \end{tabular}
    \tablabel{coupling}
      \\{\scriptsize Note: \textbf{Strong} - Strong Substitution, \textbf{Strong Addi.} - Strong Substitution + Additional Failing Tests,  \textbf{Partial} - Partial Substitution, \textbf{Partial Addi.} - Partial Substitution + Additional Failing Tests, \textbf{Not Subs.} - Not Substitution.}
\end{table}

%% file: sections/results/TDiversity-new.tex
\begin{table}[tp]
  \centering
  \scriptsize
  \caption{Ratio of Deletion Mutations and the Top-3 Newly Introduced AST Node Types for Each Approach}
    \begin{tabular}{l | l | l | l | l}
    \hline
    \textbf{Approach} & \textbf{Deletion} & \textbf{1st} & \textbf{2nd} & \textbf{3rd} \\
    \hline
    \textbf{Major} & 15.3\% & ST (78.6\%) & LT (21.4\%) & —— \\
    \textbf{LEAM} & 48.9\% & LT (23.2\%) & MR (14.4\%) & CT (9.6\%) \\
    \textbf{$\mu$BERT} & 0.4\% & LT (29.4\%) & SE (25.7\%) & MR (24.5\%) \\
    \hline
    \textbf{BugFarm (GPT-4o)} & 5.5\% & LT (41.9\%) & BO (12.2\%) & MI (9.0\%) \\
    \textbf{BugFarm (DS-671b)} & 2.1\%  & LT(45.8\%)  &  BO(14.7\%)  &  MR(8.8\%) \\
    \hline
    \textbf{LLMorpheus (GPT-4o)} & 0.3\% & LT (24.0\%) & BO (23.7\%) & MR (15.8\%) \\
    \textbf{LLMorpheus (DS-671b)} & 0.3\% & LT(29.5\%) & BO(25.0\%) & MR(15.2\%) \\
    \hline
    \textbf{\ourtool (GPT-3.5)} & 5.3\% & LT (39.2\%) & BO (17.3\%) & MI (9.9\%) \\
    \textbf{\ourtool (GPT-4o)} & 2.1\% & LT (45.4\%) & BO (16.6\%) & CC (8.8\%) \\
    \textbf{\ourtool (GPT-4o-M)} & 0.5\% & LT (43.3\%) & BO (21.6\%) & CC (7.6\%) \\
    \textbf{\ourtool (DS-671b)} & 3.8\% & LT (44.2\%) & BO (15.6\%) & CC (8.6\%) \\
    \textbf{\ourtool (DS-236b)} & 1.0\% & LT (38.8\%) & BO (19.1\%) & MI (10.1\%) \\
    \textbf{\ourtool (SC-16b)} & 0.3\% & LT (22.7\%) & BO (13.8\%) & MR (13.1\%) \\
    \textbf{\ourtool (CL-13b)} & 0.1\% & LT (47.8\%) & BO (25.7\%) & MI (9.3\%) \\
    \hline
    \end{tabular}
  \tablabel{diversity}
  \\ {\scriptsize \textbf{Deletion}: The proportion of deletion mutations. \textbf{BO}: BinaryOperation. \textbf{CC}: ClassCreator. \textbf{CT}: Cast. \textbf{LT}: Literal. \textbf{MI}: MethodInvocation. \textbf{MR}: MemberReference. \textbf{SE}: StatementExpression. \textbf{ST}: Statement.}
\end{table}

%% file: sections/results/TASTDistance-new.tex
\begin{table}[tp]
  \centering
  \scriptsize
  \caption{The Distribution of AST Edit Distance for Each Approach}
    \begin{tabular}{l|l|l|l|l|l|l}
    \hline
    \textbf{Distance} & \textbf{1} & \textbf{2} & \textbf{3} & \textbf{4} & \textbf{5} & \textbf{$\geq$6} \\
    \hline
    \textbf{Major} & 69.2\% & 27.8\% & 1.8\% & 0.6\% & 0.4\% & 0.2\% \\
    \textbf{LEAM} & 58.5\% & 23.6\% & 7.3\% & 3.4\% & 1.0\% & 6.1\% \\
    \textbf{$\mu$BERT} & 73.8\% & 8.9\% & 2.7\% & 14.2\% & 0.1\% & 0.4\% \\
    \hline
    \textbf{BugFarm (GPT-4o)} & 44.7\% & 43.5\% & 8.3\% & 1.6\% & 0.6\% & 1.4\% \\
    \textbf{BugFarm (DS-671b)} & 38.3\% & 52.1\%  &  6.7\% & 1.5\%  & 0.7  & 0.7 \\
    \hline
    \textbf{LLMorpheus (GPT-4o)} & 54.2\% & 25.7\% & 14.6\% & 3.7\% & 0.9\% & 0.8\% \\
    \textbf{LLMorpheus (DS-671b)} &  53.9\% &  31.8\% & 10.6\%  & 2.5\%  &  0.4\% &  0.8\% \\
    \hline
    \textbf{\ourtool (GPT-3.5)} & 52.4\% & 36.4\% & 6.8\% & 2.4\% & 0.9\% & 1.0\% \\
    \textbf{\ourtool (GPT-4o)} & 50.0\% & 38.8\% & 7.2\% & 2.2\% & 0.8\% & 1.0\% \\
    \textbf{\ourtool (GPT-4o-M)} & 47.3\% & 41.9\% & 6.3\% & 2.5\% & 0.8\% & 1.2\% \\
    \textbf{\ourtool (DS-671b)} & 46.1\% & 44.2\% & 6.8\% & 1.5\% & 0.4\% & 0.9\% \\
    \textbf{\ourtool (DS-236b)} & 45.1\% & 44.2\% & 6.5\% & 2.6\% & 0.8\% & 0.8\% \\
    \textbf{\ourtool (SC-16b)} & 36.0\% & 39.1\% & 8.0\% & 6.0\% & 3.7\% & 7.2\% \\
    \textbf{\ourtool (CL-13b)} & 46.4\% & 46.6\% & 3.8\% & 1.8\% & 0.5\% & 1.0\% \\
    \hline
    \end{tabular}
  \tablabel{ast-distance}
\end{table}

%% file: sections/results/TSample-new.tex
\begin{table}[tp]
  \centering
  \scriptsize
  \caption{Sample Compilable Mutations for Identifying Equivalent Mutations}
    \begin{tabular}{l|l|l|l|l}
    \hline
    \textbf{Approach} & \textbf{\#Sampled} & \textbf{\#Test Filtered} & \textbf{\#Eq. Mut.} & \textbf{EMR} \\
    \hline
    \textbf{PIT} & 97 & 26 & 1  & 1.0\% \\
    \textbf{Major} & 96 & 52 & 2  & 2.1\% \\
    \textbf{LEAM} & 96 & 48 & 2  & 2.1\% \\
    \textbf{$\mu$BERT} & 94 & 64 & 3  & 6.4\% \\
    \hline
    \textbf{BugFarm (GPT-4o)} & 96  & 67  & 3  & 3.1\% \\
    \textbf{BugFarm (DS-671b)} &  96 & 75  & 5  &  5.2\% \\
    \hline
    \textbf{LLMorpheus (GPT-4o)} & 96  & 69  & 4  & 4.2\% \\
    \textbf{LLMorpheus (DS-671b)} & 96  &  65 & 3  & 3.1\% \\
    \hline
    \textbf{\ourtool (GPT-3.5)} & 96 & 60 & 4  & 4.2\% \\
    \textbf{\ourtool (GPT-4o)} & 96 & 64 & 3  & 3.1\% \\
    \textbf{\ourtool (GPT-4o-M)} & 96 & 59 & 3  & 3.1\% \\
    \textbf{\ourtool (DS-671b)} & 96  & 65  & 6  & 6.3\% \\
    \textbf{\ourtool (DS-236b)} & 96 & 63 & 5  & 5.2\% \\
    \textbf{\ourtool (SC-16b)} & 94 & 32 & 10 & 10.6\% \\
    \textbf{\ourtool (CL-13b)} & 96 & 51 & 6  & 6.3\% \\
    \hline
    \end{tabular}
  \tablabel{sample-eq}
\end{table}

%% file: sections/results/TToken.tex
\begin{table}[tp]
  \centering
  \scriptsize
  \caption{Average Number of Used Tokens Across All LLM-Based Approaches}
    \begin{tabular}{l|l|l || l | l}
    \hline
    \textbf{Approach} & \textbf{Avg. Input Token} & \textbf{Avg. Output Token} & \textbf{All Input Token} &  \textbf{All Output Token} \\
    \hline
    \textbf{BugFarm (GPT-4o)} & 200 & 368 & 2,780,213 & 4,807,979 \\
    \textbf{BugFarm (DS-671b)} & 200 & 366 & 2,780,213 & 5,085,936 \\
    \hline
    \textbf{LLMorpheus (GPT-4o)} & 370 & 74 & 7,921,356 & 1,476,568 \\
    \textbf{LLMorpheus (DS-671b)} & 370 & 73 & 7,921,356 & 1,558,988 \\
    \hline
    \textbf{\ourtool (GPT-3.5)} & 126 & 59 & 4,353,426 & 1,855,042 \\
    \textbf{\ourtool (GPT-4o)} & 126 & 59 & 4,353,426 & 1,875,810 \\
    \textbf{\ourtool (GPT-4o-M)} & 126 & 59 & 4,353,426 & 2,027,850 \\
    \textbf{\ourtool (DS-671b)} & 126 & 58 & 4,353,426 & 1,947,224 \\
    \textbf{\ourtool (DS-236b)} & 126 & 59 & 4,353,426 & 2,064,658 \\
    \textbf{\ourtool (SC-16b)} & 126 & 63 & 4,353,426 & 1,766,537 \\
    \textbf{\ourtool (CL-13b)} & 126 & 58 & 4,353,426 & 1,599,127 \\
    \hline
    \end{tabular}
  \tablabel{tokens}
\end{table}

%% file: sections/results/TCompilationErrorTypes.tex
\begin{table}[tp]
  \centering
  \scriptsize
  %\small
  \caption{Error Types of Non-Compilable Mutations}
  \vspace{-2mm}
    \begin{tabular}{|l|l|r||l|l|r|}
    \hline
     \bf ID & \bf Error Type & \bf \% & \bf ID & \bf Error Type & \bf \% \\
    \hline
    1 & Usage of Unknown Methods & 27.3\% & 6 & Type Mismatch & 7.6\%\\
    \hline
    2 & Code Structural Destruction & 22.9\% & 7 & Incorrect Initialization & 3.1\% \\
    \hline
    3 & Incorrect Method Parameters & 12.8\% & 8 & Incorrect Location & 3.1\% \\
    \hline
    4 & Usage of Unknown Variables & 11.2\% & 9 & Incorrect Exceptions & 2.3\% \\
    \hline
    5 & Usage of Unknown Types & 9.6\% & — & — & — \\
    \hline
    \end{tabular}
  \tablabel{error-type}
\end{table}

%% file: sections/results/TSurviving.tex
\begin{table}[tp]
  \centering
  \scriptsize
  \caption{Sampled Surviving Mutants for Analyzing Survival Reasons}
    \begin{tabular}{|l|l|l||lll||lll|}
    \hline
    \textbf{Approach} & \textbf{\#Sampled} & \textbf{\#Eq. Mut.} & \textbf{NC} & \textbf{NI} & \textbf{ND} & \textbf{NC\%} &\textbf{NI\%} & \textbf{ND\%} \\
    \hline
    \textbf{PIT} & 96  & 3  & 43  & 27  & 23  & 46.24\% & 29.03\% & 24.73\% \\
    \textbf{Major} & 96  & 4  & 42  & 30  & 20  & 45.65\% & 32.61\% & 21.74\% \\
    \textbf{LEAM} & 95  & 7  & 59  & 16  & 13  & 67.05\% & 18.18\% & 14.77\% \\
    \textbf{$\mu$BERT} & 88  & 8  & 37  & 24  & 19  & 46.25\% & 30.00\% & 23.75\% \\
    \hline
    \textbf{BugFarm (GPT-4o)} & 94  & 9  & 62  & 11  & 12  & 72.94\% & 12.94\% & 14.12\% \\
    \textbf{BugFarm (DS-671b)} &  94 &  11 &  52 & 15  & 16  & 62.65\% & 18.07\% & 19.28\% \\
    \hline
    \textbf{LLMopheus (GPT-4o)} & 95  & 11  & 26  & 44  & 14  & 30.95\% & 52.38\% & 16.67\% \\
    \textbf{LLMopheus (DS-671b)} & 95  & 13  &  20 &  47 & 15  & 24.39\% & 57.32\% &  18.29\%\\
    \hline
    \textbf{\ourtool (GPT-3.5)} & 95  & 12  & 66  & 12  & 5  & 79.52\% & 14.46\% & 6.02\% \\
    \textbf{\ourtool (GPT-4o)} & 95  & 11  & 53  & 16  & 15  & 63.10\% & 19.05\% & 17.86\% \\
    \textbf{\ourtool (GPT-4o-M)} & 95  & 12  & 51  & 15  & 17  & 61.45\% & 18.07\% & 20.48\% \\
    \textbf{\ourtool (DS-671b)} & 95  & 12  & 45  & 16  & 22  & 54.22\% & 19.28\% & 26.51\% \\
    \textbf{\ourtool (DS-236b)} & 95  & 13  & 54  & 15  & 13  & 65.85\% & 18.29\% & 15.85\% \\
    \textbf{\ourtool (SC-16b)} & 93  & 16  & 62  & 11  & 4  & 80.52\% & 14.29\% & 5.19\% \\
    \textbf{\ourtool (CL-13b)} & 96  & 11  & 68  & 12  & 5  & 80.00\% & 14.12\% & 5.88\% \\
    \hline
    \end{tabular}
  \tablabel{surviving}
\end{table}

%% file: sections/discussion.tex
\section{Discussion}
\label{sec:discussion}

In this section, we examine the impact of various experiment setups for LLMs and discuss their implications.

\subsection{Sensitivity to Experimental Settings} \label{sec:diss-sentivitiy-to-settings}

We discuss the experiment setups that may affect the evaluation results, particularly the setups related to LLMs.

\subsubsection{Setup of the Prompts} 

Prompts play a critical role in the performance of LLMs.
Therefore, we aim to investigate how different prompts influence mutant generation effectiveness.
Due to budget constraints, we selected a subset of 306 bugs from our dataset, comprising 10 randomly sampled bugs from each Defects4J project and all bugs from the ConDefects dataset.

Based on the prompt design of LLMut, we investigate how different sources of contextual information affect mutant generation.
For this experiment, we select two representative LLMs—the closed-source GPT-4o and the recent open-source DeepSeek-V3-671b—and vary the prompt template used for mutation generation.
Note that based on the evaluation results, both models are leading LLMs in various metrics.
Starting from the prompt that includes the richest contextual information, we progressively remove specific sources of context in the \textit{Context} section.
This process yields five distinct prompt variants, summarized as follows:
\begin{enumerate}
    \item \textbf{Prompt 1 (P1):} P1 is the richest prompt with the code of \textit{the whole Java source file} as the \textit{Context} of the prompt.
    
    \item \textbf{Prompt 2 (P2):} P2 is the prompt with the code of \textit{the focal method} and \textit{its corresponding unit tests} as the \textit{Context} of the prompt.
    
    \item \textbf{Prompt 3 (P3):} P3 is the prompt with the code of \textit{the focal method} as the \textit{Context} of the prompt, which is the default prompt in the previous section, shown as \figref{prompt}.

    \item \textbf{Prompt 4 (P4):} P4 is the prompt with the code of the target code element to be mutated as the \textit{Context} of the prompt.
    
    \item \textbf{Prompt 5 (P5):} P5 is the prompt with the code of \textit{the focal method} as the \textit{Context}, but remove all few-shot examples.
\end{enumerate}

\input{sections/results/TPrompts-new}

\tabref{comparing-prompts} presents a comparative evaluation of different prompts on two models across the main effectiveness, validity, and efficiency metrics.

For the effectiveness metrics (i.e., real bug detectability, average Ochiai, and coupling rate), on both models, \textbf{P3}, the prompt used in the previous section, delivers the best overall performance, demonstrating its strength in generating mutants that are semantically closer to real bugs. \textbf{P5}, which removes the few-shot examples, ranks as the second-best prompt overall, suggesting that few-shot examples help guide the model toward generating mutants closer to real bugs. Interestingly, \textbf{P1}, the longest prompt with the whole Java file, performs the worst in most cases, indicating that providing more information does not necessarily help LLMs generate more bug-representative mutants. \textbf{P2}, which adds unit tests to the context compared with \textbf{P3}, shows slightly reduced effectiveness—suggesting that including unit tests does not help the model generate better mutants.

For the validity metrics, different prompts express a similar trend in both models. For the compilability rate, \textbf{P5}, the one without real-world bug few-shot examples, achieves the highest compilability rate on both models.
This indicates that removing few-shot examples reduces the syntactic complexity of the generated mutations, making them easier for LLMs to produce without introducing compilation errors.
For the duplicated mutant rate, \textbf{P1} achieves the best performance on GPT-4o, while \textbf{P5} performs best on DS-671b, suggesting that duplication is more influenced by model behavior than by the prompts.

For the efficiency costs, the results show a clear trend: prompts with more contextual information consume more time and tokens. For example, \textbf{P1} (whole file) incurs the highest cost, followed by \textbf{P3} (whole method), while \textbf{P5} (only the code snippet to be mutated) is the most efficient. Another interesting finding is that \textbf{P1} (whole file) is significantly longer than \textbf{P2} (unit tests).

We further explore the performance similarity of different prompts.
To measure their similarity in performance, we use the Spearman coefficient and Pearson correlation, as shown in the top part of \tabref{context-similarity}.
We can see that their mutual similarity exceeds 0.95, which is above the typical threshold of 0.85~\cite{siami2008sufficient,motwani2023better,kaufman2022prioritizing}, and the $p$-values are all below 0.05.
The results suggest that on both GPT-4o and DS-671b, although \textbf{P3} achieves the most balanced overall performance, the differences among prompts are not statistically significant.

\subsubsection{Experiment Setup of the Context Length}

\input{sections/results/TContextLength}

\input{sections/results/TSimilarity-new}

To compare the similarity to real defects, we deliberately select the context around the bug locations for generating mutants for all the approaches.
In our previous experiments, the context length is set to three lines around the bug location, so we conduct experiments to compare the effects of two-line and one-line contexts, as shown in \tabref{context-lengths}.
Similarly, the middle part of \tabref{context-similarity} shows the performance similarity among different contexts.
We can see that their mutual similarity exceeds 0.95, and the $p$-values are all below 0.05.
Different context lengths perform similarly, thus validating the setup of our experiment.

\subsubsection{Impact of Different Few-Shot Examples} \label{sec:discussion-diff-few-shot-examples}

\input{sections/results/TDiffExamples-new}

To evaluate how different few-shot examples influence the results, we conduct comparative experiments by varying the few-shot examples used in the default LLMut (GPT-4o) prompt.
First, we randomly select 3, 4, 5, 6, 7, 8, 9, and 10 additional examples from QuixBugs, and ensure there is no overlap with the default 6 examples.
Additionally, despite the risk of data leakage, we select 6 examples from the Defects4J dataset.
The results are presented in~\tabref{diff-examples}.

We further compute their similarity, shown as the bottom part of \tabref{context-similarity}.
We can find that their similarity exceeds 0.95, significantly exceeding the typical threshold of 0.85, and the $p$-values are all below 0.05.
Therefore, despite variations in few-shot examples, there is no statistically significant evidence indicating superior performance among the prompts.

\subsubsection{Performance of Using the Same Number of Mutants}
In Section~\ref{sec:eval}, we did not restrict the number of mutants generated by each approach.
To study the performance of each approach under fixed quantity conditions, we followed the setting of the existing studies~\cite{tian2022learning,hariri2019comparing} and limited the number of mutants generated by all methods to the minimum produced by $\mu$BERT~\cite{degiovanni2022mubert}, which is 11397.
This ensures a fair comparison across approaches by controlling for differences in mutant counts, thereby eliminating the potential bias that methods generating more mutants might appear to perform better simply due to quantity rather than quality.
For counts exceeding this number, we randomly selected the specified number of mutants for analysis.
We conducted 10 rounds of random sampling comparisons, and the results are presented in \tabref{same-number}.
%Therefore, under fixed quantity conditions, the mutants generated by two GPT-4 LLMs are still the closest to real bugs in behavior.

Under a fixed number of mutants, we observe that LLMut (DS-671b) continues to deliver the best performance in both real bug detection rate and average Ochiai coefficient, with LLMut (GPT-4o) following closely. LLMorpheus (DS-671b) and LLMorpheus (GPT-4o) achieve the highest coupling rates. In terms of validity metrics—compilability rate and duplicate mutant rate—BugFarm (DS-671b) performs the best among all LLM-based approaches. These findings indicate that, when controlling for the number of generated mutants, \textbf{different LLM-based approaches (e.g., LLMut, LLMorpheus, and BugFarm) exhibit distinct strengths rather than a single universally superior approach}.

\input{sections/results/TSameNumberMutation-new}

\begin{figure}[t]
    \centering
    \includegraphics[width=\linewidth]{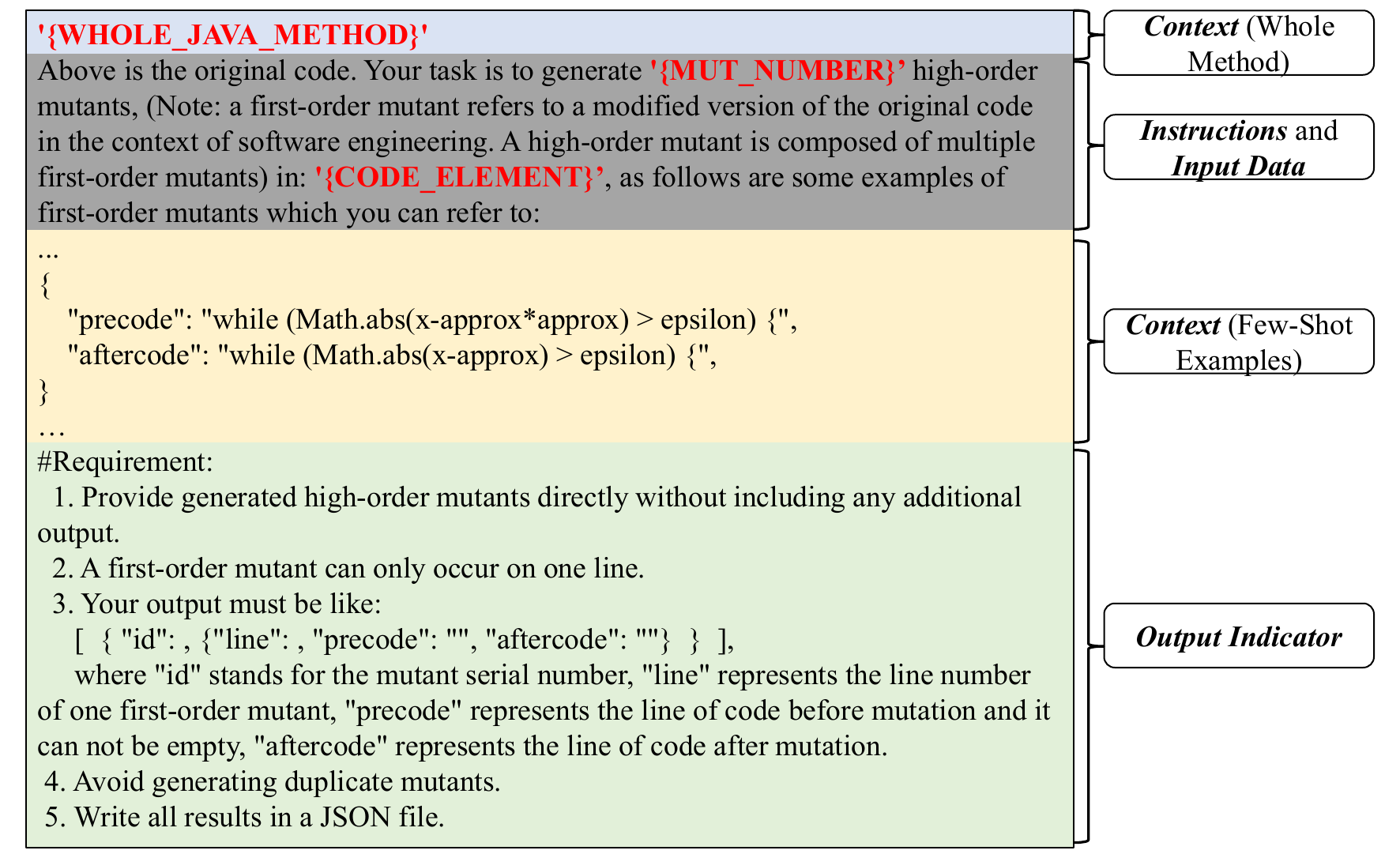}
    \caption{The Modified Prompt Template for High-Order Mutant Generation}
    \figlabel{hom-prompt}
\end{figure}

\input{sections/results/THOM}

\subsection{Effectiveness on High-Order Mutant Generation}
Previously, we only evaluated LLMs in generating first-order mutants (FOMs). To explore the effectiveness of LLMs in generating high-order mutants (HOMs)~\cite{jia2009higher,wong2020efficiently}, we conducted a preliminary experiment by evaluating LLMut on the recent mainstream models, including GPT-4o and DS-671b.

We modified the original prompt to explicitly request HOM generation, shown as \figref{hom-prompt}.
The prompts directly instruct LLMs to generate HOMs, and each HOM is composed of several first-order mutants.

\tabref{hom} shows the results.
For the effectiveness metrics, HOMs generally achieve average Ochiai scores and higher coupling rates across these models, indicating closer semantic alignment with real bugs. 
However, FOMs achieve higher real bug detectability on all models.
For the validity metrics, HOMs exhibit lower duplicate mutant rates, suggesting improved mutant diversity.
However, FOMs exhibit significantly higher compilability rates, which aligns with the expectation that larger code changes are more prone to introducing compilation errors. Our preliminary experiments suggest that strengthening LLM-based HOM generation is promising, as HOMs are semantically closer to real bugs and face more challenges in maintaining syntactic correctness.

\subsection{Implications}
Based on our findings, we discuss the pros and cons of traditional and LLM-based mutant generation approaches and suggest potential improvements.

\subsubsection{Advantages of LLM-based Approaches}
Our study demonstrates that LLMs can generate high-quality mutants for mutation testing at an acceptable cost. The key advantage of LLMs lies in their ability to produce mutants with stronger coupling relationships with real-world bugs, which is crucial for improving test effectiveness.

Additionally, LLMs could generate syntactically diverse mutants compared with traditional approaches, which is potentially the reason for generating high-coupling mutants.
For example, Listing~\ref{lst:mut-examples} presents two compilable mutants generated by GPT for Chart-1, which are theoretically beyond the capabilities of traditional approaches.
In \code{Mutant-1}, the LLM infers a relationship between variables (\code{rightBlock} implies the existence of \code{leftBlock}) and modifies the variable accordingly.
This mutant is not killed by the original bug-triggering test, highlighting its potential for guiding test enhancement.
Without natural language understanding (e.g., recognizing the correspondence between “right” and “left”), traditional approaches, both rule-based approaches and the approaches based on small-scale models, struggle to generate such context-aware mutants.
In \code{Mutant-2}, the LLM transforms an \code{else} branch into an \code{else if} condition by synthesizing a logical condition using the integer variable \code{seriesCount} from the context, demonstrating an ability to introduce structurally meaningful changes that traditional approaches cannot achieve.

\input{figs/MutExample}

Additionally, our study explores how different prompt designs, model choices, and contextual information influence mutant generation performance.
We find that selecting models with stronger code generation capabilities (e.g., GPT-4o) and designing prompts carefully are critical to achieving optimal results.

Moreover, our study demonstrates that LLMs have clear potential in generating HOMs. Compared with first-order mutants (FOMs), HOMs generally achieve higher coupling rates and average Ochiai scores, suggesting that they are semantically closer to real bugs, while suffering from lower compilability rates and reduced real bug detectability. The results highlight that strengthening LLM-based HOM generation is therefore a promising research direction.

\subsubsection{Advantages of Traditional Approaches}
Despite the advantages of LLM-generated mutants, traditional approaches still exhibit strengths in reliability and efficiency.
They produce fewer non-compilable and duplicate mutants, with PIT achieving a 100\% compilability rate and a 0\% duplicate mutant rate, while Major follows closely with a 97.6\% compilability rate and 0\% duplicate mutant rate.
In contrast, LLM-based approaches exhibit significantly lower compilability and higher duplicate mutant rates.
Additionally, overall, traditional rule-based approaches generate a slightly lower equivalent mutant rate, further avoiding redundant costs.
This suggests that traditional approaches remain highly practical when strict compilability and low redundancy are prioritized.

\subsubsection{Challenges and Future Directions}
Although LLMs demonstrate superior behavioral similarity to real bugs, our findings indicate that there is still substantial room for improvement. One major limitation is the high rate of non-compilable mutants, which limits their practical application in mutation testing.
Our analysis reveals that common causes include syntactic errors in generated expressions, incorrect method invocations, and type mismatches.
To address these challenges, future research should explore the integration of LLM-based approaches with program analysis and repair techniques to refine mutant generation. Specifically, leveraging static analysis to filter or adjust syntactically incorrect mutants before execution can enhance their reliability.

Moreover, exploring the applications of LLM-generated mutants in the downstream tasks, such as mutation-based test case prioritization (MBTCP) and mutation-based fault localization (MBFL). Our study has shown that LLM-generated mutants often have a higher similarity with real bugs (e.g., have larger Ochiai coefficients), highlighting the potential of adopting them for these tasks.

Another important direction is cross-language generalization. While our study primarily focuses on Java, mutation testing is equally critical in languages such as C/C++, Python, JavaScript, and Rust. Traditional mutation testing tools usually require language-specific implementations, leading to duplicated engineering efforts and limited portability. In contrast, modern LLMs possess inherent multilingual capabilities, as they are pre-trained on large-scale, cross-language code corpora. This enables them to naturally transfer mutation strategies across programming languages without retraining from scratch. Future work should investigate how well LLMs can exploit this cross-language knowledge, and how prompts, training data, or filtering strategies can be adapted to different language paradigms (e.g., strongly typed vs. dynamically typed). Advancing towards multilingual mutant generation would not only broaden the applicability of LLMs in software testing but also provide stronger evidence for their robustness in heterogeneous, real-world development environments.

\subsection{Limitations of Our Study}
Our study has the following limitations that are left for further exploration.

First, while we evaluated a comprehensive set of static and behavioral metrics, we did not investigate more advanced dynamic indicators. For instance, the phenomenon of \textit{subsuming mutants}~\cite{garg2022cerebro,diniz2021dissecting}, where one mutant is always killed by any test that kills another, has been studied in the mutant testing literature as a way to assess better redundancy and the true fault-detection power of test suites. Incorporating such dynamic analyses would provide deeper insights into the effectiveness of LLM-generated mutants beyond the metrics reported in this paper.
Additionally, while in the context of test suite enhancement, two mutants with identical failing test sets may be considered redundant, in mutation-based fault localization~\cite{wang2025systematic}, the number of mutants per statement directly influences suspiciousness scores, since these are typically computed as averages across all associated mutants.

Second, our experiments are limited to the Java programming language. Although Java is the most widely studied language in mutation testing, it does not cover the diversity of programming paradigms. Mutant behaviors may differ significantly in languages such as C/C++, Python, JavaScript, and Rust, due to differences in type systems, memory models, and runtime environments. Future research should examine the generalizability of LLM-based mutant generation across multiple languages. Notably, modern LLMs are inherently multilingual, trained on large-scale cross-language corpora, which makes cross-language mutant generation a promising direction for broadening the applicability of our findings.

Third, our study relies on a fixed few-shot examples, which restricts LLMs to producing context-specific mutants for the given code snippets. 
Moreover, constructing few-shot examples requires manual effort and may introduce unintended biases into the model’s behavior during inference.
In addition, we impose no restrictions on the mutation operators applied by LLMs, whereas rule-based approaches rely on predefined and relatively simple operators. Consequently, it is still an open question how LLMs would perform if constrained to the same operator set as rule-based approaches.

%% file: sections/results/TPrompts-new.tex
\begin{table}[tp]
  \centering
  \scriptsize
  \caption{Comparative Evaluation of Different Prompts for \ourtool}
    \begin{tabular}{|l|c|c|c|c|c||c|c|c|c|c|}
    \hline
    \multicolumn{1}{|c|}{\multirow{2}{*}{\textbf{Metric}}} & \multicolumn{5}{c||}{\textbf{ \ourtool (GPT-4o) }} & \multicolumn{5}{c|}{\textbf{\ourtool (DS-671b)}} \\
\cline{2-11}       & \textbf{P1} & \textbf{P2} & \textbf{P3} & \textbf{P4} & \textbf{P5} & \textbf{P1} & \textbf{P2} & \textbf{P3} & \textbf{P4} & \textbf{P5} \\
    \hline
    \textbf{Mut. \#} & 3696 & 3692 & 3781 & \cellcolor{gray!40}3807 & 3788 & 3878 & 3844 & \cellcolor{gray!40}3851 & 3834 & 3847 \\
    \hline
    \textbf{Mut. Score} & \cellcolor{gray!40}0.841  &  0.801  & 0.773  & 0.832 & 0.785  & \cellcolor{gray!40}0.776  & 0.753  & 0.765  & 0.760  & 0.750  \\
    \hline
    \hline
    \textbf{Real Bug Detec.} & 83.1\% & 83.9\% & \cellcolor{gray!40}85.0\% & 83.1\% & 84.7\% & 83.1\% & 84.7\% & 85.5\% & 83.9\% & \cellcolor{gray!40}85.8\% \\
    \hline
    \textbf{Avg. Ochiai} & 64.8\% & 65.0\% & \cellcolor{gray!40}66.4\% & 64.8\% & 65.3\% & 66.4\% & 65.8\% & \cellcolor{gray!40}67.5\% & 66.4\% & 65.6\% \\
    \hline
    \textbf{Coupling Rate} & 47.9\% & 49.7\% & \cellcolor{gray!40}52.2\% & 48.4\% & 52.1\% & 49.1\% & 50.0\% & \cellcolor{gray!40}53.2\% & 50.9\% & 51.7\% \\
    \hline
    \hline
    \textbf{Compilability Rate} & 78.4\% & 81.6\% & 80.1\% & 77.4\% & \cellcolor{gray!40}82.5\% & 79.5\% & 82.0\% & 82.8\% & 78.0\% & \cellcolor{gray!40}83.5\% \\
    \hline
    \textbf{Dup. Mut. Rate} & \cellcolor{gray!40}7.7\% & 7.9\% & 7.9\% & 8.7\% & 8.1\% & 8.2\% & 7.0\% & 7.3\% & 8.5\% & \cellcolor{gray!40}6.4\% \\
    \hline
    \hline
    \textbf{Avg. Gen. Time} & 1.607 & 1.584 & 1.389 & \cellcolor{gray!40}1.017 & 1.282 & 2.596 & 2.601 & 2.573 & \cellcolor{gray!40}2.488 & 2.571 \\
    \hline
    \textbf{Avg. \# Token} & 392 & 218 & 105 & \cellcolor{gray!40}44 & 94 & 392 & 219 & 104 & \cellcolor{gray!40}43 & 92 \\
    \hline
    \end{tabular}
    \\{\scriptsize Note: \textbf{P3} is the prompt used in the previous section. The prompt template that achieves the best performance is highlighted in grey background color.}
  \tablabel{comparing-prompts}
\end{table}

%% file: sections/results/TContextLength.tex
\begin{table}[tp]
  \centering
  \scriptsize
  \caption{ Performance of Different Context Lengths of \ourtool (GPT-4o)}
  \vspace{-3mm}
    \begin{tabular}{|l|c|c|c|}
    \hline
    \textbf{Metric } & \multicolumn{1}{l|}{\textbf{1-Line Context }} & \multicolumn{1}{l|}{\textbf{2-Line Context}} & \multicolumn{1}{l|}{\textbf{3-Line Context}} \\
    \hline
    \textbf{Real Bug Dectc.} & 93.3\% & 91.7\% & 93.3\% \\
    \hline
    \textbf{Avg. Ochiai} & 46.7\% & 50.0\% & 48.3\% \\
    \hline
    \textbf{Coupling Rate} & 37.3\% & 36.8\% & 37.7\% \\
    \hline
    \textbf{Compilability Rate} & 80.7\% & 81.4\% & 82.1\% \\
    \hline
    \textbf{Dup. Mut. Rate} & 4.9\% & 6.6\% & 5.3\% \\
    \hline
    \end{tabular}
  \tablabel{context-lengths}
\end{table}

%% file: sections/results/TSimilarity-new.tex
\begin{table}[tp]
  \centering
  \scriptsize
  \caption{Performance Similarity of Different LLM Configurations of \ourtool}
    \begin{tabular}{|l|r|r||r|r|}
    \hline
    \textbf{Settings} & \multicolumn{1}{l|}{\textbf{Spea. Coef}} & \multicolumn{1}{l||}{\textbf{$p$-value}} & \multicolumn{1}{l|}{\textbf{Pear. Coef}} & \multicolumn{1}{l|}{\textbf{$p$-value}} \\
    \hline
    \textbf{GPT-4o P3 v.s. GPT-4o P1} & 0.9500  & 8.76E-05 & 0.9969  & 5.37E-09 \\
    \hline
    \textbf{GPT-4o P3 v.s. GPT-4o P2} & 1.0000  & 0.00E+00 & 0.9995  & 8.56E-12 \\
    \hline
    \textbf{GPT-4o P3 v.s. GPT-4o P4} & 0.9500  & 8.76E-05 & 0.9998  & 8.32E-14 \\
    \hline
    \textbf{GPT-4o P3 v.s. GPT-4o P5} & 1.0000  & 0.00E+00 & 0.9999  & 5.61E-19 \\
    \hline
    \textbf{DS-671b P3 v.s. DS-671b P1} & 1.0000  & 0.00E+00 & 0.9999  & 1.51E-30 \\
    \hline
    \textbf{DS-671b P3 v.s. DS-671b P2} & 1.0000  & 0.00E+00 & 0.9999  & 4.95E-31 \\
    \hline
    \textbf{DS-671b P3 v.s. DS-671b P4} & 1.0000  & 0.00E+00 & 0.9999  & 2.93E-29 \\
    \hline
    \textbf{DS-671b P3 v.s. DS-671b P5} & 1.0000  & 0.00E+00 & 0.9999  & 1.08E-31 \\
    \hline
    \hline
    \textbf{1-Line vs. 2-Line} & 0.9999  & 1.40E-24 & 0.9989  & 4.54E-05 \\
    \hline
    \textbf{2-Line vs. 3-Line} & 0.9999  & 1.40E-24 & 0.9995  & 1.46E-05 \\
    \hline
    \textbf{1-Line vs. 3-Line} & 0.9999  & 1.40E-24 & 0.9998  & 2.99E-06 \\
    \hline
    \hline
    \textbf{QB-6a (Default) v.s. QB-6b} & 1.0000  & 0.00E+00 & 0.9999  & 2.99E-15 \\
    \hline
    \textbf{QB-6a (Default) v.s. QB-3} & 1.0000  & 0.00E+00 & 0.9999  & 1.82E-14 \\
    \hline
    \textbf{QB-6a (Default) v.s. QB-4} & 1.0000  & 0.00E+00 & 0.9999  & 3.97E-15 \\
    \hline
    \textbf{QB-6a (Default) v.s. QB-5} & 1.0000  & 0.00E+00 & 0.9999  & 7.62E-14 \\
    \hline
    \textbf{QB-6a (Default) v.s. QB-7} & 1.0000  & 0.0E+00 & 0.9999  & 1.12E-14 \\
    \hline
    \textbf{QB-6a (Default) v.s. QB-8} & 1.0000  & 0.0E+00 & 0.9999  & 1.28E-13 \\
    \hline
    \textbf{QB-6a (Default) v.s. QB-9} & 1.0000  & 0.00E+00 & 0.9999  & 7.03E-15 \\
    \hline
    \textbf{QB-6a (Default) v.s. QB-10} & 1.0000  & 0.00E+00 & 0.9999  & 7.03E-15 \\
    \hline
    \textbf{QB-6a (Default) v.s  D4J-6} & 1.0000  & 0.00E+00 & 0.9999  & 9.81E-17 \\
    \hline
    \end{tabular}
  \tablabel{context-similarity}
   \\{\scriptsize \textbf{QB-N}: Randomly samples N Examples from QuixBug. QB-6a is the set of default examples, while QB-6b randomly samples 6 other examples. D4J-6 randomly samples 6 examples from Defects4J.}
\end{table}

%% file: sections/results/TDiffExamples-new.tex
\begin{table}[tp]
  \centering
  \scriptsize
  \caption{Impact on Different Few-Shot Examples by \ourtool (GPT-4o)}
    \begin{tabular}{|l|c|c|c|c|c|c|c|c|c|c|}
    \hline
    \textbf{Metric} & \textbf{QB-6a} & \textbf{QB-3} & \textbf{QB-4} & \textbf{QB-5} & \textbf{QB-6b} & \textbf{QB-7} & \textbf{QB-8} & \textbf{QB-9} & \textbf{QB-10} & \textbf{D4J-6} \\
    \hline
    \textbf{Real Bug Detec} & 0.933  & 0.900  & 0.917  & 0.933  & 0.917  & 0.900  & 0.883  & 0.933  & 0.933  & 0.933  \\
    \hline
    \textbf{ Avg. Ochiai} & 0.483  & 0.483  & 0.450  & 0.417  & 0.450  & 0.450  & 0.433  & 0.467  & 0.483  & 0.483  \\
    \hline
    \textbf{Coupling Rate} & 0.379  & 0.379  & 0.385  & 0.380  & 0.373  & 0.362  & 0.348  & 0.367  & 0.352  & 0.388  \\
    \hline
    \textbf{Comp. Rate} & 82.1\% & 82.4\% & 80.8\% & 81.7\% & 81.6\% & 81.4\% & 81.8\% & 82.2\% & 81.4\% & 83.1\% \\
    \hline
    \textbf{Dup.Mut.Rate} & 5.3\% & 5.6\% & 6.4\% & 5.6\% & 5.8\% & 5.6\% & 7.5\% & 6.4\% & 6.2\% & 5.8\% \\
    \hline
    \textbf{Avg. \# Token} & 106 & 98 & 102 & 103 & 106 & 108 & 109 & 110 & 113 & 107 \\
    \hline
    \end{tabular}
   \tablabel{diff-examples}

   {\scriptsize \textbf{QB-N}: Randomly samples N Examples from QuixBug. QB-6a is the set of default examples, while QB-6b randomly samples 6 other examples. D4J-6 randomly samples 6 examples from Defects4J.}
\end{table}

%% file: sections/results/TSameNumberMutation-new.tex
\begin{table}[tp]
  \centering
  \scriptsize
  \caption{Performance Under the Same Number of Mutations among All Studied Approaches}
    \begin{tabular}{l|l|l|l|l|l}
    \hline
    \textbf{Name} & \textbf{R. B. Detec.} & \textbf{Avg. Ochiai} & \textbf{Coup. Rate} & \textbf{Comp. Rate} & \textbf{Dup. Rate} \\
    \hline
    \textbf{PIT} & 31.6\% & 12.6\% & 17.4\% & —  & — \\
    \textbf{Major} & 68.8\% & 22.2\% & 45.4\% & \cellcolor{gray!40}96.9\% & \cellcolor{gray!40}0.0\% \\
    \textbf{LEAM} & 64.6\% & 14.1\% & 37.2\% & 54.1\% & 2.6\% \\
    \textbf{$\mu$BERT} & 66.0\% & 19.0\% & 43.6\% & 26.1 & 1.9\% \\
    \hline
    \textbf{BugFarm (GPT-4o)} & 58.6\% & 17.7\% & 47.9\% & 74.2\% & 7.5\% \\
    \textbf{BugFarm (DS-671b)} & 58.9\% & 17.9\% & 48.0\% & 77.4\% & 4.6\% \\
    \hline
    \textbf{LLMorpheus (GPT-4o)} & 67.5\% & 23.6\% & 50.2\% & 68.2\% & 7.3\% \\
    \textbf{LLMorpheus (DS-671b)} & 68.5\% & 24.9\% & \cellcolor{gray!40}51.3\% & 67.8\% & 6.8\% \\
    \hline
    \textbf{\ourtool (GPT-3.5)} & 82.2\% & 26.5\% & 45.6\% & 61.2\% & 9.9\% \\
    \textbf{\ourtool (GPT-4o)} & 87.4\% & 36.0\% & 48.8\% & 75.5\% & 7.7\% \\
    \textbf{\ourtool (GPT-4o-M)} & 88.6\% & 31.8\% & 47.9\% & 74.3\% & 7.1\% \\
    \textbf{\ourtool (DS-671b)} & \cellcolor{gray!40}89.2\% & \cellcolor{gray!40}39.4\% & 49.4\% & 76.3\% & 7.7\% \\
    \textbf{\ourtool (DS-236b)} & 87.0\% & 34.1\% & 47.9\% & 74.6\% & 7.6\% \\
    \textbf{\ourtool (SC-16b)} & 33.9\% & 15.4\% & 27.1\% & 12.4\% & 9.6\% \\
    \textbf{\ourtool (CL-13b)} & 68.5\% & 16.9\% & 43.6\% & 69.5\% & 38.3\% \\
    \hline
    \end{tabular}
  \tablabel{same-number}
\end{table}

%% file: sections/results/THOM.tex
\begin{table}[tp]
  \centering
  \scriptsize
  \caption{Comparison Between First-Order Mutants and High-Order Mutants}
    \begin{tabular}{|l|c|c|c|c|c|c|}
    \hline
    \multirow{2}{*}{\textbf{Metric}} & \multicolumn{2}{c|}{\textbf{\ourtool (GPT-4o)}} & \multicolumn{2}{c|}{\textbf{\ourtool (DS-671b)}} \\
\cline{2-5}      & \textbf{FOM} & \textbf{HOM} & \textbf{FOM} & \textbf{HOM} \\
    \hline
    \textbf{Real Bug Detec.}   & \cellcolor{gray!40}0.933  & \cellcolor{gray!40}0.933  & \cellcolor{gray!40}0.950  & 0.933  \\
    \hline
    \textbf{Avg. Ochiai}   & \cellcolor{gray!40}0.483  & \cellcolor{gray!40}0.483  & \cellcolor{gray!40}0.500  & 0.483  \\
    \hline
    \textbf{Coupling Rate}  & 0.379  & \cellcolor{gray!40}0.507  & 0.377  & \cellcolor{gray!40}0.506  \\
    \hline
    \textbf{Compilability Rate} & \cellcolor{gray!40}82.1\% & 70.2\% & \cellcolor{gray!40}84.1\% & 73.3\% \\
    \hline
    \textbf{Dup. Mut. Rate}  & 5.3\% & \cellcolor{gray!40}3.5\% & 5.1\% & \cellcolor{gray!40}3.1\% \\
    \hline
    \end{tabular}
    \\{\scriptsize Note: The order achieving the best performance in terms of a metric is highlighted in a grey background color.}
  \tablabel{hom}
\end{table}

% \begin{table}[tp]
%   \centering
%   \scriptsize
%   \caption{\bocomment{Comparison Between First-Order Mutations and High-Order Mutations}}
%     \begin{tabular}{|l|c|c|c|c|c|c|}
%     \hline
%     \multirow{2}{*}{\textbf{Metric}} & \multicolumn{2}{c|}{\textbf{\ourtool (GPT-3.5)}} & \multicolumn{2}{c|}{\textbf{\ourtool (GPT-4o)}} & \multicolumn{2}{c|}{\textbf{\ourtool (DS-671b)}} \\
% \cline{2-7}       & \textbf{FOM} & \textbf{HOM} & \textbf{FOM} & \textbf{HOM} & \textbf{FOM} & \textbf{HOM} \\
%     \hline
%     \textbf{Real Bug Detec.} & \cellcolor{gray!40}0.917  & 0.900  & 0.933  & 0.933  & \cellcolor{gray!40}0.950  & 0.933  \\
%     \hline
%     \textbf{Avg. Ochiai} & 0.467  & \cellcolor{gray!40}0.483  & 0.467  & \cellcolor{gray!40}0.483  & \cellcolor{gray!40}0.500  & 0.483  \\
%     \hline
%     \textbf{Coupling Rate} & 0.136  & \cellcolor{gray!40}0.249  & 0.151  & \cellcolor{gray!40}0.307  & 0.169  & \cellcolor{gray!40}0.306  \\
%     \hline
%     \textbf{Compilability Rate} & \cellcolor{gray!40}67.5\% & 63.4\% & \cellcolor{gray!40}76.3\% & 70.2\% & \cellcolor{gray!40}78.4\% & 73.3\% \\
%     \hline
%     \textbf{Dup. Mut. Rate} & 13.7\% & \cellcolor{gray!40}5.2\% & 5.3\% & \cellcolor{gray!40}3.5\% & 5.1\% & \cellcolor{gray!40}3.1\% \\
%     \hline
%     \end{tabular}
%     \\{\scriptsize Note: The order achieving the best performance in terms of a metric is highlighted in grey background color.}
%   \tablabel{hom}
% \end{table}

%% file: figs/MutExample.tex
\begin{codelist}
\begin{lstlisting}[style=Java,caption=Two Example Mutations of GPT on Chart-1,label=lst:mut-examples] 
<@\hlc{gray!15}{\qquad // Mutant-1}@>
<@\hlc{red!15}{- \qquad Size2D size = this.rightBlock.arrange(g2, c4); }@>
<@\hlc{green!15}{+ \qquad   Size2D size = this.leftBlock.arrange(g2, c4); }@>
<@\hlc{gray!15}{\qquad // Mutant-2}@>
<@\hlc{red!15}{- \qquad else  \{ }@>
<@\hlc{red!15}{- \qquad \enspace \enspace g2.setStroke(getItemOutlineStroke(row, column, selected)); }@>
<@\hlc{red!15}{- \qquad \} }@>
<@\hlc{green!15}{+ \qquad else if (seriesCount == 1) \{ }@>
<@\hlc{green!15}{+ \qquad \enspace \enspace g2.setStroke(getItemOutlineStroke(row, column, selected)); }@>
<@\hlc{green!15}{+ \qquad \} }@>
\end{lstlisting}
\end{codelist}

%% file: sections/threats.tex
\section{Threats to Validity} \label{sec:threats}

In this section, we illustrate the threats to the validity of our study.

The selected programming language, mutant generation approaches, datasets, and LLMs could be a threat to the validity of our results. To mitigate this threat, employ the most widely studied mutation testing approaches as baselines, including learning-based and rule-based, the most widely used language (i.e., Java), and the most widely used dataset Defects4J, and we adopt the widely studied models covering a range of base models (i.e., GPT models, StarCoder, CodeLlama, and DeepSeek models).
Moreover, our study further covers two state-of-the-art LLM-based mutant generation approaches, BugFarm and LLMorpheus, as baselines, both of which are equipped with the recent mainstream models GPT-4o and DS-671b.

The metrics we used may threaten the conclusion of our study. Each metric has inherent limitations. For instance, \textit{Real Bug Detection Rate} and \textit{Coupling Rate} capture only partial aspects of similarity to real bugs: a method may detect many bugs but with weak coupling on each bug, or achieve strong coupling concentrated on a few bugs.
Likewise, \textit{AST Edit Distance} reflects structural differences, but a higher syntactic distance does not necessarily imply larger semantic changes. Additionally, the measurement of \textit{Average Generation Time} may be affected by external factors such as GPU type, service provider implementation, and network fluctuations, introducing unavoidable noise.
To mitigate the bias of individual metrics, we report a comprehensive set of complementary measures. Particularly, besides time cost, we also report token consumption, which provides a more stable and model-independent estimate of computational and financial cost.

Another validity threat may be due to data leakage, i.e., the fact that the data in Defects4J~\cite{just2014defects4j} may be covered in the training set of the studied LLMs. To mitigate this threat, we employed another dataset, ConDefects~\cite{wu2023condefects}, which includes programs and faults that were made after the release time of most LLMs we use (except DeepSeek-V3-671b only) and thus have limited data leakage risk. Additionally, to increase confidence in our results, we also checked whether the tools can introduce exact matches (syntactically) with the studied faults. We hypothesize that, in case the tools have been tuned based on specific fault instances, the tools would introduce at least one mutant that is an exact match with the faults we investigate. On Defects4J, the results are LLMut (GPT-3.5) (133), LLMut (GPT-4o) (63), LLMut (GPT-4o-M) (138), LLMut (DS-671b) (72), LLMut (DS-236b) (66), LLMut (SC-16b) (13), LLMut (CL-13b) (28), BugFarm (GPT-4o) (59), BugFarm (DS-671b) (66), LLMorpheus (GPT-4o) (26), LLMorpheus (DS-671b) (31), LEAM (287), $\mu$Bert (12), and Major (22), respectively.
On ConDefects, the results are LLMut (GPT-3.5) (22), LLMut (GPT-4o) (25), LLMut (GPT-4o-M) (31), LLMut (DS-671b) (35), LLMut (DS-236b) (32), LLMut (SC-16b) (35), LLMut (CL-13b) (56), BugFarm (GPT-4o) (22), BugFarm (DS-671b) (29), LLMorpheus (GPT-4o) (19), LLMorpheus (DS-671b) (28), LEAM (56), $\mu$Bert (5), and Major (92), respectively.
The comparison indicates that on Defects4J, LEAM tends to produce significantly more exact matches than the other approaches.
Interestingly, Major produces a similar number of exact matches with LLMut (CL-13b). $\mu$Bert yields significantly the fewest number of exact matches, indicating a minimal or no advantage for all these approaches (except for LLM-based approaches with the GPT models and LEAM) due to exact matches (in the case of Defects4J). Perhaps more interestingly, on the ConDefects dataset, which has not been seen by any of the tools, Major has the highest number of exact matches, indicating a minor influence of any data leakage on the reported results. Nevertheless, the LLMs we studied exhibit the same trend on both datasets, achieving a Spearman coefficient of 0.943 and a Pearson correlation of 0.944, both with a $p$-value less than 0.05, indicating their performance is similar on both datasets.

The different experimental settings may also threaten the validity of our results.
To address the threat, we explore the impacts of prompt design, context length, few-shot examples, and mutant numbers on the performance of LLMs. The results show that different settings are highly similar.
One potential threat to validity is that the metrics used in this study may not fully reflect the actual bug detection in practice.
Our evaluation is conducted on the fixed versions of programs, rather than the faulty versions where real-world testing would typically be applied, which may introduce potential differences.
In other words, our study operates under the \textit{clean program assumption}, which may not fully capture the dynamics of bug detection in practical scenarios.
However, to mitigate this threat, we use the widely adopted Defects4J 2.0 dataset, which consists of real bugs from well-maintained open-source projects, ensuring the relevance of our findings to real bugs.

The subjective nature of human decisions in labeling for equivalent mutants, non-compilation errors, and surviving mutants is another potential threat. To mitigate this threat, we follow a rigorous annotation process where two co-authors independently annotated each mutant. The final Cohen’s Kappa coefficient indicates a relatively high level of agreement between the two authors.

%% file: sections/related.tex
\section{Related Work} \label{sec:related}
In this section, we introduce the related work of our study. 

\subsection{Mutation Testing and Mutant Generation}
Mutation testing is a systematic testing approach that aims to guide the generation of high-utility test suites. It works by introducing syntactic modifications (i.e., \textit{mutants}) into the source code of the programs under test. By observing if test suites can detect these changes~\cite{hamlet1977testing,demillo1978hints}, one can identify weaknesses of their test suites, i.e., mutants that are not killed. Besides testing, mutation analysis is a tool that can support multiple applications in software engineering~\cite{shi2019mitigating,linares2017enabling}.
Most commonly, mutants are employed as substitutes for real-world bugs~\cite{just2014mutants,andrews2005mutation,daran1996software,namin2011use}, to guide fault localization~\cite{moon2014ask,papadakis2015metallaxis,wang2025systematic}, test prioritization~\cite{shin2019empirical} and program repair~\cite{hanna2025reinforcement,xiao2023expressapr,xiao2024accelerating}.

All these tasks require high-utility mutants. This is because it is required to compile and execute every mutant with tests, which is computationally expensive. Therefore, generating redundant or non-compilable mutants, as done by most of the existing learning-based approaches, increases the computational cost involved~\cite{zhang2016predictive,zhang2018predictive,wang2017faster,wang2021faster,garg2022cerebro}. Additionally, when mutants are used as substitutes for real defects, the mutants must be as syntactically close to the real bugs as possible, making the mutants more natural~\cite{hindle2016naturalness,jimenez2018mutants,hariri2019comparing}. Moreover, when performing tasks such as fault localization~\cite{moon2014ask,papadakis2015metallaxis} and program repair~\cite{le2011genprog}, the behavior of mutants must be coupled with real faults/bugs to make these techniques effective. 

Traditional mutation testing approaches generate mutants by pre-defined program transformation rules (i.e., \textit{mutation operators}), which are referred to as \textit{rule-based}.
For example, one widely used mutation operator is to change one arithmetic binary operator with another one (e.g., $a+b \mapsto a-b$).
During the last two decades, the majority of existing mutation testing approaches have been rule-based~\cite{jia2010analysis,amalfitano2022java}.
Researchers primarily focus on approaches in Java (e.g., PIT~\cite{coles2016pit}, Major~\cite{just2011major}, JavaLanche~\cite{schuler2009javalanche}, MuJava~\cite{ma2005mujava}, IBIR~\cite{khanfir2023ibir}, MIN ~\cite{zhang2025enriching}, etc.) and C/C++ (e.g., Proteum~\cite{delamaro1996proteum}, Milu~\cite{jia2008milu}, WinMut~\cite{wang2021faster}, etc.), leveraging the mature ecosystems and tooling chains of these languages for mutation testing.

Recently, with the emergence of machine learning techniques, researchers proposed using deep learning models to generate mutants.
For instance, DeepMutation~\cite{tufano2019learning} utilizes machine sequence-to-sequence neural translation methods, while LEAM~\cite{tian2022learning} employs learning-guided rule-based expansion.
LEAM++, the extension of LEAM, further explores learning-based mutation selection~\cite{tianleam++}.
However, these approaches suffer from challenges in generating syntactically correct mutants and mutants with diverse forms, often failing to produce syntactically valid mutants. Perhaps the closest to our work is that of  Degiovanni \emph{et al.}, who proposed $\mu$BERT~\cite{degiovanni2022mubert}. $\mu$BERT generates mutants by masking the code elements to be mutated and then predicting the mutants via BERT~\cite{devlin2018bert}.

Several recent studies have also explored the use of LLMs for mutant generation.
Tip \emph{et al.} concurrently investigated the use of LLMs for mutant generation in JavaScript~\cite{tip2025llmorpheus}, primarily focusing on the feasibility of leveraging LLMs to produce mutants. In contrast, our study conducts a broader comparison by evaluating LLMs against a diverse set of baselines, including both rule-based and learning-based approaches. Furthermore, we explore different prompt strategies, such as incorporating few-shot examples and enforcing structured JSON output, and we systematically examine multiple prompt templates in the context of mutation testing.
Similarly, Endres \emph{et al.} employed large models for mutant generation to aid deduction~\cite{endres2024can}, while Deb \emph{et al.} used GPT as one of the baselines in their approach~\cite{deb2024syntax}.
In contrast to their projects, our work systematically explores a variety of LLMs and prompt designs, offering a thorough comparison with state-of-the-art approaches.
Dakhel \emph{et al.}~\cite{dakhel2024effective} and Ibrahimzada \emph{et al.}~\cite{ibrahimzada2023automated} also proposed generating artificial bugs using LLMs.
These approaches mainly evaluate from a certain perspective, such as mutation score or program repair.
BugFarm~\cite{ibrahimzada2023automated} intends to generate artificial bugs that are hard to kill, by first choosing code locations based on a trained small model and then mutating the code via an LLM.
Compared with these approaches, our study is more comprehensive and evaluates existing mutant generation approaches from many aspects.
Specifically, we measure the behavioral similarity between real bugs and mutants using established metrics, evaluate the syntactic diversity of mutants, and analyze the non-compilable mutants produced by LLMs.
Our findings illustrate both the strengths and limitations of LLMs for mutant generation, and highlight directions for future improvement.

\subsection{LLMs for Code}
Recently, Large Language Models (LLMs) have shown remarkable potential in a wide range of natural language processing (NLP)~\cite{brown2020language} and coding tasks~\cite{wang2024software,fan2023large,schafer2023empirical}.
Existing LLMs are mostly based on Transformer~\cite{vaswani2017attention} and are trained by a large corpus of data, including texts and source codes.
LLMs are driven by instructions in natural languages, i.e., \textit{prompts}, which guide the models to perform target tasks and are crucial to performance.
The design of prompts plays a crucial role in obtaining accurate and useful responses from these models.

Closed-source LLMs are proprietary models developed and maintained by companies or organizations, often offering advanced features and requiring a subscription or payment for access, while open-source LLMs are freely available models that allow users to view and modify them.
For the models used in our study, GPT-3.5-Turbo, GPT-4o, and GPT-4o-Mini are closed-source LLMs released by OpenAI~\cite{achiam2023gpt}, while DeepSeek-Coder, StarChat, and CodeLlama are open-source.
An open-source LLM is usually fine-tuned against a base model for specified tasks.
DeepSeek-Coder~\cite{zhu2024deepseek} is based on DeepSeek.
CodeLlama~\cite{roziere2023code} is derived from Llama~\cite{touvron2023llama} by fine-tuning with extensive code datasets.
StarChat is built on StarCoder~\cite{li2023starcoder} to enhance conversational abilities.

%% file: sections/conclusion.tex
\section{Conclusion} \label{sec:conclusion}

In this paper, we systematically investigate the performance of existing mutant generation approaches across multiple dimensions, including their effectiveness, validity, and efficiency.
Our study conducts extensive and detailed comparative experiments, covering both widely studied rule-based mutant generation approaches and LLM-based methods.
Particularly, we evaluate not only existing LLM-based approaches but also our new prompt for mutant generation.
Our evaluation results highlight that LLMs significantly outperform traditional approaches in generating diverse s that closely mimic real bugs.
However, the results also expose key limitations of LLMs, particularly their tendency to generate a high proportion of non-compilable mutants.
Therefore, we analyze the error types of the LLM-generated non-compilable mutants and identify that member assessment and method invocation are more likely to lead LLMs to generate non-compilable mutants.
Based on our findings, we emphasize the need for further research to enhance LLM-based mutation testing, addressing its current shortcomings while leveraging its strengths.

Our implementation and the experimental data are available at our artifact: \textbf{\url{https://github.com/da-123-snake/Mutantion-Generation-Study}}.